\documentclass[a4paper,11pt]{article}

\usepackage{jinstpub} 

\usepackage{csquotes} 
\usepackage{amsmath} 		
\usepackage{multirow} 		
\usepackage{multicol} 		
\usepackage{caption} 		
\usepackage{bm} 			
\usepackage{siunitx} 		
\usepackage{makecell}
\graphicspath{{./plots/}}                

\newcommand{\flowOne}{\textsc{energy distribution flow}}
\newcommand{\flowTwo}{\textsc{causal flows}}

\title{\textsc{L$\boldsymbol{2}$LFlows}: Generating High-Fidelity $\boldsymbol{3}$D Calorimeter Images}

\author[a]{Sascha Diefenbacher}
\author[b]{Engin Eren}
\author[b,c]{Frank Gaede}
\author[a,c]{Gregor Kasieczka}
\author[d,e,1]{Claudius Krause\note{Corresponding author.}}
\author[a,f,g]{Imahn Shekhzadeh}
\author[d]{David Shih}
\affiliation[a]{Institut f\"ur Experimentalphysik, Universit\"at Hamburg, Germany}
\affiliation[b]{Deutsches Elektronen-Synchrotron DESY, Germany}
\affiliation[c]{Center for Data and Computing in Natural Sciences (CDCS), Hamburg, Germany}
\affiliation[d]{NHETC, Department of Physics \& Astronomy, Rutgers University, Piscataway, NJ USA}
\affiliation[e]{Institut f\"ur Theoretische Physik, Universit\"at Heidelberg, Germany}
\affiliation[f]{Computer Science Department, University of Geneva, Switzerland}
\affiliation[g]{Department of Information Systems, Geneva School of Business Administration (HES-SO Geneva), Switzerland}

\emailAdd{c.krause@thphys.uni-heidelberg.de}

\abstract{We explore the use of normalizing flows to emulate Monte Carlo detector simulations of photon showers in a high-granularity electromagnetic calorimeter prototype for the International Large Detector (ILD). Our proposed method --- which we refer to as \enquote{Layer-to-Layer Flows} (\textsc{L$2$LFlows}) --- is an evolution of the CaloFlow architecture adapted to a higher-dimensional setting ($30$ layers of $10\times 10$ voxels each). The main innovation of \textsc{L$2$LFlows} consists of introducing $30$ separate normalizing flows, one for each layer of the calorimeter, where each flow is conditioned on the previous five layers in order to learn the layer-to-layer correlations.  
We compare our results to the BIB-AE, a state-of-the-art generative network trained on the same dataset and find our model has a significantly improved fidelity.}


\arxivnumber{2302.11594}
\begin{document}
\maketitle
\flushbottom

%
\section{Introduction}
\label{sec:Intod}
In order to study Nature at the fundamental level and rigorously test the Standard Model (SM) of particle physics, current and future collider experiments need accurate and plentiful simulations of the detector response.
The most precise simulation toolkit in high-energy physics is \textsc{Geant4}  \cite{geant4_2003,geant4_update,geant4_update_2016}; however its precision comes at an enormous computational cost. The bulk of this cost is borne by the simulation of individual particle showers in the calorimeter --- so much so that it is currently a major bottleneck at the LHC, and is forecast to overwhelm the available computational resources without further R\&D~\cite{Collaboration:2802918,Software:2815292}. This has motivated, in recent years, a growing interest into using deep generative models as fast and accurate emulators of \textsc{Geant4} simulations~\cite{Paganini:2017hrr,Paganini_2018,deOliveira:2017rwa,Erdmann:2018kuh,Erdmann:2018jxd,Belayneh:2019vyx,Buhmann:2020pmy,ATL-SOFT-PUB-2020-006,Krause:2021ilc,Krause:2021wez,Buhmann:2021lxj,buhmann2021fast,Buhmann:2021caf,ATLAS:2021pzo,Mikuni:2022xry,ATLAS:2022jhk,Krause:2022jna,Cresswell:2022tof}. For this purpose, a variety of generative architectures, including generative adversarial networks (GANs)~\cite{Goodfellow:2014upx}, variational autoencoders (VAEs)~\cite{2013arXiv1312.6114K}, normalizing flows (NFs)~\cite{rezende2015variational}, and score-based generative models~\cite{https://doi.org/10.48550/arxiv.1907.05600,ho2020denoising,Song2021ScoreBasedGM} have been considered.\footnote{For a recent review see Ref.~\cite{hep_generative_models_review}.}

Especially NFs have shown promising fidelity when applied to the simulation of comparatively low-dimensional ($d\sim 500$) calorimeter datasets~\cite{Krause:2021ilc,Krause:2021wez,Krause:2022jna}. NFs are diffeomorphisms between the data space $\mathbb R^{d}$ and a latent space with a tractable distribution such as a Gaussian in $\mathbb R^{d}$. They are trained by minimizing the negative log-likelihood (NLL), which gives them a more meaningful loss function than the commonly used GANs or VAEs, and tends to result in higher-quality generated samples for calorimeter simulations. However, the drawback of NFs is that they are very memory-intensive, requiring many parameters to encode a sufficiently expressive invertible transformation between the data space and a latent space of the same dimensionality. This has made it challenging to generalize the CaloFlow approach of ~\cite{Krause:2021ilc,Krause:2021wez,Krause:2022jna} to higher-dimensional calorimeter datasets. Going to higher dimensionalities in calorimeter read-outs is desired to improve the accuracy of particle reconstructions and to aid in separating overlapping showers. This motivates future detector concepts such as the International Large Detector~(ILD)~\cite{ILD-TDR_V4,ILD-IDR}, which is one of the proposed detectors at the International Linear Collider~(ILC); or the CMS HGCAL at the HL-LHC \cite{hgcal_cms}. 

This work explores the steps needed to adapt NFs to a higher-dimensional calorimeter dataset, resulting in the new \textsc{L$2$LFlows} architecture\footnote{Our code is publicly available at \url{https://gitlab.com/Imahn/l2lflows} and~\cite{shekhzadeh_imahn_2023_8174200}. Our training data is available at~\cite{training_data}.}.
Although the methods we devise are fairly general and should have many potential future applications (including to datasets 2 and 3 of the CaloChallenge \cite{calo_challenge}, which are higher-dimensional than the considered dataset in this work), we will focus on using photon showers in an electromagnetic calorimeter (ECal) prototype for ILD as a testbed, for concreteness. These showers are simulated using \textsc{Geant4},
and projected to a regular grid of $30\times 30 \times 30$ voxels (there are $30$ layers in total, each layer having $30$ voxels in $x$- and $y$-direction respectively). 
Unlike previous works based on this dataset~\cite{Buhmann:2020pmy, Buhmann:2021lxj}, here we further reduce the transverse dimensionality to $10\times 10$ (by retaining only the central voxels in each layer) resulting in a $30\times 10 \times 10$ dataset shape. This is appropriate, since the particle impinges on the center of the $30\times 30\times 30$ cube. The reduction in dimsensionality is done to shorten the computation times needed for this first proof-of-concept demonstration of NFs for higher-dimensional calorimeter simulation. As we discuss further in Sec.~\ref{sec:conclusion}, we expect the generalization to the full $30\times 30\times 30$-dimensional dataset to be straightforward.

As in the CaloFlow~\cite{Krause:2021ilc,Krause:2021wez,Krause:2022jna} approach, we choose a two-step strategy for the architecture, where we generate the total energy depositions and the shower shapes in each layer separately. The first step --- total energy depositions per layer --- is extremely lightweight and essentially unchanged from original CaloFlow. We will refer to this first step as \flowOne\ (in \cite{Krause:2021ilc,Krause:2021wez,Krause:2022jna} it was called \enquote{Flow I}). 
The second step --- describing shower shapes in each layer --- is where we have innovated beyond the original CaloFlow algorithm. There, a second NF (called \enquote{Flow II}) was trained to generate the full shower across all layers, but in our experiments this did not generalize well to the higher-dimensional setting in terms of memory consumption. So instead, here we choose to train $30$ separate NFs, where each NF generates the shower in one specific calorimeter layer, but is \textit{conditioned on the voxel energies in the five previous layers}. We refer to this step as \flowTwo\ and this is the key innovation 
that allows us to generalize to higher-dimensional datasets. By splitting it into $30$ separate NFs, with conditioning from layer to layer, we keep both the memory requirements and the fidelity of the generated showers commensurate with original CaloFlow.

We will see that this new approach yields superior performance along several performance metrics compared to the state-of-the-art Bounded Information Bottleneck Autoencoder (BIB-AE)~\cite{Buhmann:2020pmy,Buhmann:2021lxj,Buhmann:2021caf} architecture. In addition, this approach generalizes naturally to more irregularly-shaped detector voxelizations and allows for parallel training on multiple GPUs. 

The structure of this paper is as follows:~In Sec.~\ref{sec:dataset}, the dataset is introduced in more detail; Sec.~\ref{sec:l2lflows} describes our architecture; Sec.~\ref{sec:results} shows our results; and finally Sec.~\ref{sec:conclusion} concludes and gives an outlook for future work.

\section{Dataset}\label{sec:dataset}

The ILD~\cite{ILD-TDR_V4,ILD-IDR} is one of two proposed detector concepts for ILC. As a modern detector concept, the ILD is specifically optimized for particle flow algorithms (PFAs)~\cite{Thomson_2009,CMS:2017yfk}, which aim at the correct reconstruction of every individual particle created in the event. One key requirement for PFA is a precise and highly granular set of hadronic and electromagnetic calorimeters.  

The ILD ECal that is used as the basis of this study is a sampling calorimeter with $30$ alternating layers with passive tungsten absorbers and active silicon sensors. The first 20 absorber layers have a thickness of $2.1~\textrm{mm}$ with the subsequent $10$ layers being twice as thick. Each silicon layer features individual cells with a size of $5 \times 5$ mm$^{2}$. 

We utilize a dataset containing $950$k photon showers simulated in the detailed and realistic detector model of ILD, implemented in DD4hep~\cite{dd4hep} in the iLCSoft~\cite{ilcsoft} framework. This dataset was used in past work on generative calorimeter simulation \cite{Buhmann:2020pmy}, where it is described in more detail. For this work, the $950$k showers are split into $760$k training, $95$k validation and $95$k test showers. These showers originate from photons with energies uniformly distributed between $10$ and $100$ GeV and were simulated using \textsc{Geant4} version $10.4$ (with the QGSP\_BERT physics list). All photons hit the calorimeter at the same position and perpendicularly to the calorimeter layers. The coordinate system used in this work defines the $z$-axis to be parallel to the trajectory of the photons, with the $xy$-plane being parallel to the calorimeter layers. For the classifier tests in Sec.~\ref{sec:classifier}, further $665$k independent test showers are available.  

In addition, for some comparison plots, we also use independent test sets containing $4$k showers with discrete energies between $20$ and $100$ GeV in $10$ GeV steps. These discrete incident energies will be used to study the linearity and width amongst other quantities. 

While previous generative projects on this dataset~\cite{Buhmann:2020pmy, Buhmann:2021lxj} used a data shape of $30 \times 30 \times 30$, this work focuses on the core of the showers located in a $10 \times 10$ cell region in the $xy$-plane around the impact point to reduce the computation times and the memory footprint of the generative models. The cores of the showers still contain $92\%$ of the shower energy. This results in a data shape of $30\times 10 \times 10$ where the first dimension indicates the depth along the propagation direction of the shower.

\section{\texorpdfstring{\textsc{L$2$LFlows}}{L2LFlows}} 
\label{sec:l2lflows}

Our approach uses NFs~\cite{rezende2015variational} to learn the probability density of showers in the calorimeter conditioned on the incident energy, $p(\text{shower}|\text{incident energy})$. NFs efficiently learn a change-of-variables transformation
\begin{align}\label{eq:cog_formula}
    p_X(x) = p_Z\left(f^{-1}(x)\right) \cdot \left\vert \det \frac{\partial f^{-1}(x)}{\partial x} \right\vert ,  
\end{align}
where $p_{X}$ and $p_{Z}$ are probability density functions in data and latent space, respectively, $f: \mathbb R^{d}\rightarrow \mathbb R^{d}$ is a diffeomorphism between the two spaces, $x = f(z)$ and $\partial f^{-1}/\partial x$ denotes the Jacobian matrix of $f^{-1}$. Since Eq.~\eqref{eq:cog_formula} gives us access to the negative log-likelihood (NLL) of data points, NFs can be trained by minimizing the NLL directly. 

In our case, $X$ will be the voxelized energy deposits of individual calorimeter showers and $Z$ a multivariate Gaussian distribution. In order to compute the Jacobian determinant efficiently, we use autoregressive transformations realized as masked autoregressive flows (MAFs)~\cite{2017arXiv170507057P}, built with rational quadratic splines (RQS) \cite{NEURIPS2019_7ac71d43,10.1093/imanum/2.2.123} and Masked Autoencoder for Distribution Estimation (MADE) blocks~\cite{2015arXiv150203509G}.  
The NFs are implemented with the help of the \textsc{nflows} package \cite{nflows} in \textsc{PyTorch} \cite{pytorch}. 

In the following subsections, we will describe the two parts of \textsc{L$2$LFlows}, the \flowOne\ and \flowTwo, which generate the layer energies and shower shapes respectively.

\subsection{\flowOne}\label{subsec:layer_flow}

The task of the \flowOne\ is to learn the total energy depositions per ECal layer\footnote{Note that we start counting the ECal layers from $0$ instead of $1$.}, which is described by the following conditional PDF:
\begin{align}\label{pdf_layer_flow}
    p(E_{0}, \dots, E_{29}\vert E_{\text{inc}}) , 
\end{align}
where $E_\text{inc}$ denotes the incident particle energy, and $E_{i}$ is the energy deposited by the shower in layer $i$, obtained by summing over all voxels in the given layer $i$. 

Within a sampling calorimeter, it is necessary to apply an energy threshold to account for the fact that calorimeters have inherent electronic noise, and thus depositions that are too small become unreliable. We, therefore, apply a cutoff to the individual voxel energies with a threshold of $10^{-4}$~GeV before calculating the layer energies $E_{i}$. This threshold corresponds to half the energy loss of a minimum-ionizing particle in the ILD ECal \cite{Buhmann:2020pmy}. 

The \flowOne\ is lightweight and, as such, does not present a bottleneck on computation times. Therefore, the model closely follows the original CaloFlow approach. The most noteworthy change is the modified preprocessing. Since the ILD ECAL is a sampling calorimeter, only a fraction of the energy of a particle is recorded. Thus, it was not necessary to enforce a strict energy upper limit  $\sum_i E_i \leq E_{\text{inc}}$ through preprocessing, as was the case in the original CaloFlow. Instead, we choose a simpler preprocessing, outlined in App.~\ref{new_pp_appendix}. 

\begin{figure}[t!]
    \centering
    \includegraphics[width=\textwidth]{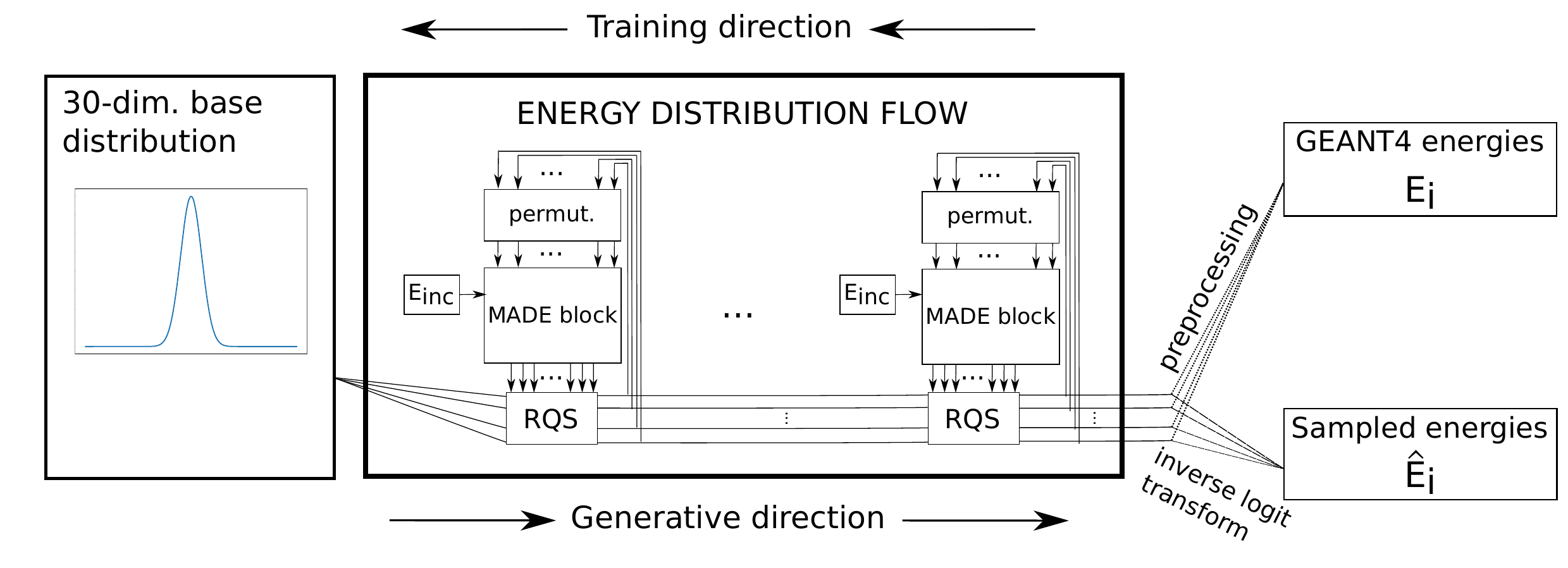}
    \caption{Architecture of the \flowOne.}
    \label{layer_flow_architec}
\end{figure}

The \flowOne\ architecture is shown in Fig.~\ref{layer_flow_architec} and details of the model and its training can be found in App.~\ref{appendix:models}. By construction, generation happens recursively over the dimensionality of MAFs. In total, the \flowOne\ has about $200$k parameters and was trained for $200$ epochs on a single NVIDIA\textsuperscript{\textregistered} V100\textsuperscript{\textregistered} with $32$ GB VRAM, which took less than $8$ hours. We subsequently use the validation NLL to select the best checkpoint among the $200$ epochs.

\subsection{\flowTwo}
\label{subsec:multiple_flow}

Next, we turn to the second step of the generation process: generating shower shapes conditioned on the total incident energy and the total deposited energies in each layer. Our overarching goal here, as in the original CaloFlow, is to learn
\begin{equation}\label{eq:flowIIgoal}
p({\mathcal I}_{0},\dots,{\mathcal I}_{29}|E_{0},\dots,E_{29},E_{\text{inc}})
\end{equation}
where the ECal voxel energy depositions of layer $i$ are denoted by $\mathcal I_{i}\in{\mathbb R}^{100}$. Unlike in Sec.~\ref{subsec:layer_flow}, no cutoff is applied to the voxel energy depositions used in the \flowTwo\ training.  This prevents potential sharp edges in the voxel data, which would be caused by the cutoff, from interfering with the training of the \flowTwo. (For the \flowOne, this issue was already circumvented, as each layer energy is the aggregate of multiple voxels, lessening any potential edges.) The voxel energy depositions are preprocessed similarly to the layer energies used in the \flowOne. The precise nature of the preprocessing is outlined in App.~\ref{new_pp_appendix}. 

In the original CaloFlow, a single NF was trained on all the calorimeter voxels of every layer together, to directly learn (\ref{eq:flowIIgoal}). Since 
the number of parameters of a single NF scales quadratically with the dimensionality $d$ of the samples, the single-NF approach of original CaloFlow applied to the ILD dataset (which has $d=3000$) would lead to a prohibitive number of parameters ($> 1$B). 
One can attempt to reduce the number of parameters by decreasing the number of MADE blocks as well as RQS bins, but this leads to a significantly reduced fidelity. 

To reduce the number of parameters without sacrificing quality, our key idea here is to instead train one NF per ECal layer. Since the evolution of a shower in layer $i$ depends on what happened in the previous layers, NF $i$ has to be conditioned on the voxel energy depositions of the previous layers. In other words, we endeavor to train $30$ separate NFs to learn the distributions:
\begin{equation}\label{eq:flowIIcond}
    p_i({\mathcal I}_i| {\mathcal I}_{0}, \dots, {\mathcal I}_{i-1}, E_{0}, \dots, E_{29}, E_{\text{inc}}),\qquad i=0,\dots,29
\end{equation}
If each distribution $p_i$ could be learned perfectly, then they could be multiplied together to reconstruct the full joint distribution (\ref{eq:flowIIgoal}). This would be in effect its own kind of autoregressive model. However, in later layers, there are a lot of conditioning features, and we observed that attempting to model the full conditional likelihood (\ref{eq:flowIIcond}) resulted in suboptimal performance. 

Instead, we found it beneficial to approximate the full conditional distribution (\ref{eq:flowIIcond}) with:
\begin{equation}\label{eq:flowIIcond_trunc}
    p_i({\mathcal I}_i|{\mathcal I}_{i-n_{\text{cond}}}, \dots, {\mathcal I}_{i-1}, E_{0}, \dots, E_{29},E_{\text{inc}})
\end{equation}
i.e.~to truncate the conditional at $n_{\text{cond}}$ previous layers. Due to the computational cost, a complete scan over $n_{\text{cond}}$ was not possible, but some small trials convinced us that $n_{\text{cond}} = 5$ gave a reasonably good balance between number of parameters and performance. 

In an effort to further reduce the number of parameters in the NFs, the models are not directly conditioned on the full previous layers. Instead, these layers are passed through an embedding network $G_{i}$:
\begin{equation}\label{pdf_flow_i_embed_full_cond}
    p_i({\mathcal I}_i|G_{i}({\mathcal I}_{i-n_{\text{cond}}}, \dots, {\mathcal I}_{i-1},E_{0}, \dots, E_{29}, E_{\text{inc}}))
\end{equation}
In an ablation study, we found the performace difference between conditioning on only $E_i$ and $E_{\text{inc}}$ and conditioning on all $E_{0}$, $\dots$, $E_{29}$ and $E_{\text{inc}}$ to be small, and hence for simplicity, we only condition on $E_i$ and $E_{\text{inc}}$; hence, our PDF from Eq.~\eqref{pdf_flow_i_embed_full_cond} simplifies to 
\begin{equation}\label{pdf_flow_i_embed}
    p_i({\mathcal I}_i|G_{i}({\mathcal I}_{i-n_{\text{cond}}}, \dots, {\mathcal I}_{i-1}, E_{i}, E_{\text{inc}})). 
\end{equation}
The embedding network $G_{i}$ takes in the context features ${\mathcal I}_{i-n_{\text{cond}}}$, $\dots$, ${\mathcal I}_{i-1}$, $E_{i}$, $E_{\text{inc}}$ and learns a representation of them that minimizes the NLL loss of the NFs. It is trained jointly with the NF, and there are different kinds of architectures one can consider, e.g.~a fully-connected network, a recurrent-neural network, etc. We use a fully-connected embedding network (having two hidden layers with $256$ and $128$ nodes each), with an output that is $64$-dimensional. With this embedding network, we observed no loss in performance, with a reduction of $1.8$M in the number of parameters when comparing Eq.~\eqref{pdf_flow_i_embed} with Eq.~\eqref{eq:flowIIcond_trunc}. Table \ref{context_multiple_flows_n_5} shows the context features for each NF. For NFs $0$ to $4$, there are less than $5$ preceding ECal layers, thus they have less than $502$ context features.\footnote{NF $0$, which learns the distribution of the voxel energies of layer $0$, does not use an embedding network, since it is only conditioned on $E_{0}$ and $E_{\text{inc}}$.} 

\begin{figure}[h!]
    \centering
    \includegraphics[width=\textwidth]{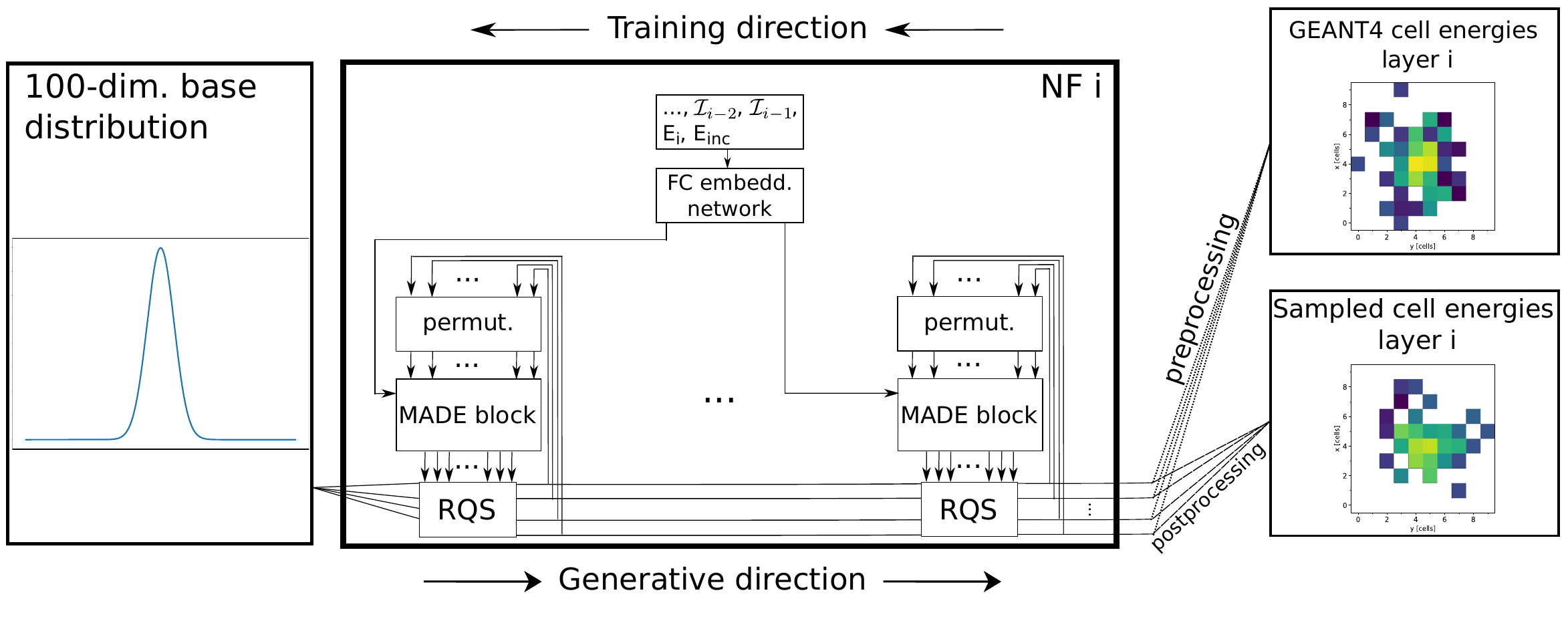}
    \caption{Architecture of the \textsc{\flowTwo}. As mentioned in the main text, NF $0$ does not make use of an embedding network for the conditioning. The postprocessing is
    explained in detail in App.~\ref{new_pp_appendix}.}
    \label{multiple_flows_architec_plot}
\end{figure}

Because of this conditioning scheme, generation happens recursively; however, training can happen in parallel on multiple GPUs, since all required context features are derived from the training data. For example, to generate the voxel energies in layer $2$, those of layers $0$ and $1$ must be generated first, and then NF $2$ is conditioned on the voxel energies of the previous layers as well as $E_{2}$ and $E_{\text{inc}}$. In this way, the whole calorimeter can be traversed. The architecture of the \flowTwo\ is visualized in Fig.~\ref{multiple_flows_architec_plot}.  A more detailed description of the model can again be found in App.~\ref{appendix:models}.

\begin{table}[h!]
\centering 
    \begin{tabular}{c|c|c}
	\hline
	NF $i$ & Context features & Context shape \\
	\hline 
	$0$ & $E_0$, $E_{\text{inc}}$ & [$N$, $2$] \\ \hline 
	$1$ & $\mathcal I_0$, $E_1$, $E_{\text{inc}}$ & [$N$, $102$]  \\ \hline 
	$2$ & $\mathcal I_0$, $\mathcal I_1$, $E_2$, $E_{\text{inc}}$ & [$N$, $202$]  \\ \hline
	$3$ & $\mathcal I_0$, $\mathcal I_1$, $\mathcal I_2$, $E_3$, $E_{\text{inc}}$ & [$N$, $302$]  \\ \hline
	$4$ & $\mathcal I_0$, $\mathcal I_1$, $\mathcal I_2$, $\mathcal I_3$, $E_4$, $E_{\text{inc}}$ & [$N$, $402$]  \\ \hline
	$\geq 5$ & $\mathcal I_{i-5}$, $\mathcal I_{i-4}$, $\mathcal I_{i-3}$, $\mathcal I_{i-2}$, $\mathcal I_{i-1}$,  $E_{i}$, $E_{\text{inc}}$ & [$N$, $502$]  
	\end{tabular}
\captionof{table}{For the conditioning on the previous $5$ ECal layers, i.e.~$n_{\text{cond}}=5$, this table shows the context features each NF gets and their shape before being fed into an embedding network. Here, $N$ denotes the batch size used during training or sampling.}
\label{context_multiple_flows_n_5}
\end{table}

During generation, it turns out that the conditioning on the $E_{i}$ is not sufficient to guarantee that the energies per layer by the sampled showers equal $E_{i}$. Hence, some postprocessing like rescaling to the $E_{i}$ of the \flowOne\ and a thresholding of low-energy voxels is necessary. We detail our method in App.~\ref{new_pp_appendix} and illustrate its effect in Fig.~\ref{fig:app:pp}. 

The \flowTwo\ have in total $44.8$M parameters, and they were trained on a single NVIDIA\textsuperscript{\textregistered} V100\textsuperscript{\textregistered} with $32$ GB VRAM for about $55$ GPU-hours. As for the \flowOne, we use the validation NLL to select the best checkpoint among the $200$ epochs. 

\section{Results}
\label{sec:results}

We now evaluate the performance of the \textsc{L$2$LFlows} approach. We benchmark it against a state-of-the-art shower generation model based on the BIB-AE framework, adapted from Ref.~\cite{Buhmann:2021caf} and modified to operate on the photon showers with shape $30\times10\times10$ by retraining it. The BIB-AE consists of an encoder and a decoder pair, which is trained using a set of adversarial critics. The BIB-AE generation process employs an additional post-processing step and a Kernel-Density-Estimation--based latent sampling, as described in Ref.~\cite{Buhmann:2021caf}. The BIB-AE model and PostProcessor model have a combined total of $9.3$M parameters, while the critics used to train them have an additional $3.7$M parameters. 

\subsection{Distributions}\label{subsec:distributions}

\begin{figure}[b!]
    \begin{minipage}[c]{0.31\textwidth}
        \centering 
        \includegraphics[width=\textwidth]{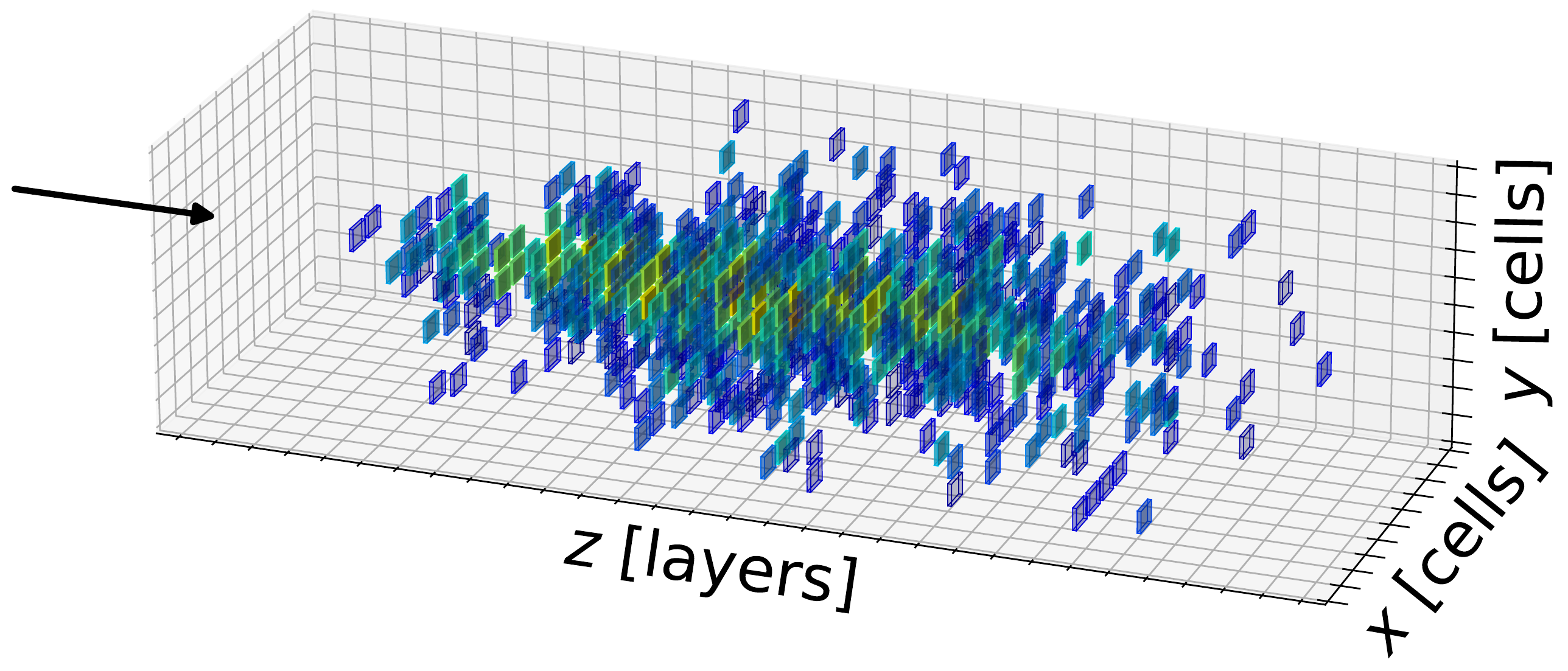}
    \end{minipage}
    \begin{minipage}[c]{0.31\textwidth}
        \centering
        \includegraphics[width=\textwidth]{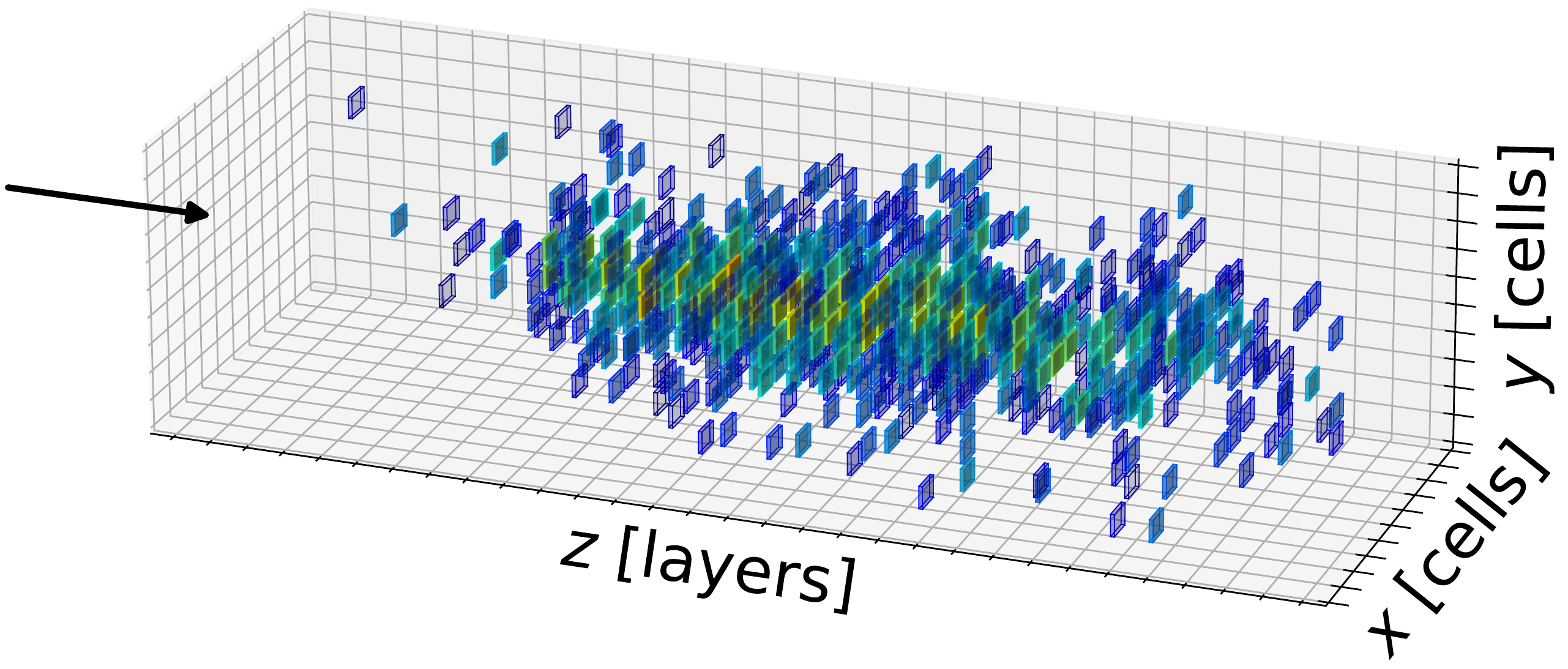}
    \end{minipage}
    \begin{minipage}[c]{0.37\textwidth}
        \centering 
        \includegraphics[width=\textwidth]{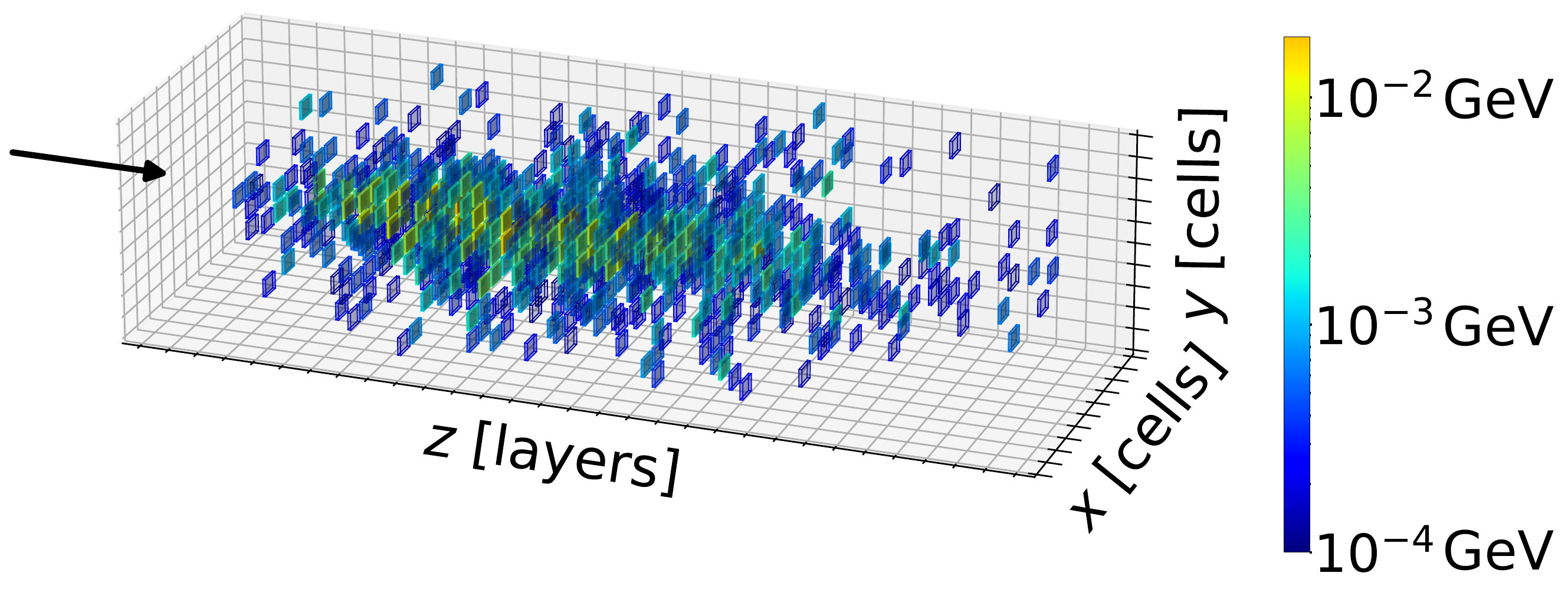}
    \end{minipage}
    \caption{BIB-AE--generated shower (left), \textsc{Geant4} test shower (middle) and \textsc{L$2$LFlows}-generated shower (right). The black arrow indicates the (hypothetical) direction of an incoming particle.}
    \label{results_single_showers}
\end{figure}

Figure \ref{results_single_showers} shows a single test shower of \textsc{Geant4} as well as a generated shower from the BIB-AE and \textsc{L$2$LFlows} each. All single showers have an incident energy $E_{\text{inc}} \approx 50$ GeV. We see that the individual shower from \textsc{L$2$LFlows} looks reasonable, with a broadly realistic morphology of voxels and energy depositions. 

\begin{figure}[t!]
    \begin{minipage}[c]{0.33\textwidth}
        \centering 
        \includegraphics[width=\textwidth]{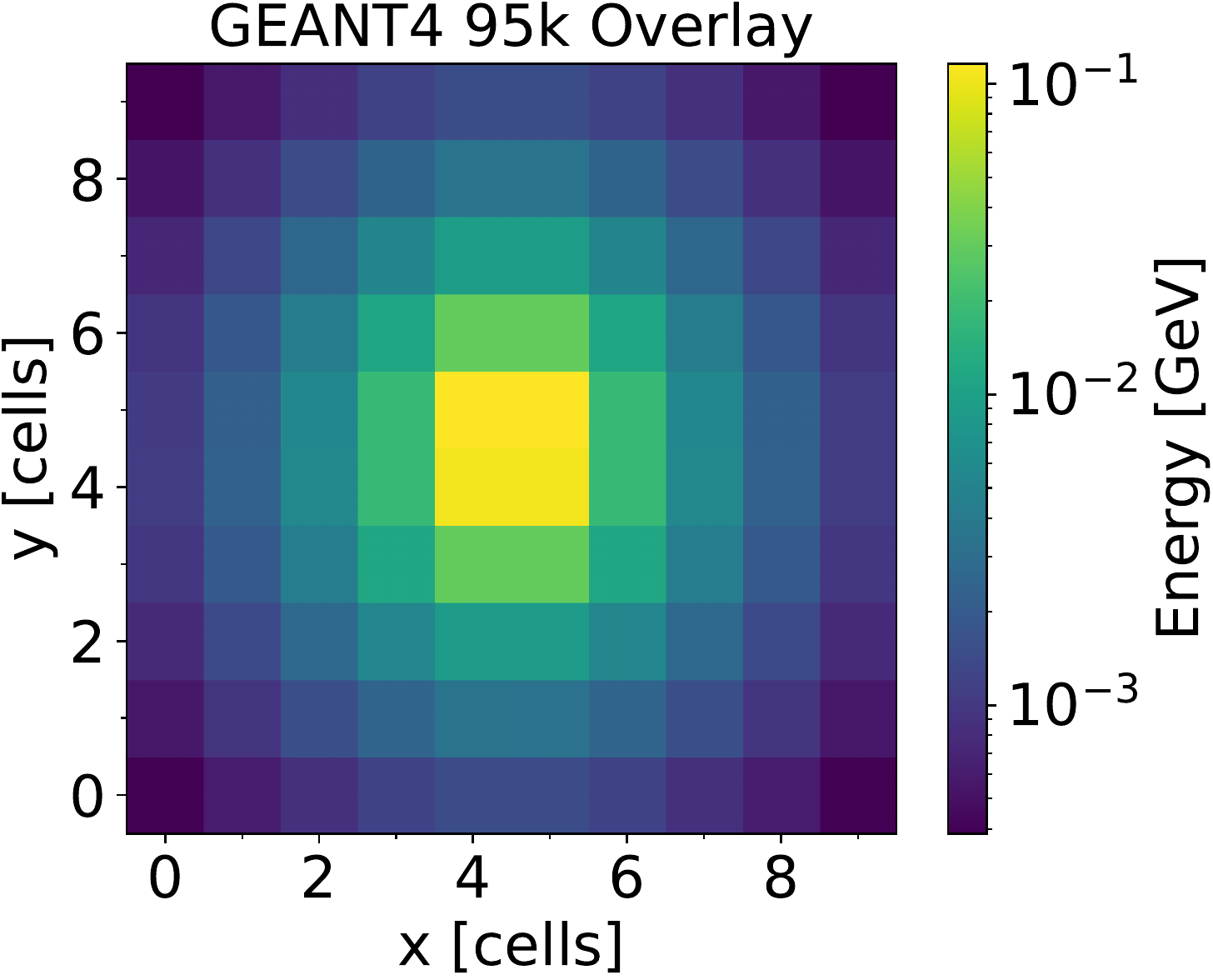} 
    \end{minipage}
    \begin{minipage}[c]{0.33\textwidth}
        \centering 
        \includegraphics[width=0.79\textwidth]{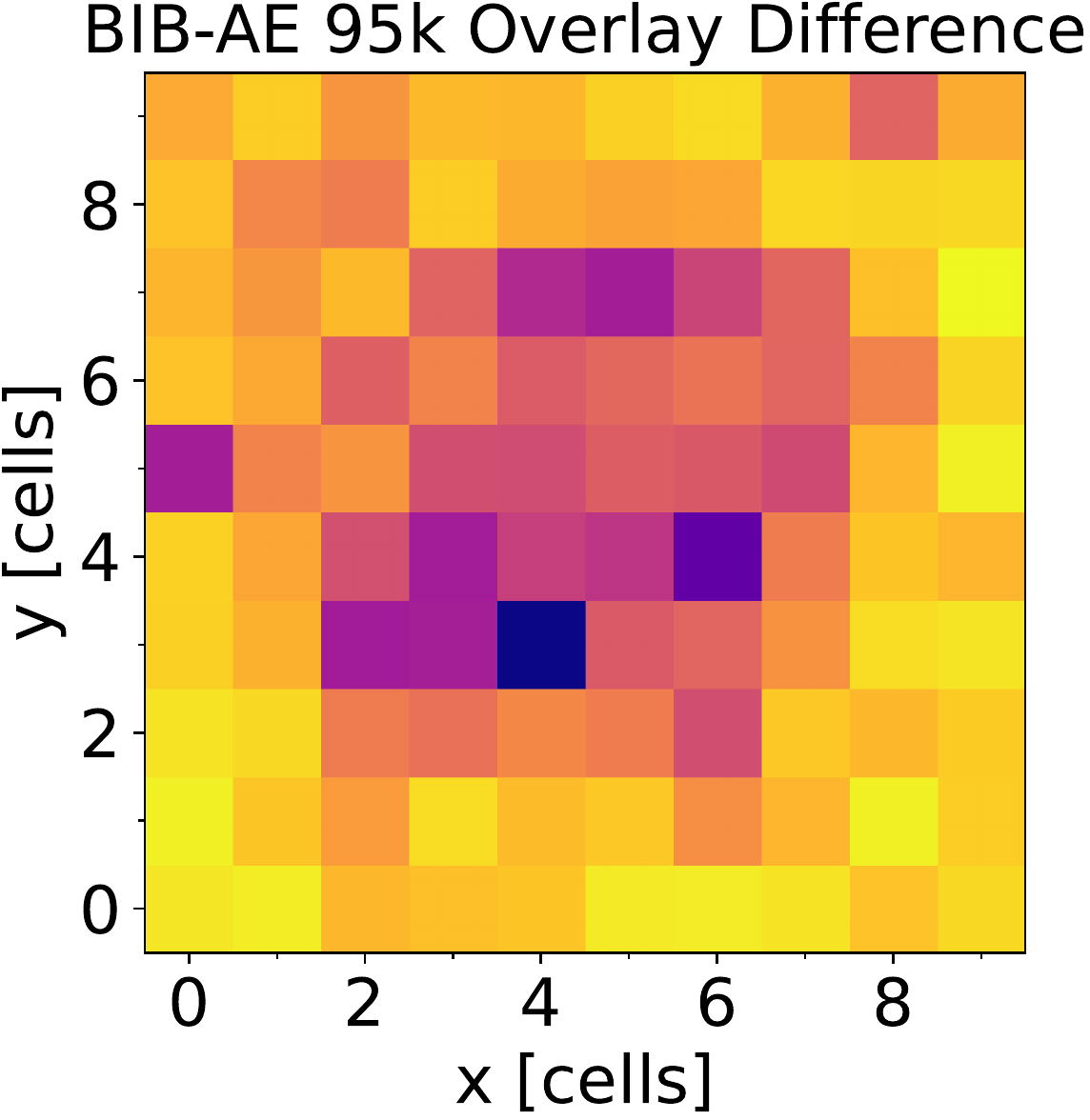} 
    \end{minipage}
    \begin{minipage}[c]{0.33\textwidth}
        \centering 
        \includegraphics[width=1\textwidth]{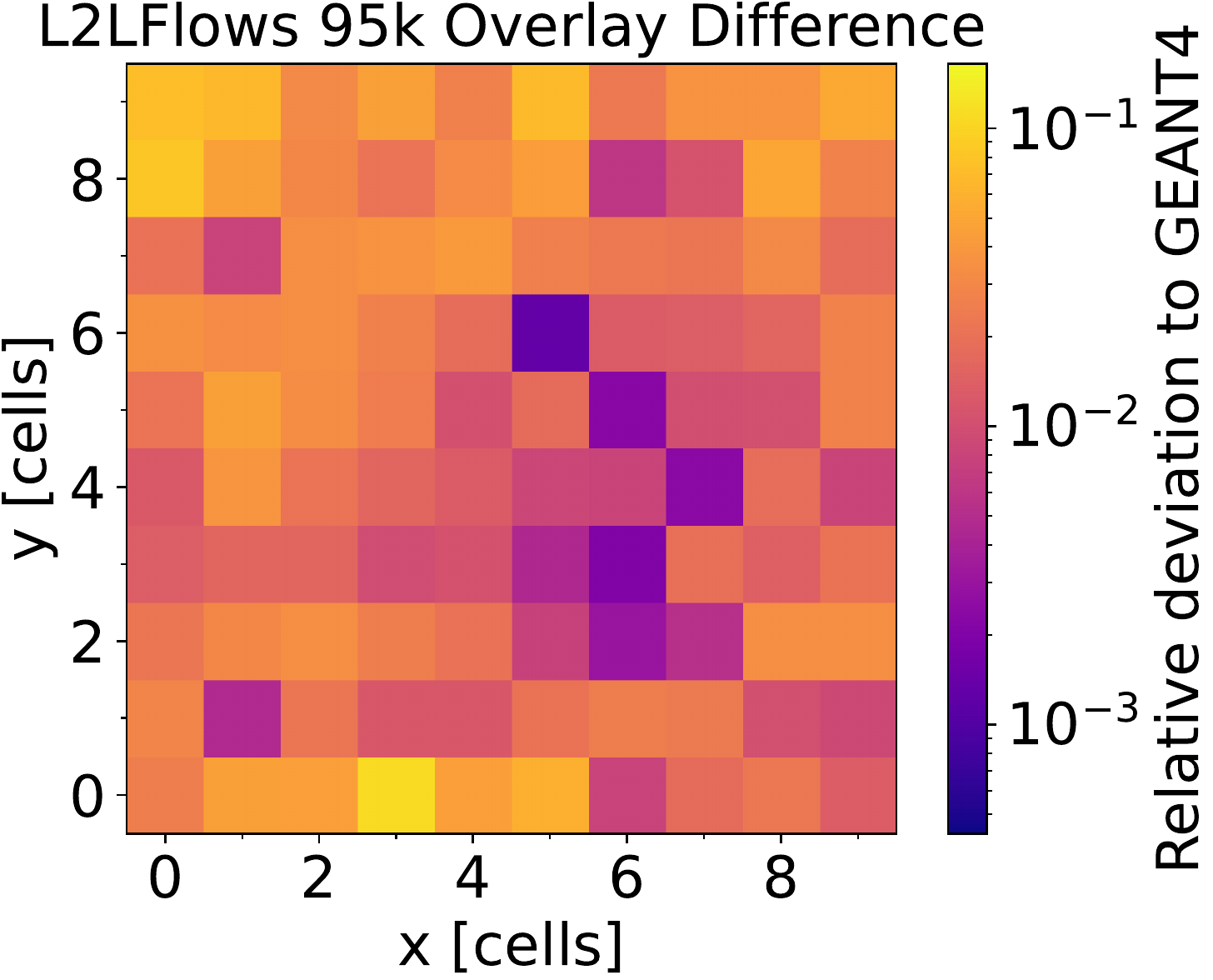} 
    \end{minipage}\vspace{0.5cm}
    \begin{minipage}[c]{0.33\textwidth}
        \centering 
        \includegraphics[width=0.8\textwidth]{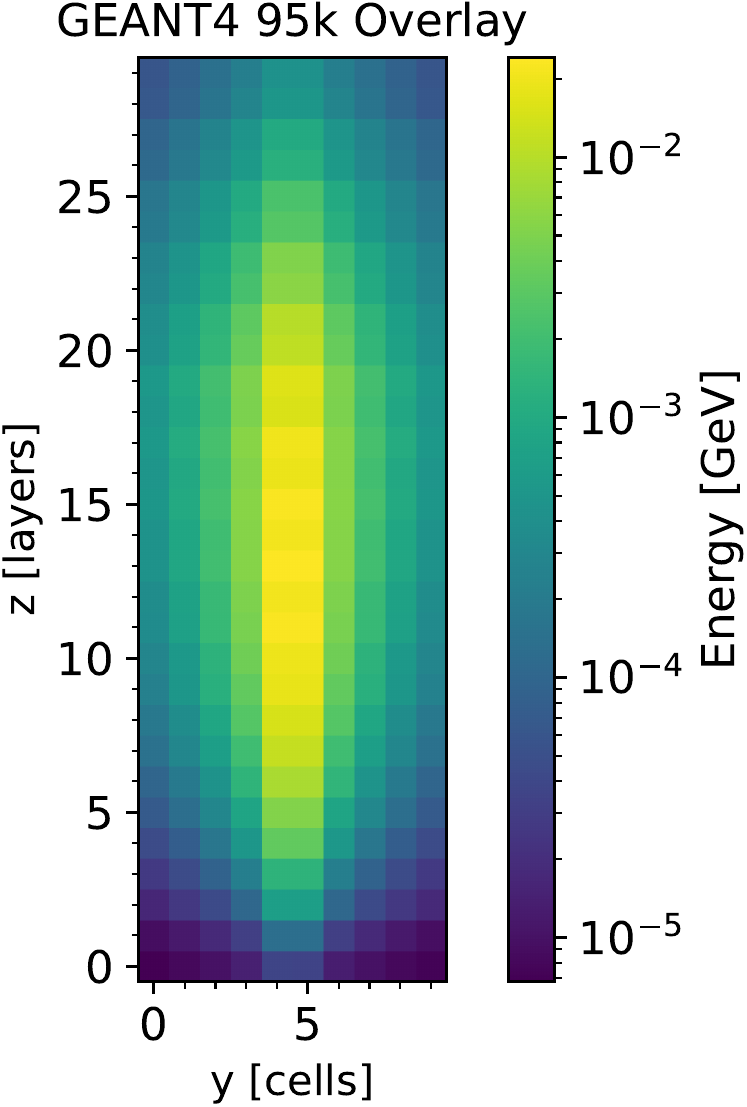} 
    \end{minipage}
    \begin{minipage}[c]{0.33\textwidth}
        \centering 
        \includegraphics[width=0.75\textwidth]{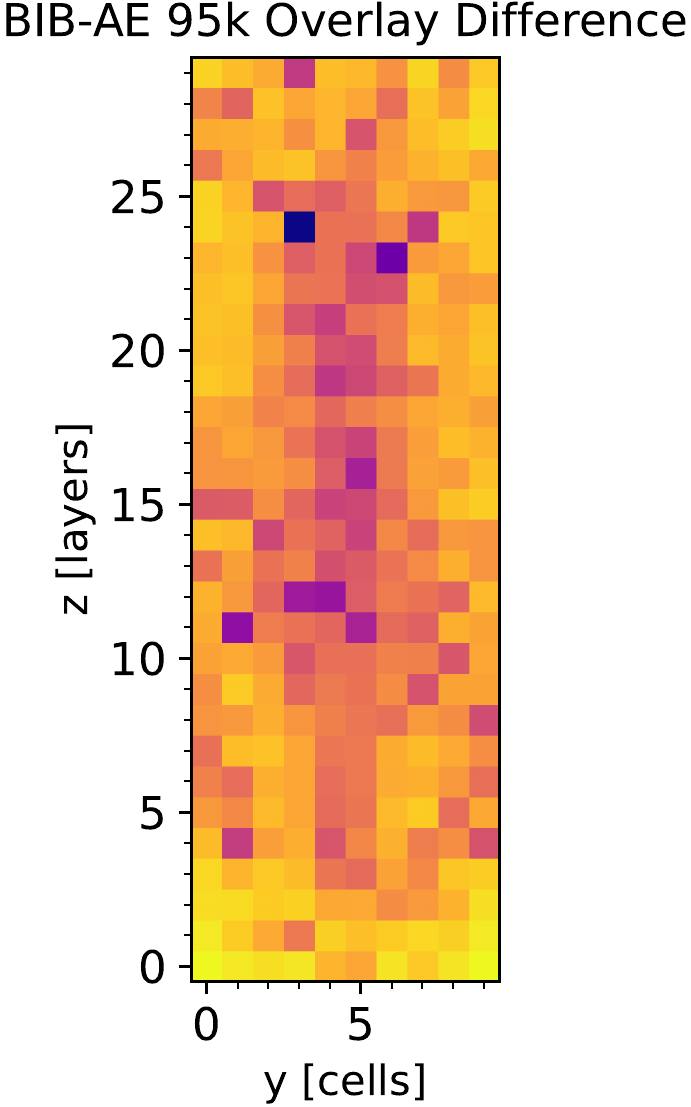} 
    \end{minipage}
    \begin{minipage}[c]{0.33\textwidth}
        \centering 
        \includegraphics[width=0.9\textwidth]{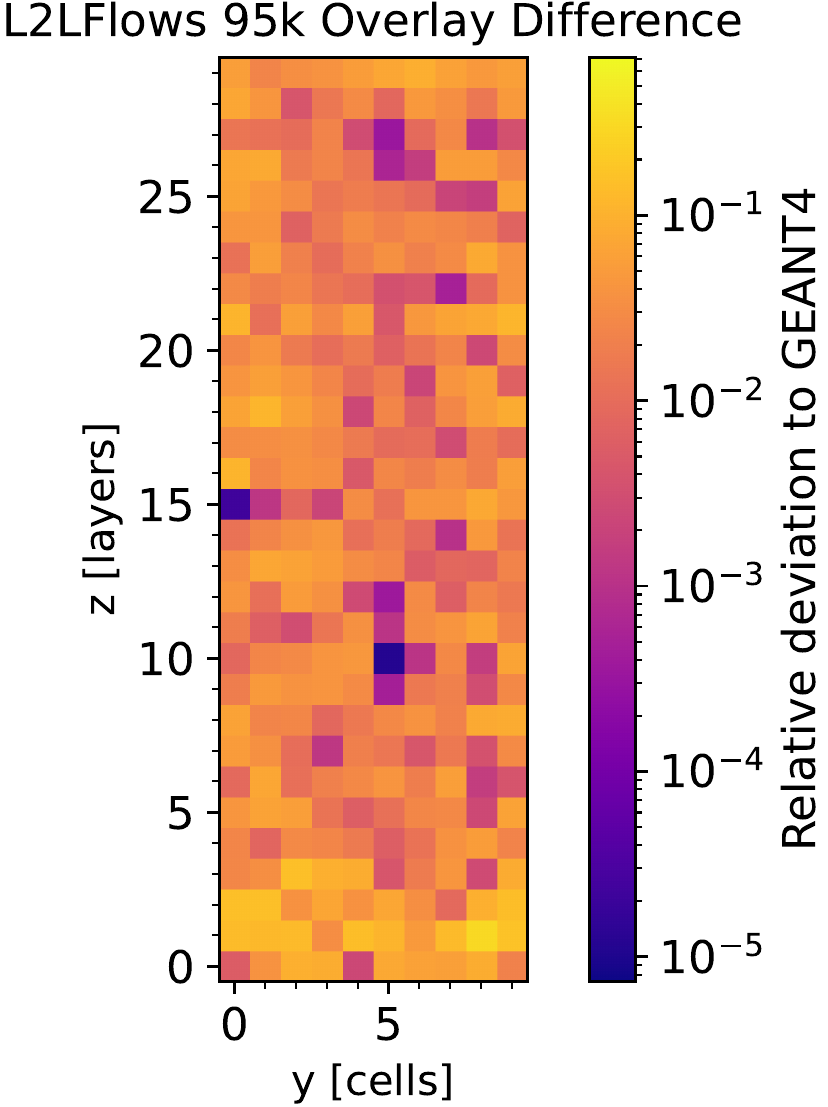}
    \end{minipage}\vspace{0.5cm}
    \begin{minipage}[c]{0.33\textwidth}
        \centering 
        \includegraphics[width=0.8\textwidth]{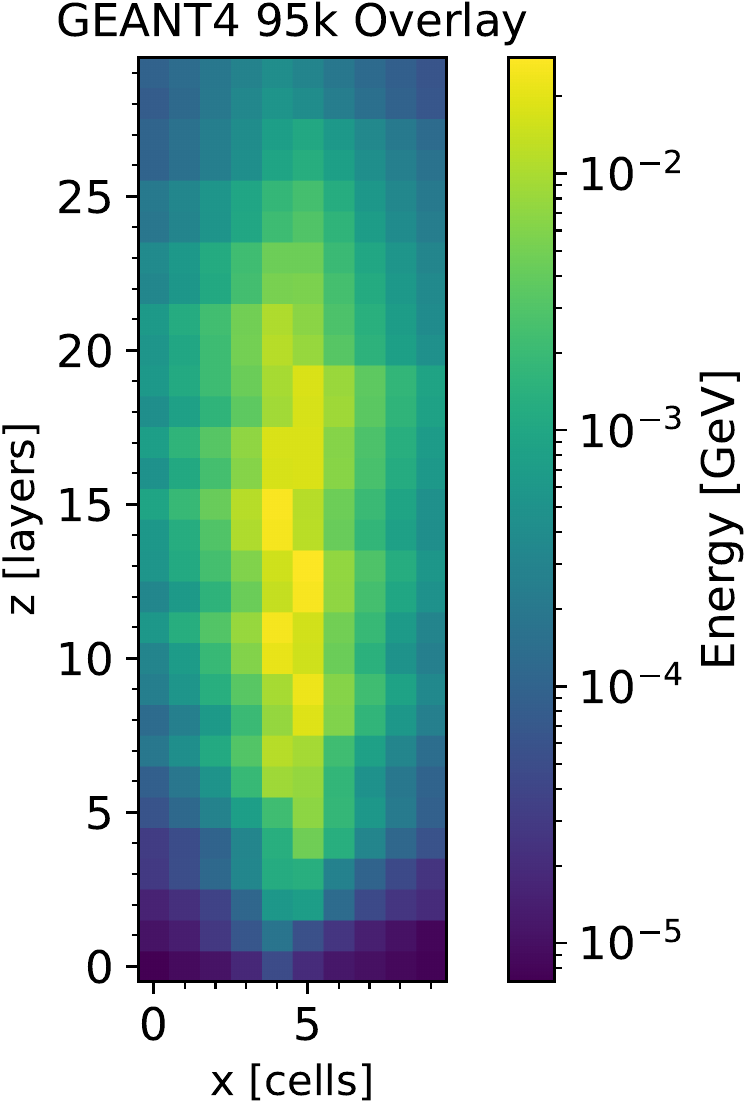}
    \end{minipage}
    \begin{minipage}[c]{0.33\textwidth}
        \centering 
        \includegraphics[width=0.75\textwidth]{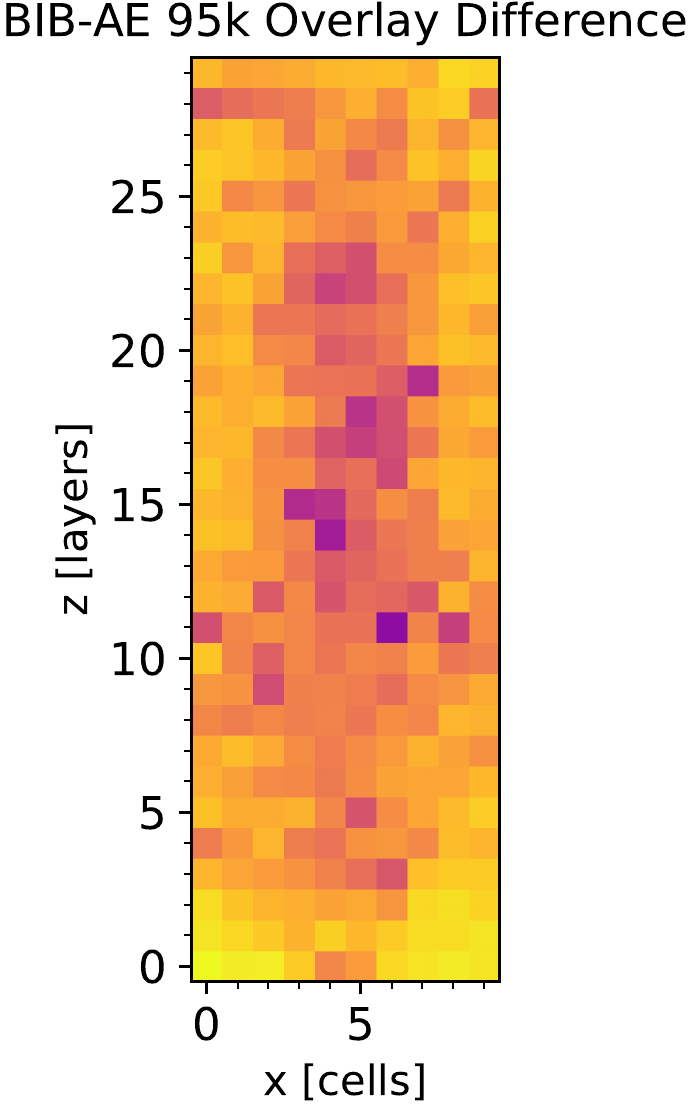}
    \end{minipage}
    \begin{minipage}[c]{0.33\textwidth}
        \centering 
        \includegraphics[width=0.9\textwidth]{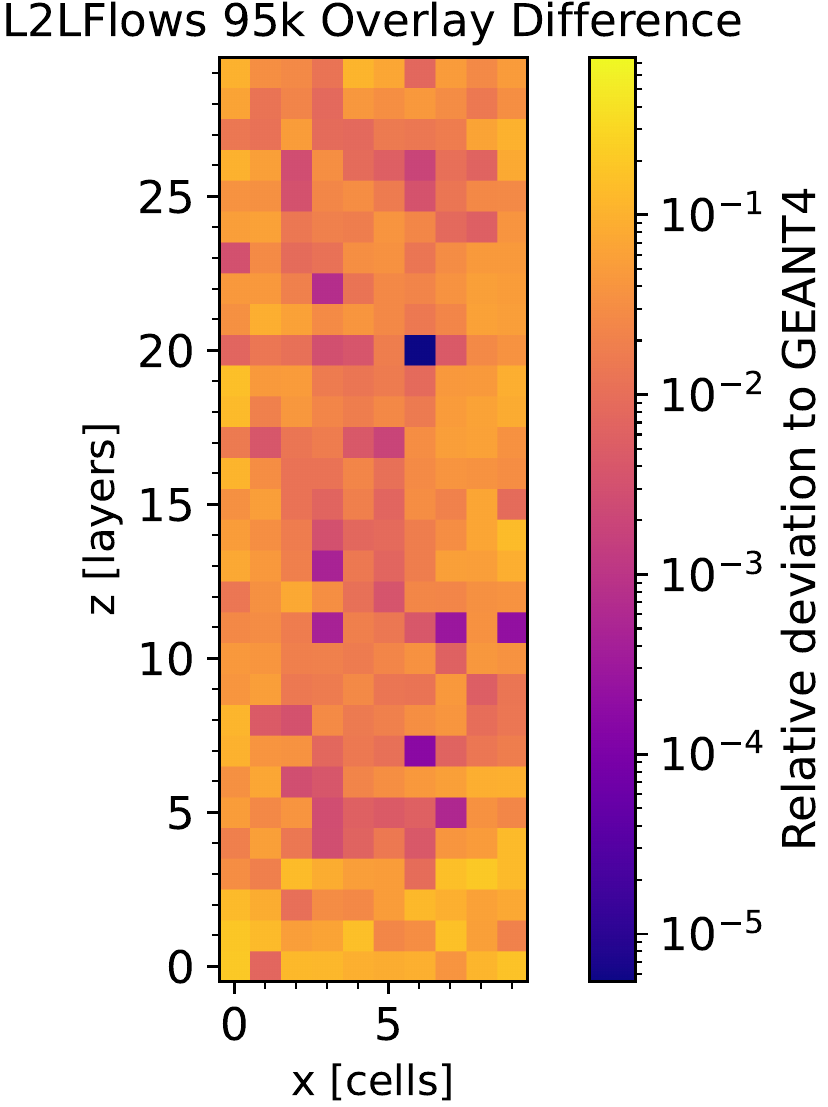}
    \end{minipage}
    \caption{Overlay of $95$k showers for all simulators for the full spectrum, where the voxel energies are summed along the $z$- (top), $x$- (middle) and $y$-axis (bottom). In all plots, the mean over the number of showers is taken. For \textsc{Geant4}, the shown colormap is the energy scale, whereas for the BIB-AE and \textsc{L$2$LFlows}, the colormap (both generative networks make use of the same one) corresponds to the relative deviations to \textsc{Geant4}, defined in Eqs.~\ref{eq:L2Lflowsrel} and \ref{eq:BIBAErel}.}
    \label{overlay_proj_ax_1}
\end{figure}

Figure \ref{overlay_proj_ax_1} shows the \textit{overlay} of $95$k showers, i.e.~the mean of the voxel energies of $95$k showers. In order to create two-dimensional plots, the voxel energies are summed over the $z$-, $x$- or $y$-axis. For \textsc{Geant4}, the $95$k test showers are used. To highlight potential differences for the BIB-AE and \textsc{L$2$LFlows}, we show the absolute relative deviation to \textsc{Geant4} for both generative networks per voxel:
\begin{align}
\label{eq:L2Lflowsrel}
    \text{\textsc{L$2$LFlows}}^{\text{relative}}_{i, j} &:= \frac{\left\vert\text{\textsc{L$2$LFlows}}^{\text{overlay}}_{i, j} - \textsc{Geant4}^{\text{overlay}}_{i, j}\right\vert}{\textsc{Geant4}^{\text{overlay}}_{i, j}}, 
                                \\ 
\label{eq:BIBAErel}
    \text{BIB-AE}^{\text{relative}}_{i, j} &:= \frac{\left\vert\text{BIB-AE}^{\text{overlay}}_{i, j} - \textsc{Geant4}^{\text{overlay}}_{i, j}\right\vert}{\textsc{Geant4}^{\text{overlay}}_{i, j}}, 
\end{align} 
where $i$ and $j$ denote voxel positions. We observe that in general the generative models capture the overlay quite well, with \textsc{L$2$LFlows} having smaller deviations from \textsc{Geant4} than the BIB-AE.

\begin{figure}[t!]
    \begin{minipage}[c]{0.49\textwidth} 
        \centering 
        \includegraphics[width=\textwidth]{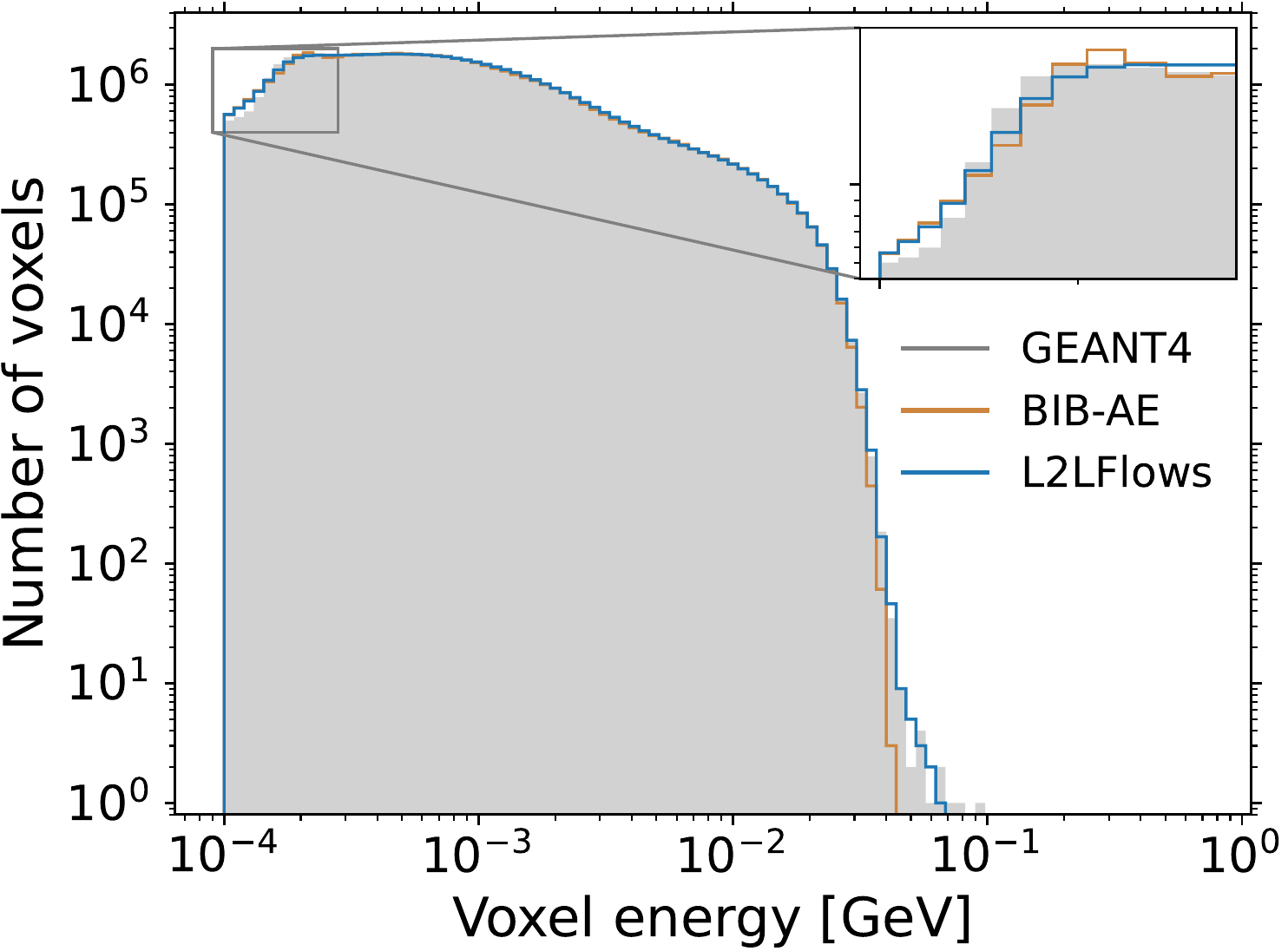}
    \end{minipage}\hspace{0.1cm}
    \begin{minipage}[c]{0.49\textwidth} 
        \centering 
        \includegraphics[width=\textwidth]{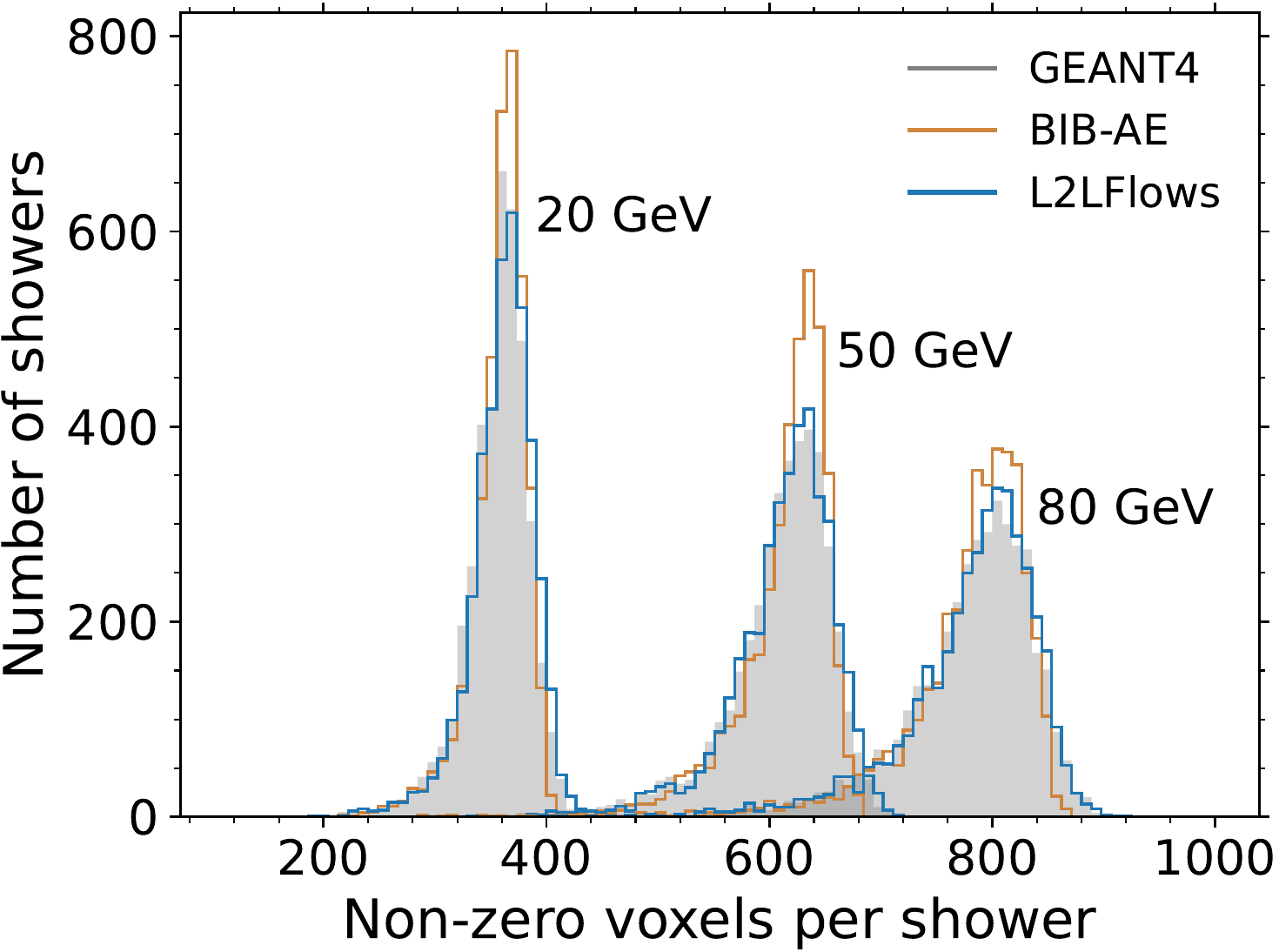}
    \end{minipage}
    \caption{Distributions comparing \textsc{Geant4} (grey), the BIB-AE (orange) and \textsc{L$2$LFlows} (blue). Left:~Distribution of voxel energies with shower incident energies uniformly distributed between $10$ and $100$ GeV, based on $95$k showers for every model. Right:~Number of voxels above half the MIP cutoff for $4$k showers of $20$, $50$, and $80$ GeV photons each for every model.}
    \label{fig:voxel_dist}
\end{figure}

To compare the performance of the generative models in more detail, we start by looking at the showers on the voxel level. Figure~\ref{fig:voxel_dist} shows the distributions of voxel energies as well as the sparsity, i.e.~the number of non-zero voxels per shower. One characteristic that repeats itself in several histograms is that the BIB-AE is not capable of capturing the full \textsc{Geant4} distribution, which can e.g.~be seen in the sparsity plot. \textsc{L$2$LFlows} is much better in this regard. Further, the energy deposited around
the energy of a minimum ionizing particle (MIP)
in the voxel distribution is better modeled by \textsc{L$2$LFlows} in comparison to the BIB-AE, which slightly overshoots it. While \textsc{L$2$LFlows} does not learn the \textsc{Geant4} distribution perfectly, it learns the distributions much better than the BIB-AE. 

\begin{figure}[h!]
    \begin{minipage}[c]{0.49\textwidth}
        \centering 
        \includegraphics[width=\textwidth]{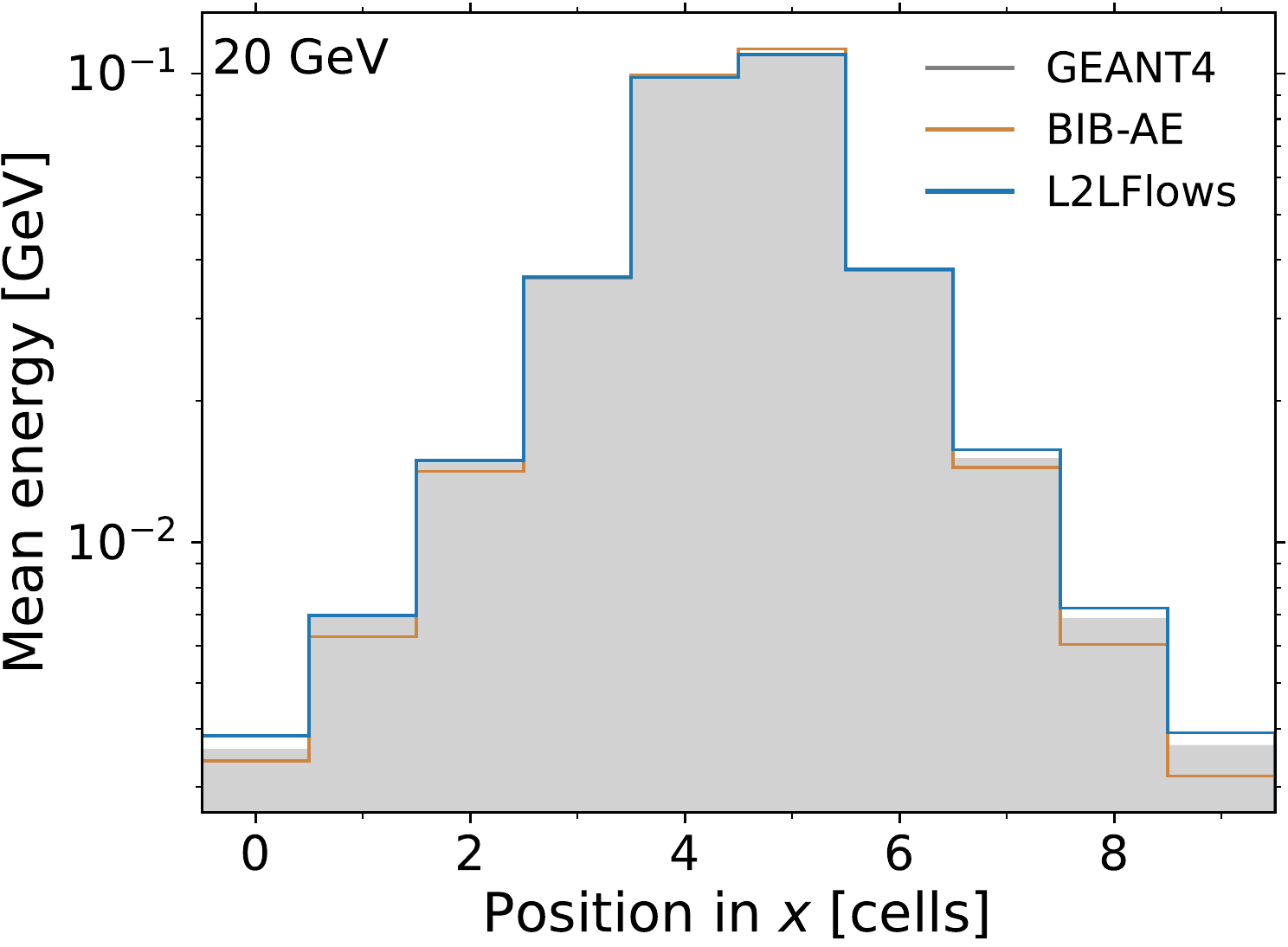}
    \end{minipage}
    \begin{minipage}[c]{0.49\textwidth}
        \centering 
        \includegraphics[width=\textwidth]{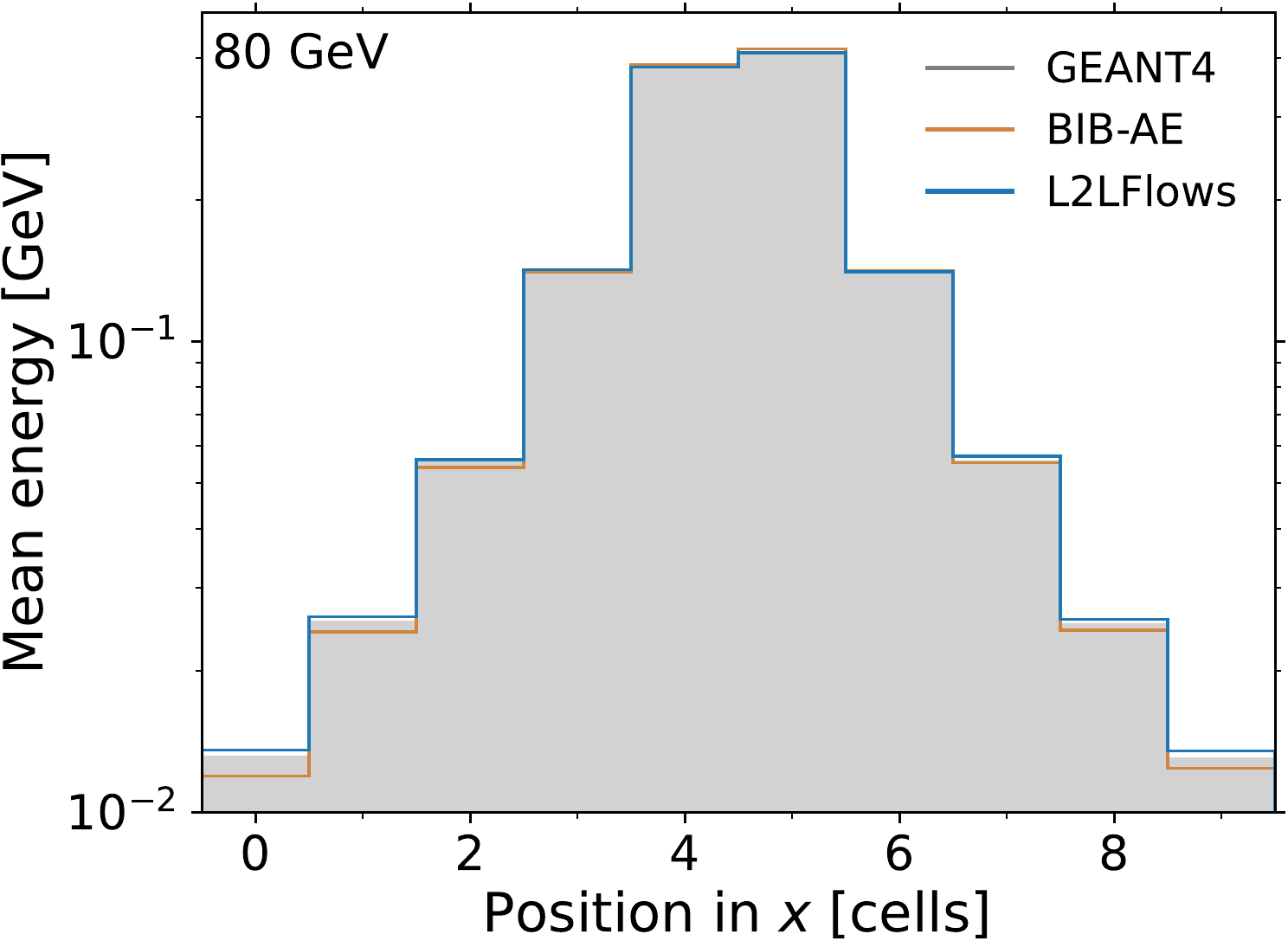}
    \end{minipage}\vspace{0.5cm}
    \begin{minipage}[c]{0.49\textwidth}
        \centering 
        \includegraphics[width=\textwidth]{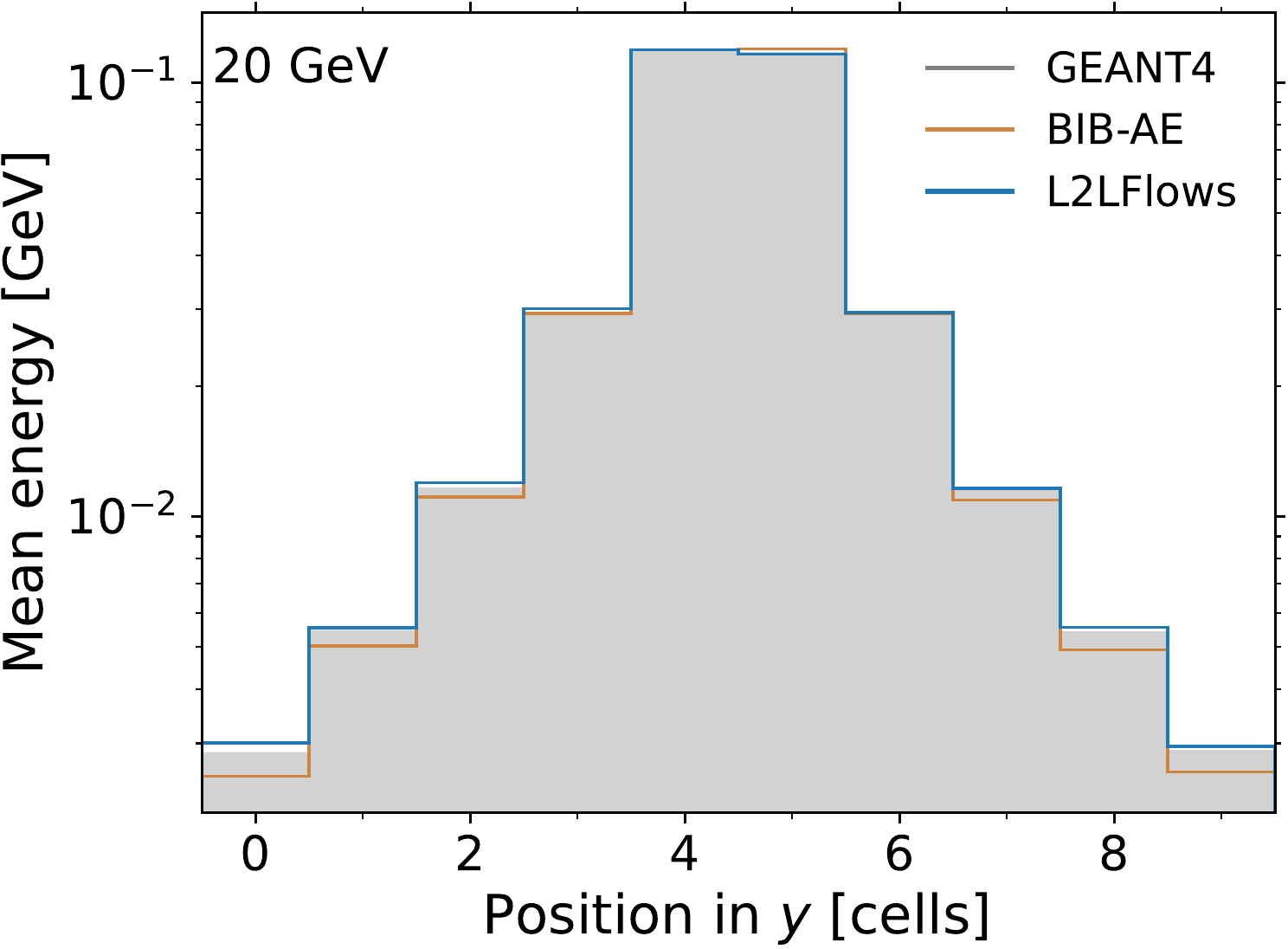}
    \end{minipage}
    \begin{minipage}[c]{0.49\textwidth}
        \centering 
        \includegraphics[width=\textwidth]{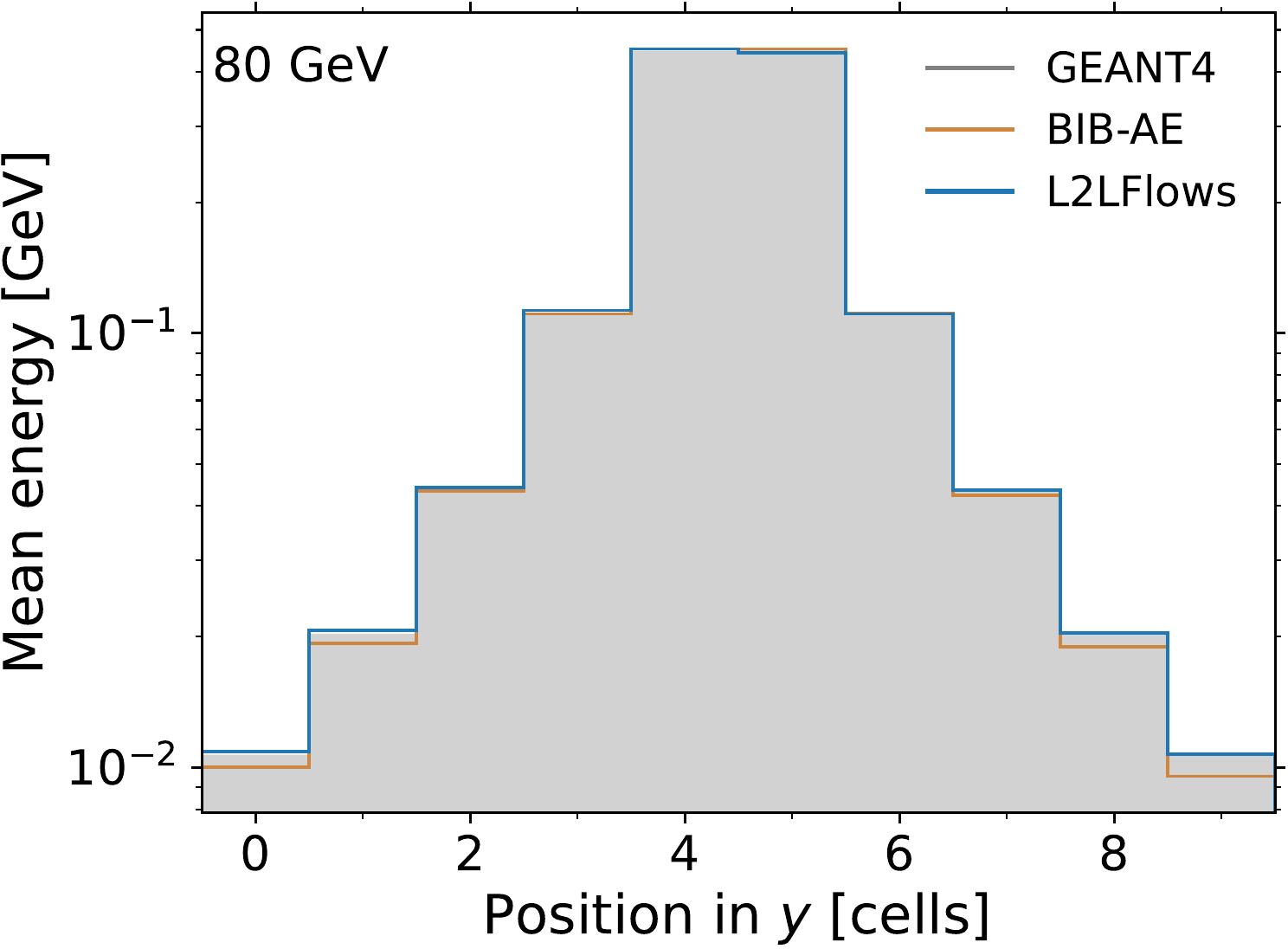}
    \end{minipage}\vspace{0.5cm}
    \begin{minipage}[c]{0.49\textwidth} 
        \centering 
        \includegraphics[width=\textwidth]{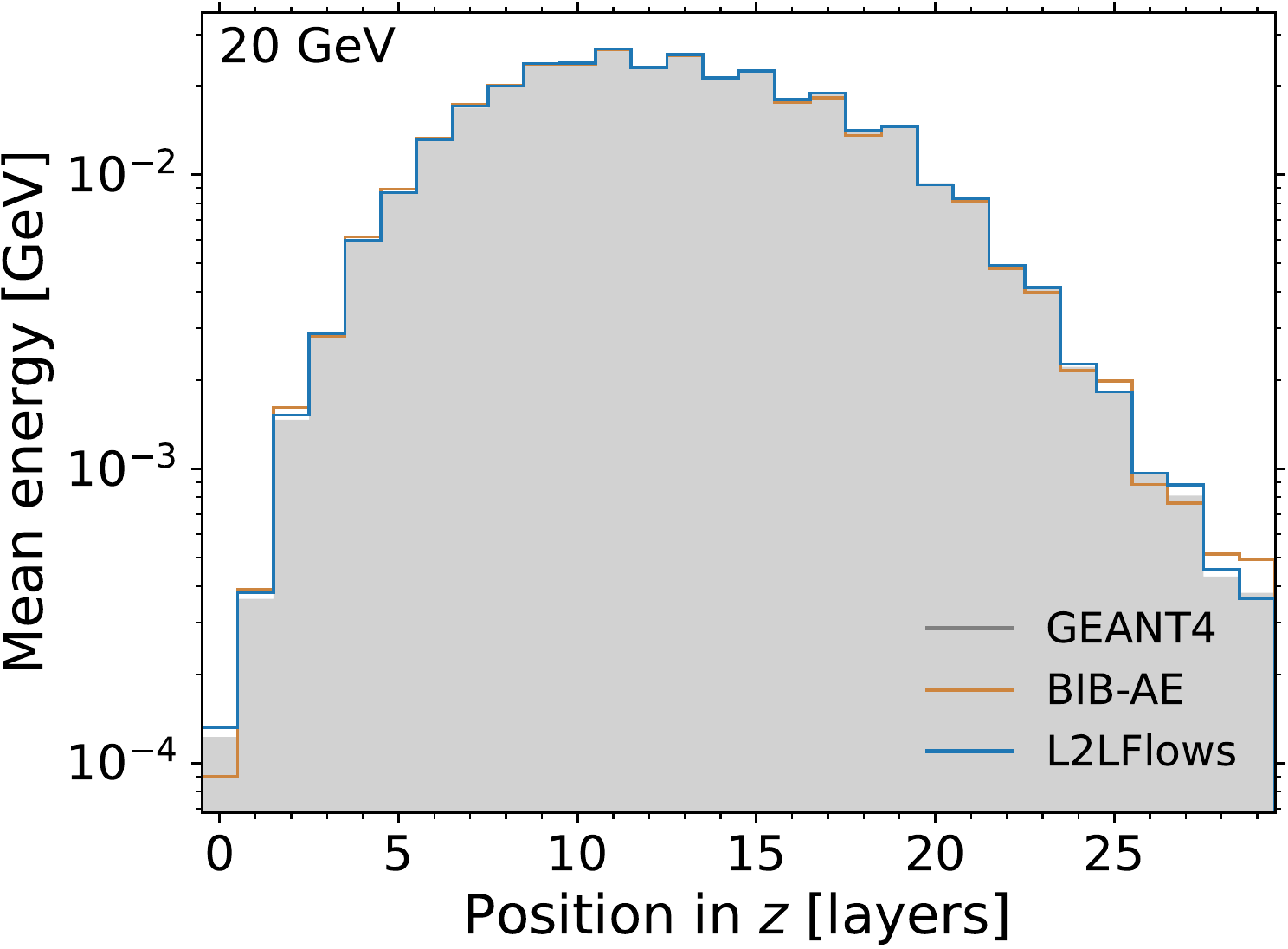}
    \end{minipage}\hspace{0.1cm}
    \begin{minipage}[c]{0.49\textwidth}
        \centering 
        \includegraphics[width=\textwidth]{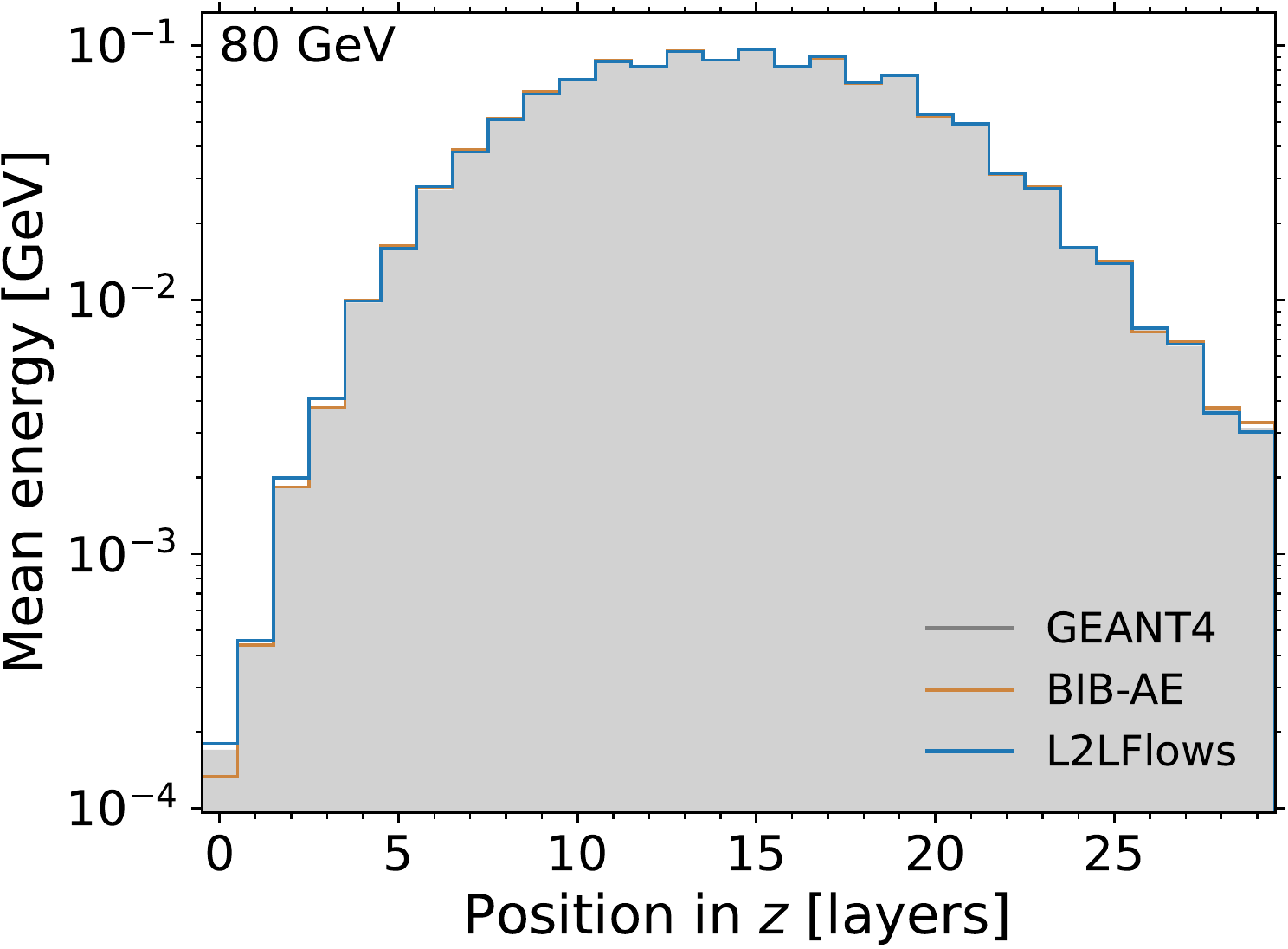}
    \end{minipage}
    \caption{Comparisons of the energy profiles of \textsc{Geant4} (grey), the BIB-AE (orange) and \textsc{L$2$LFlows} (blue). The upper row of plots shows the energy profiles in the $x$-direction, the center row shows the profile along the $y$-direction, and the lower row shows the profile along the $z$-direction. The left-hand side plot shows the profile for showers caused by $20$ GeV photons, and the right-hand side shows the profile for $80$ GeV photons, covering both the high- and low-energy regions of the data set. At each discrete energy, \textsc{Geant4}, the BIB-AE and \textsc{L$2$LFlows} are plotted with $4$k samples each.}
    \label{results:energy_flow_hists_discrete}
\end{figure}

For $E_{\text{inc}} \in \{20, 80\}$ GeV, Fig.~\ref{results:energy_flow_hists_discrete} shows the energy profiles in $x$-, $y$- and $z$-direction. As can be seen, the larger the incident energy $E_{\text{inc}}$, the more the maximum in the energy profiles shifts to later layers, which both the BIB-AE and \textsc{L$2$LFlows} are able to learn. Deviations for both simulators mainly exist in a few initial and final layers. 

\begin{figure}[t!]
    \begin{minipage}[c]{0.47\textwidth} 
        \centering 
        \includegraphics[width=\textwidth]{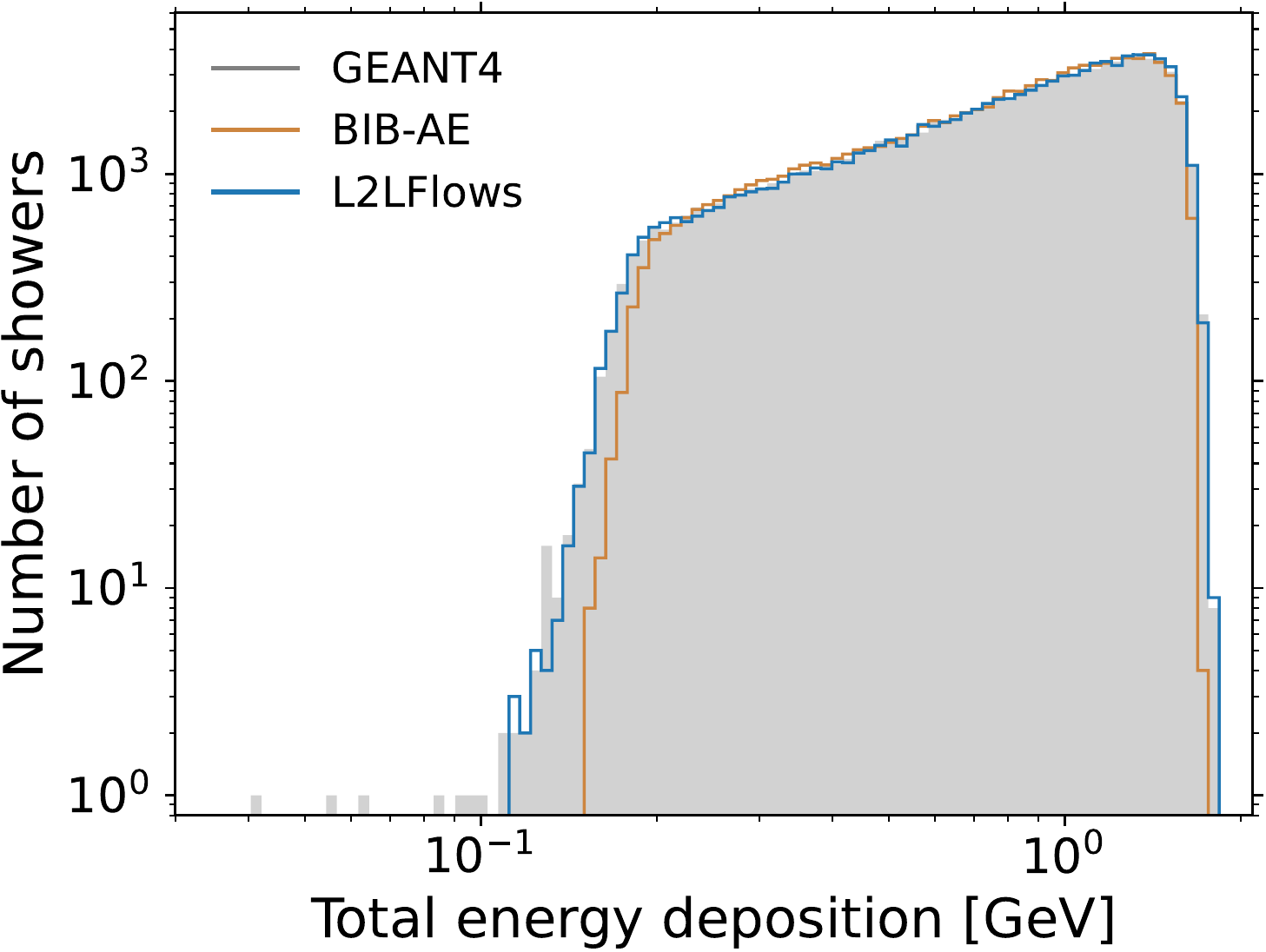}
    \end{minipage}\hspace{0.1cm}
    \begin{minipage}[c]{0.47\textwidth} 
        \centering 
        \includegraphics[width=\textwidth]{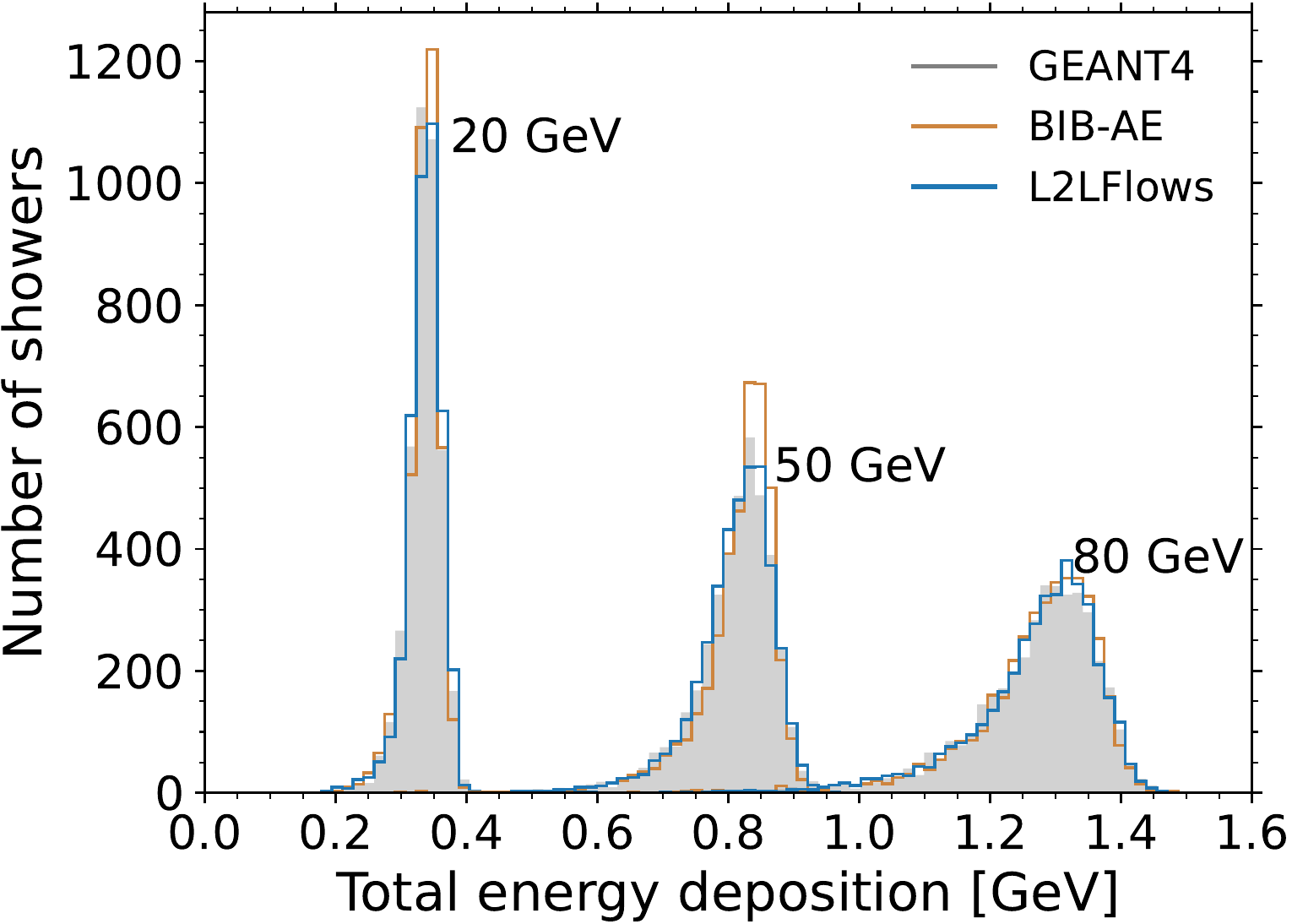}
    \end{minipage}
    \caption{Comparisons between \textsc{Geant4} (grey), the BIB-AE (orange) and \textsc{L$2$LFlows} (blue). Left:~Total deposited energy per shower with shower incident energies uniformly distributed between $10$ and $100$ GeV ($95$k showers are used for every model). Right:~Same for discrete incident energies of $20$, $50$ and $80$ GeV ($4$k showers are used for every model).}
    \label{results:energy_flow_hists}
\end{figure}

The distributions in Fig.~\ref{results:energy_flow_hists} show the total energy depositions ($E_{\text{depos}} := \sum_{i}E_{i}$), both for continuous incident energies uniformly distributed in $[10, 100]$ GeV (left) and for discrete incident energies $E_{\text{inc}} \in \{20, 50, 80\}$ GeV (right). In both of these distributions we observe that \textsc{L$2$LFlows} is much closer to the \textsc{Geant4} distribution than the BIB-AE.

\begin{figure}[t!]
    \begin{minipage}[c]{0.47\textwidth} 
        \centering 
        \includegraphics[width=\textwidth]{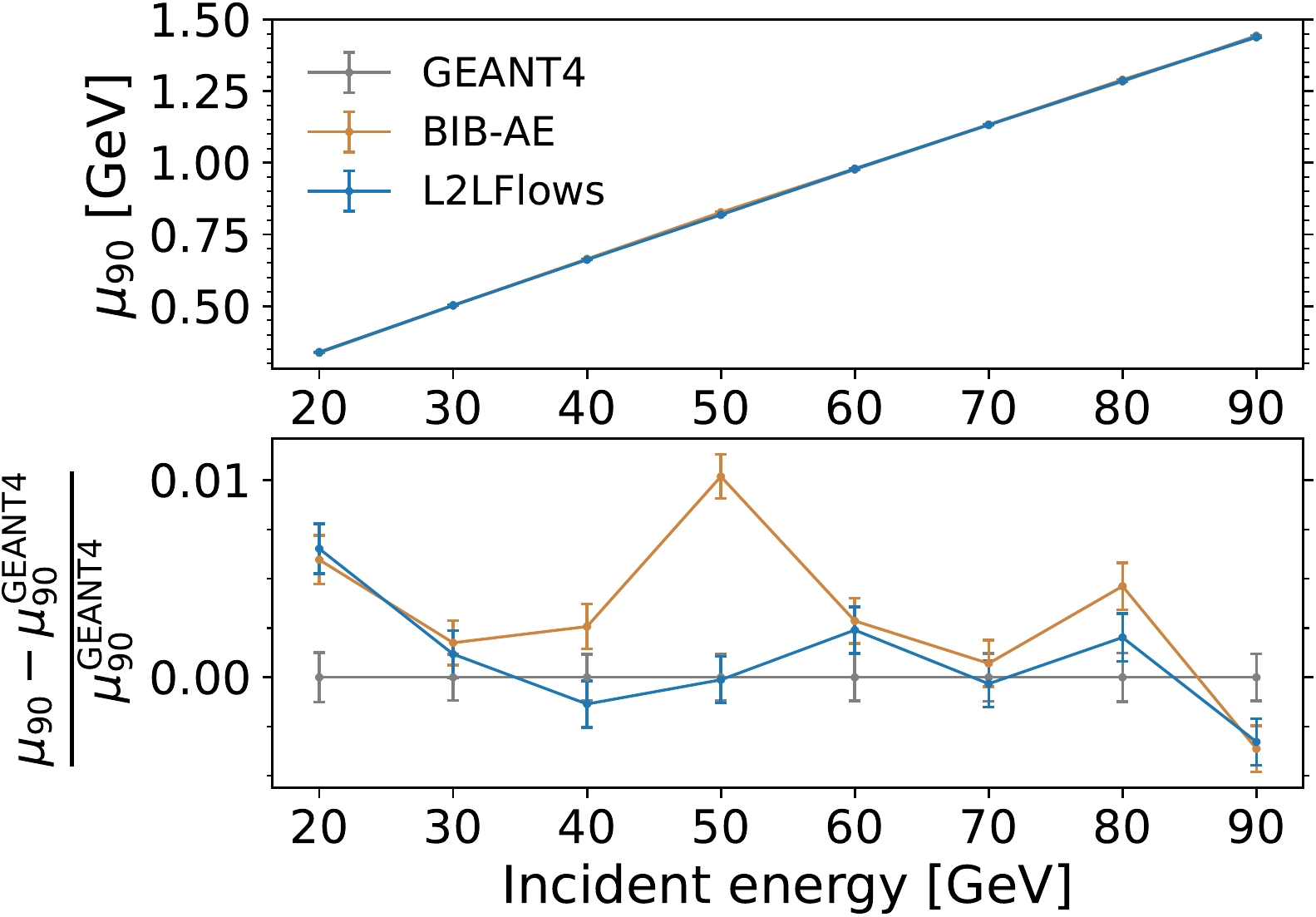}
    \end{minipage}\hspace{0.1cm}
    \begin{minipage}[c]{0.47\textwidth}
        \centering 
        \includegraphics[width=\textwidth]{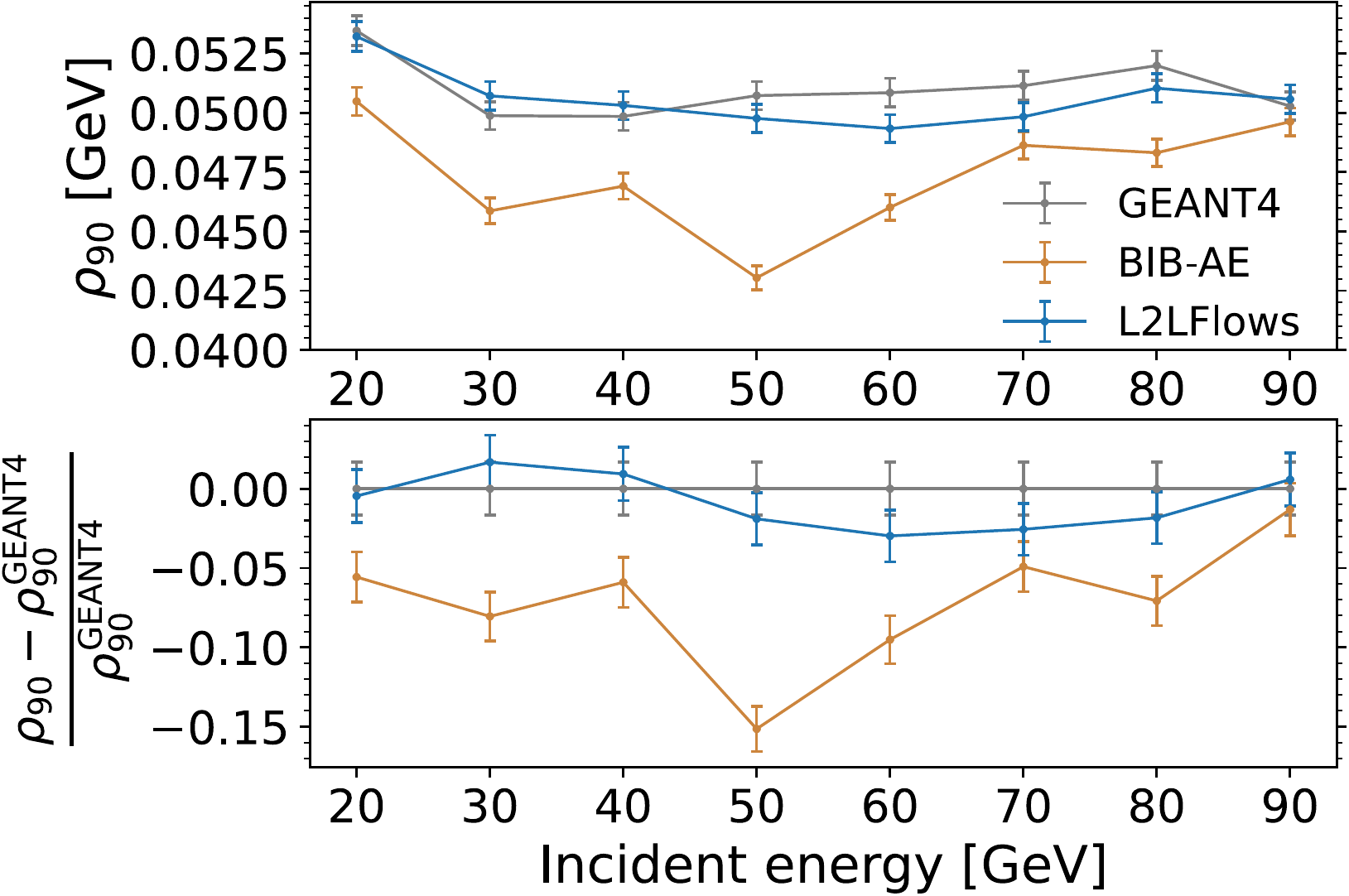}
    \end{minipage}
    \caption{Comparisons between \textsc{Geant4} (grey), the BIB-AE (orange) and \textsc{L$2$LFlows} (blue) for discrete energies. The plots show the linearity (left) and the ratio of $\rho_{90} := \sigma_{90}/\mu_{90}$ of the ECal (right) for discrete incident energies between $10$ and $100$ GeV in $10$ GeV steps. At each discrete energy, \textsc{Geant4}, the BIB-AE and \textsc{L$2$LFlows} are plotted with $4$k samples each.}
    \label{results:linearity}
\end{figure}

Figure~\ref{results:linearity} shows the linearity\footnote{This does not correspond to the actual calorimeter linearity or resolution, as the increased thickness of the last $10$ ECal layers is not calibrated for. It is, however, still a vital means for determining the performance of the generative approaches.} (and its relative deviation to \textsc{Geant4}) as well as the width (again with its relative deviation).\footnote{The linearity $\mu_{90}$ is defined as the mean deposited energy over the ECal for discrete $E_{\text{inc}}$ of a $90\%$ subset of the samples that have the smallest range. The width $\rho_{90}$ is defined as $\rho_{90} := \mu_{90}/\sigma_{90}$, where $\sigma_{90}$ is the standard deviation of the $90\%$ subset of the energy deposition samples that have the smallest range.} 
For the linearity, the relative deviation is for the BIB-AE maximally about $1\%$, for \textsc{L$2$LFlows} the deviation is everywhere below $0.75\%$. For the width plot,\footnote{One might be tempted to call $\rho_{90}$ the \enquote{resolution}, but because of the different thicknesses of the tungsten absorber layers, cf.~Sec.~\ref{sec:dataset}, this is not the case \cite{Buhmann:2020pmy}.} the relative deviation for \textsc{L$2$LFlows} is everywhere below $5\%$, whereas for the BIB-AE, the maximum deviation is about $15\%$. 

\begin{figure}[h!]
    \begin{minipage}[c]{0.325\textwidth}
        \centering 
        \includegraphics[trim={0cm 0cm 2.06cm 0cm}, clip, width=\textwidth]{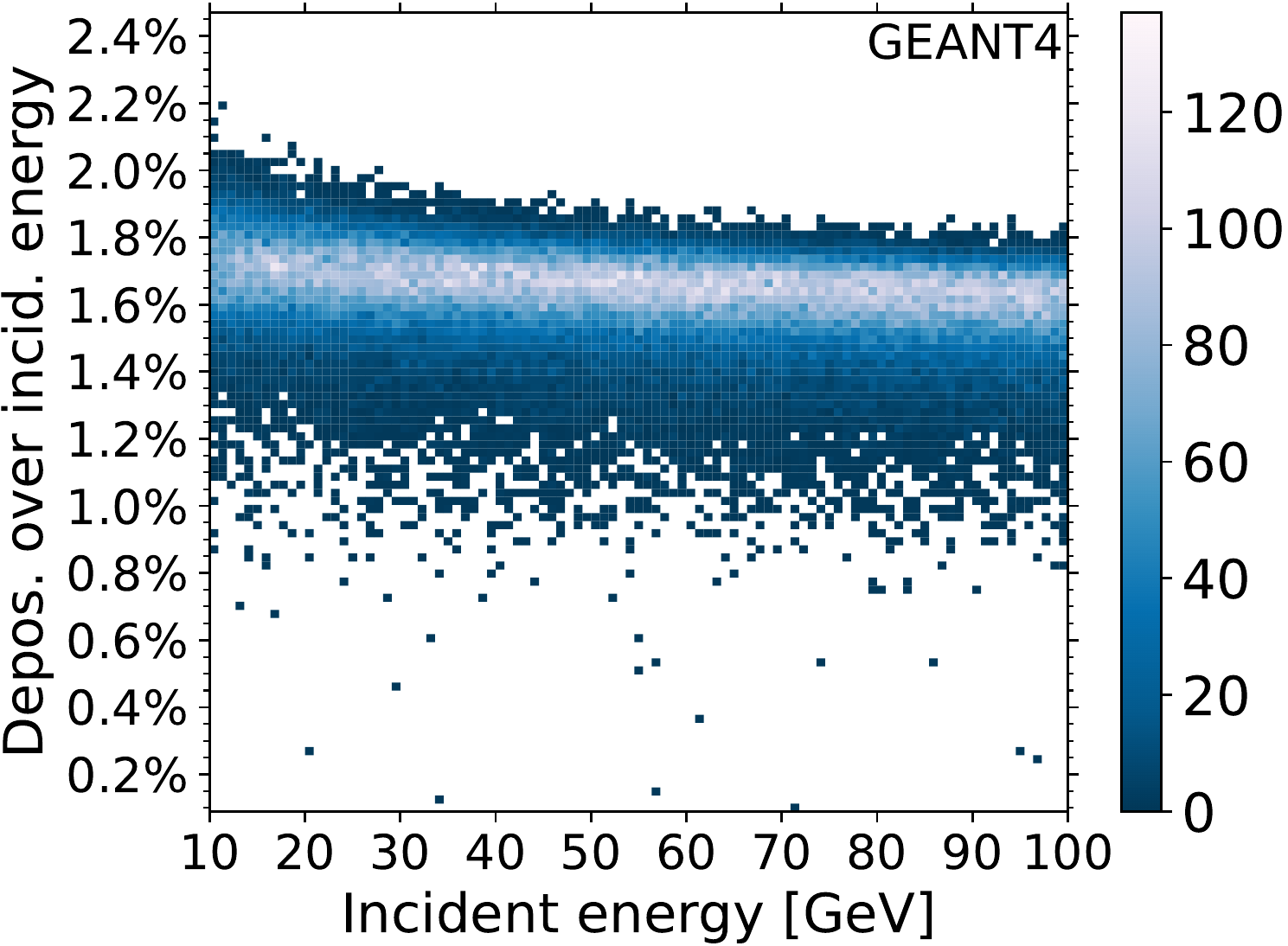}
    \end{minipage}
    \begin{minipage}[c]{0.325\textwidth}
        \centering 
        \includegraphics[trim={0cm 0cm 2.06cm 0cm}, clip, width=\textwidth]{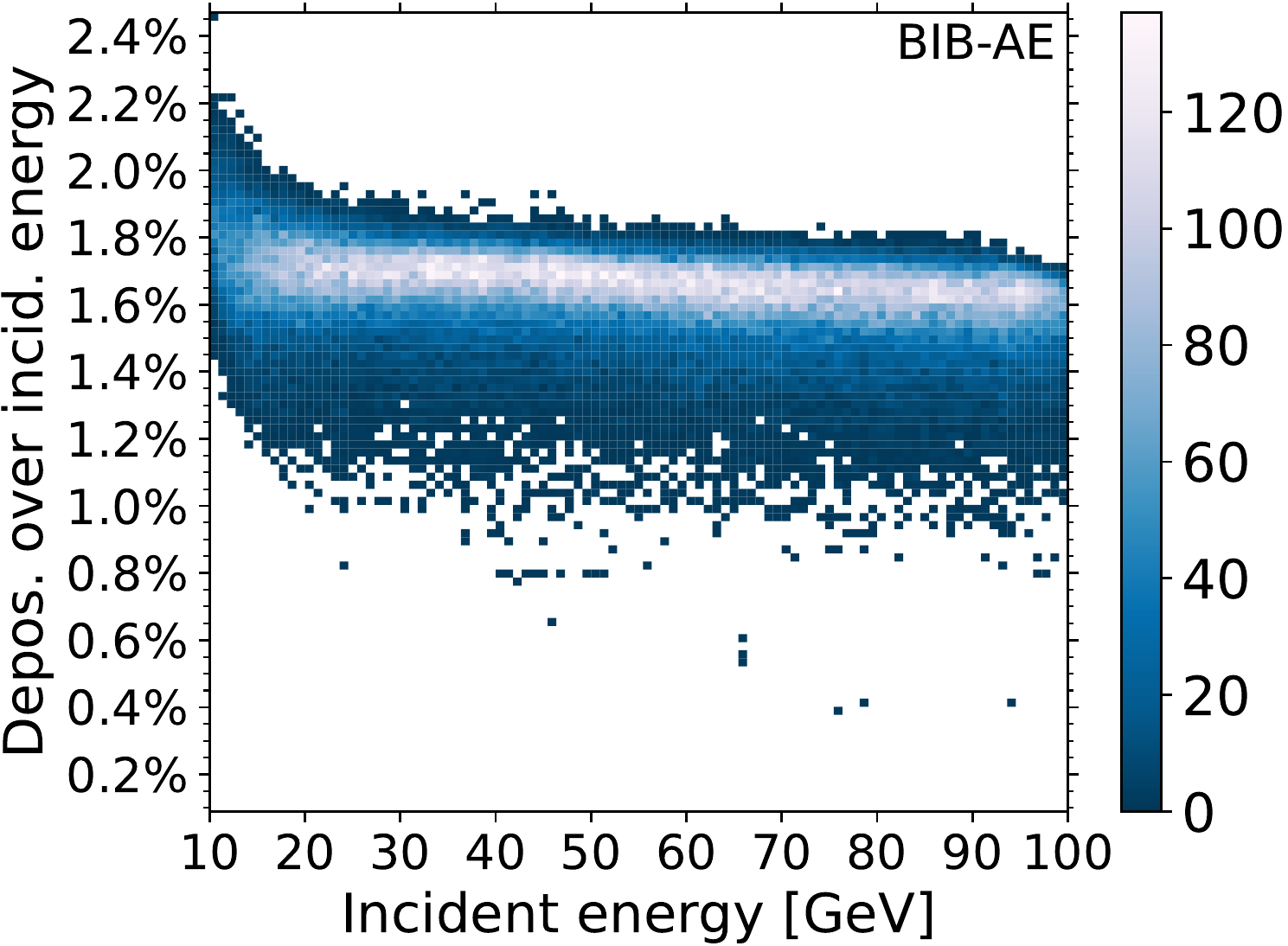}
    \end{minipage}
    \begin{minipage}[c]{0.325\textwidth}
        \centering 
        \includegraphics[width=1.13\textwidth]{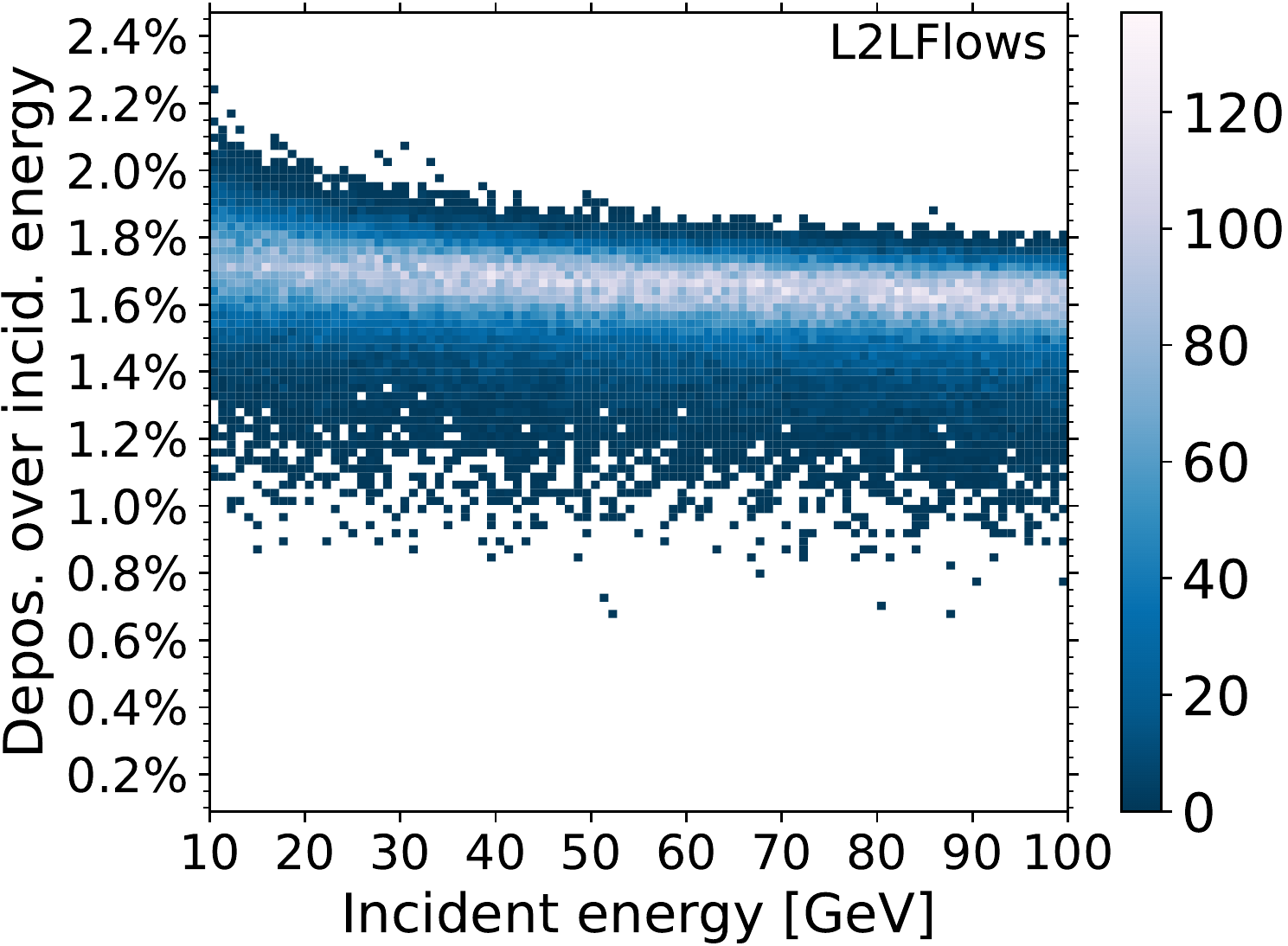}
    \end{minipage}\vspace{0.5cm}
    \begin{minipage}[c]{0.325\textwidth}
        \centering 
        \includegraphics[trim={0 0 2.06cm 0}, clip, width=\textwidth]{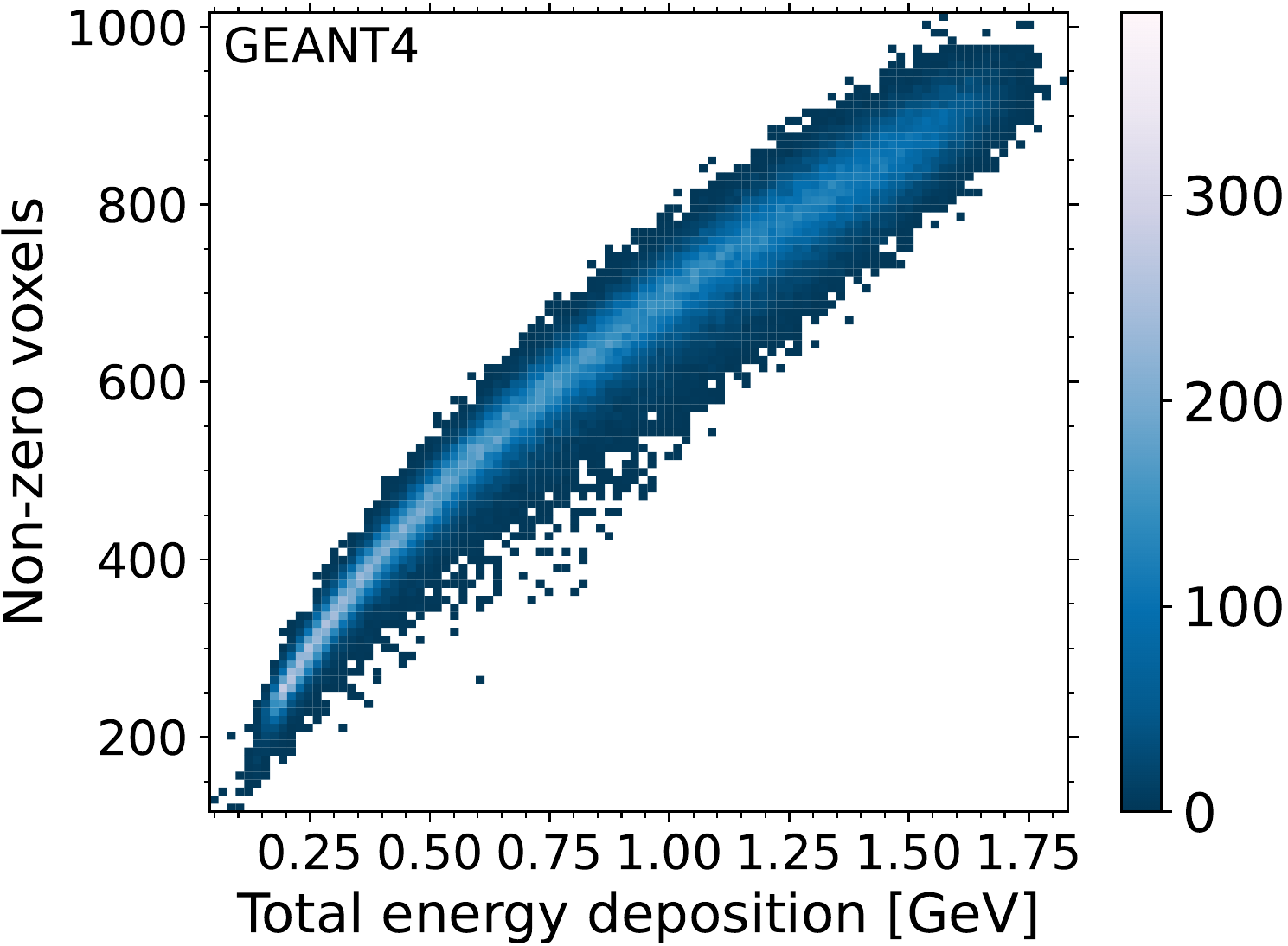}
    \end{minipage}
    \begin{minipage}[c]{0.325\textwidth}
        \centering 
        \includegraphics[trim={0 0 2.06cm 0}, clip, width=\textwidth]{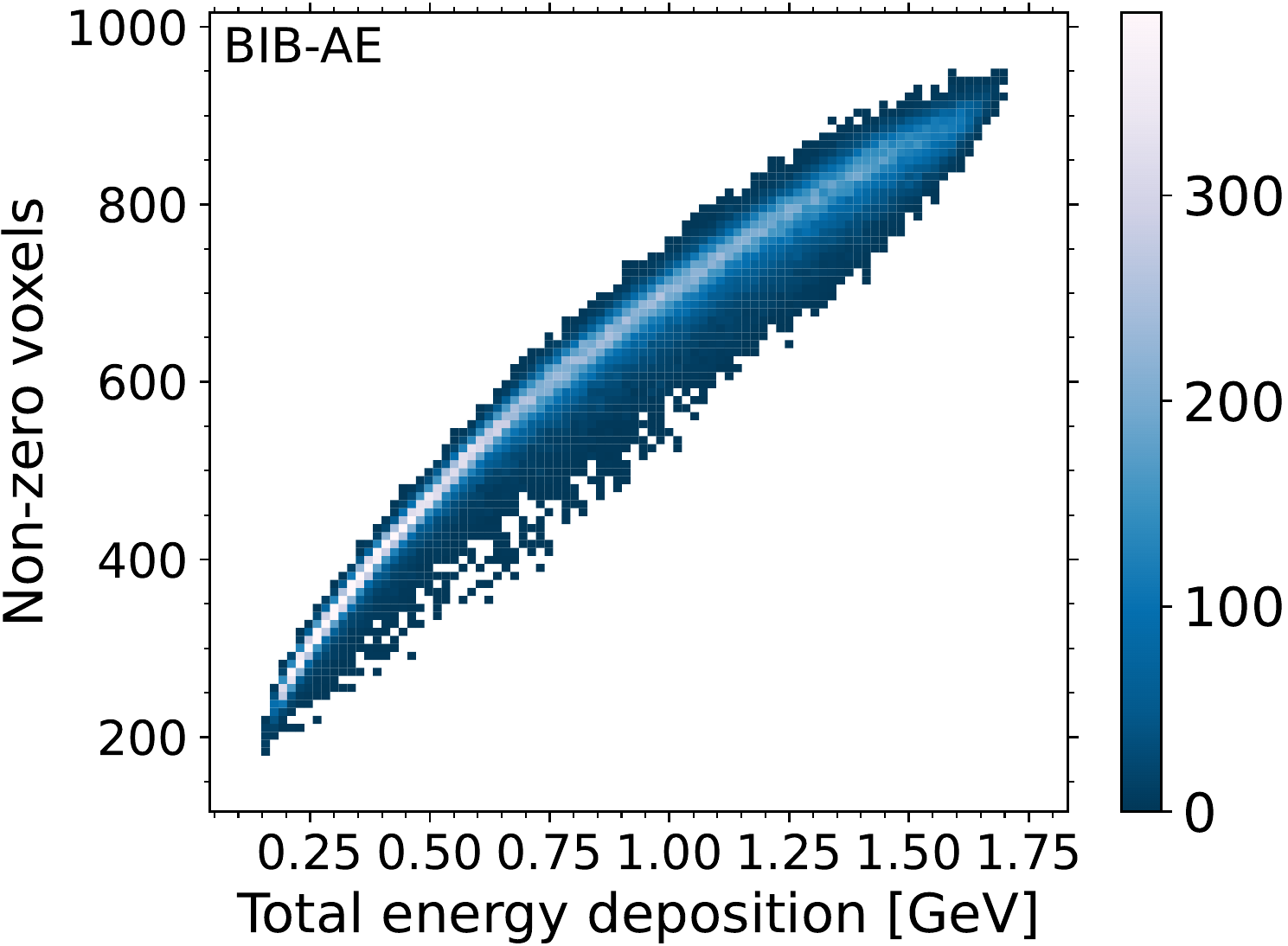}
    \end{minipage}
    \begin{minipage}[c]{0.325\textwidth}
        \centering 
        \includegraphics[width=1.13\textwidth]{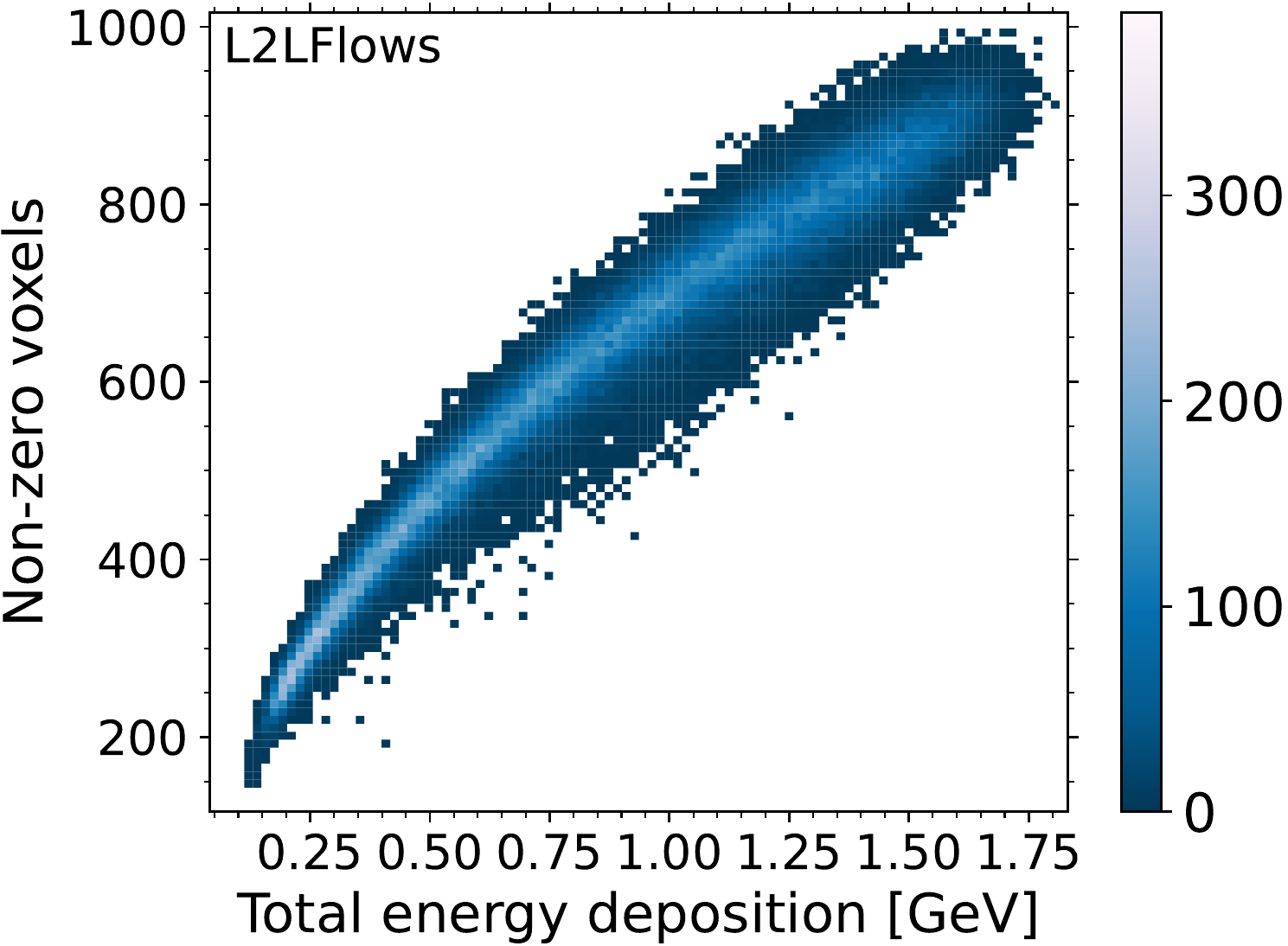}
    \end{minipage}\vspace{0.5cm}\hspace{-0.1cm}
    \begin{minipage}[c]{0.325\textwidth}
        \centering 
        \includegraphics[trim={0 0 2.06cm 0}, clip, width=0.97\textwidth]{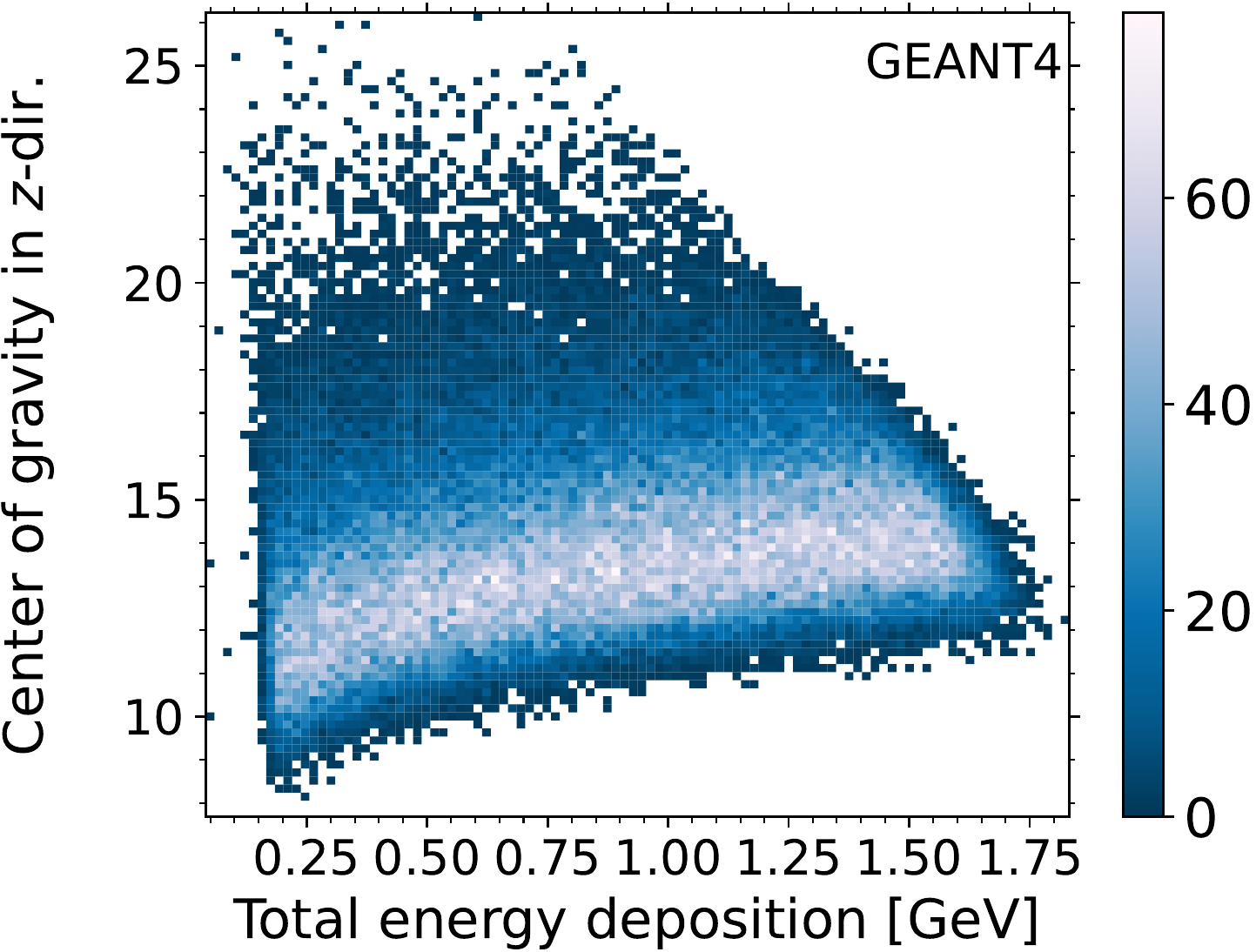}
    \end{minipage}
    \begin{minipage}[c]{0.325\textwidth}
        \centering 
        \includegraphics[trim={0 0 2.06cm 0}, clip,  width=0.97\textwidth]{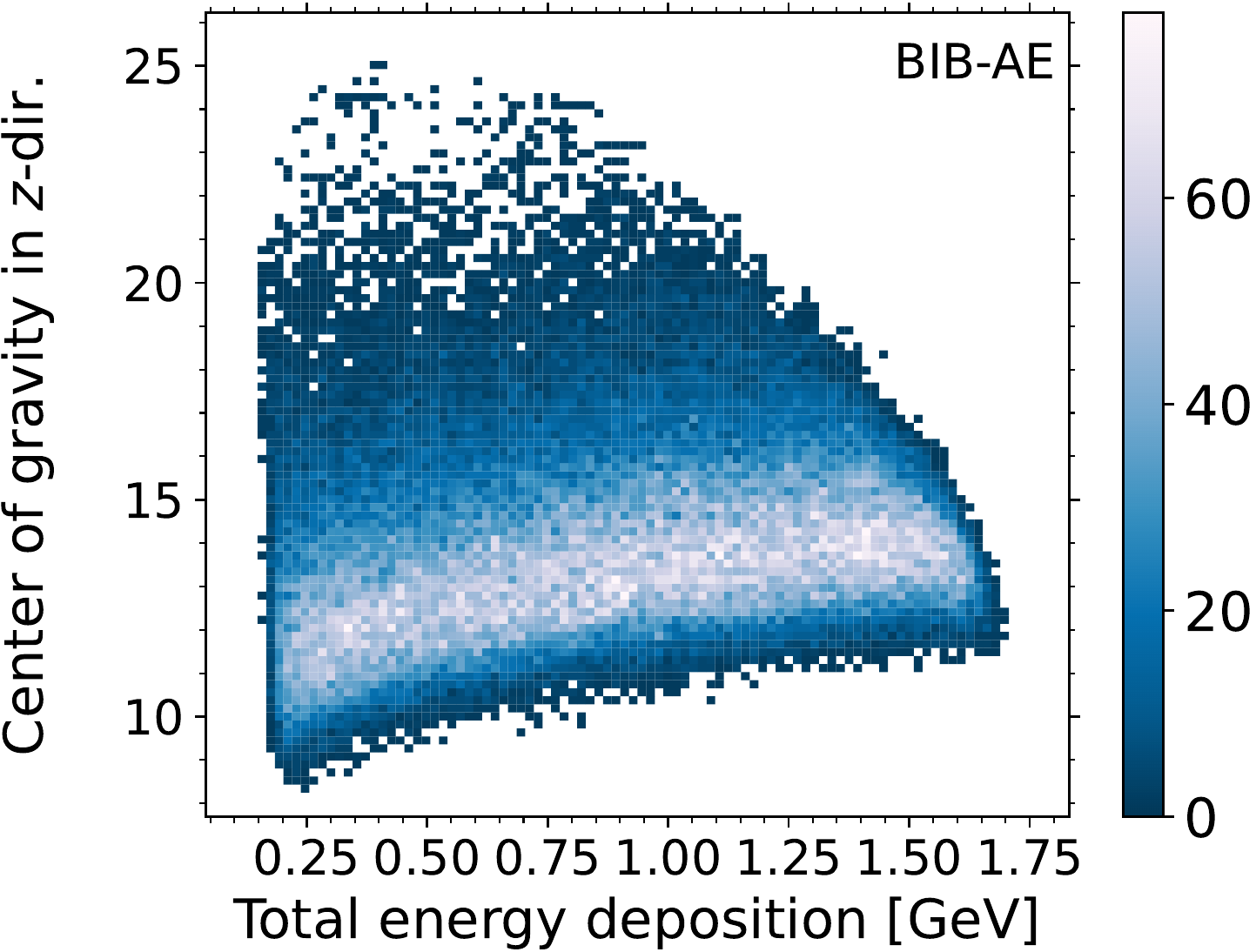}
    \end{minipage}
    \begin{minipage}[c]{0.325\textwidth}
        \centering 
        \includegraphics[width=1.1\textwidth]{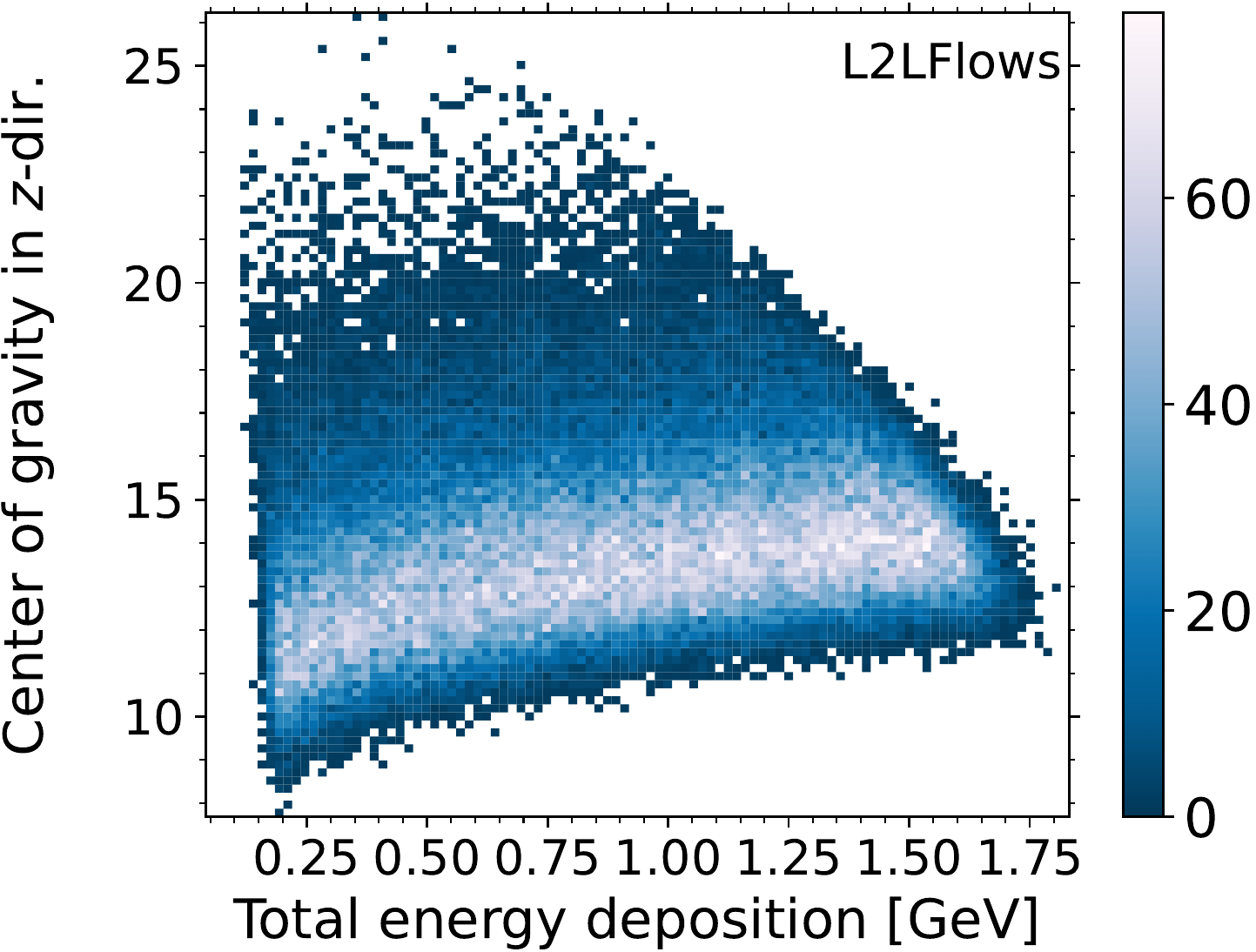}
    \end{minipage}
    \centering
    \caption{$2$D histograms comparing correlations between selected sets of variables for \textsc{Geant4}, the BIB-AE and \textsc{L$2$LFlows}. The upper row of plots shows the ratio of the deposited energy $E_{\text{depos}}$ to the incident energy $E_{\text{inc}}$ as a function of $E_{\text{inc}}$. While a perfect calorimeter would have a constant ratio for \textsc{Geant4}, in practice, because of leakage, the curve falls off over the range. The center row shows the number of voxels in which energy was deposited versus the total deposited energy. The lower row shows the center of gravity in $z$-direction versus the total deposited energy. All plots are shown for the full spectrum with $95$k showers for every model.}
    \label{results:energy_flow_plots_ratios}
\end{figure}

It is also interesting to examine the ratio of $E_{\text{depos}}$ over $E_{\text{inc}}$ plotted as a function of $E_{\text{inc}}$. The upper row of Fig.~\ref{results:energy_flow_plots_ratios} shows that the functional form of the ratio is not constant for \textsc{Geant4}. While a perfect calorimeter would yield a constant ratio for \textsc{Geant4}, in practice, because of leakage and the increased thickness of the last ten absorber layers, the curve falls off over the range. The fact that the ratio of the deposited over the incident energy is only $\mathcal O(1 \%)$ is expected, as the ILD ECal is a sampling calorimeter. As becomes apparent from Fig.~\ref{results:energy_flow_plots_ratios}, \textsc{L$2$LFlows} learns the functional form much better than the BIB-AE. In particular, the BIB-AE has problems at the edges. At the left edge, i.e.~for \mbox{$E_{\text{inc}} \approx 10$ GeV}, ratios of $2\%$ and more are too populated compared to \textsc{Geant4}, yet ratios of around $1.5\%$ and less are too thinly populated. At the right edge, i.e.~for $E_{\text{inc}} \approx 100$ GeV, the functional form falls off too quickly. Further, in the middle row of Fig.~\ref{results:energy_flow_plots_ratios}, we show the sparsity plotted against $E_{\text{depos}}$. The BIB-AE learns a distribution that is thinner compared to the one from \textsc{Geant4}, and its core has too many occurrences. For \textsc{L$2$LFlows}, the agreement to the \textsc{Geant4} distribution is much better, and differences are barely visible by eye. Finally, the last row of Fig.~\ref{results:energy_flow_plots_ratios} shows the $2$D correlations for the center of gravity in $z$-direction versus the total deposited energy. It can be seen that the BIB-AE is yet again not capturing the full distribution, as its $2$D plot is more compact compared to \textsc{Geant4}. In contrast, \textsc{L$2$LFlows} exhibits a superb performance.   

\begin{figure}[h!]
    \begin{minipage}[c]{0.495\textwidth} 
        \centering
        \includegraphics[width=\textwidth]{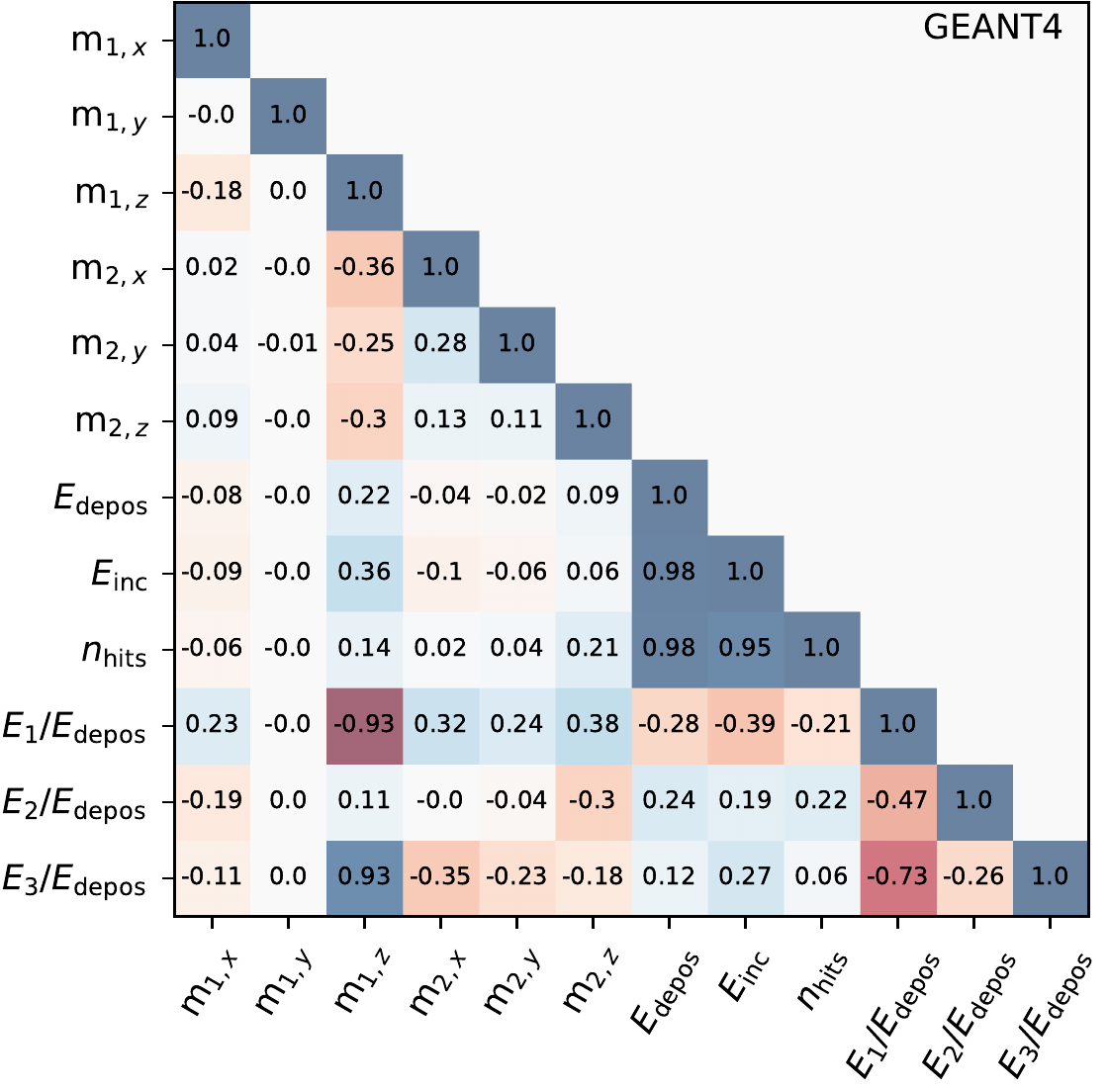}
    \end{minipage}
    \begin{minipage}[c]{0.495\textwidth} 
        \centering
        \includegraphics[width=\textwidth]{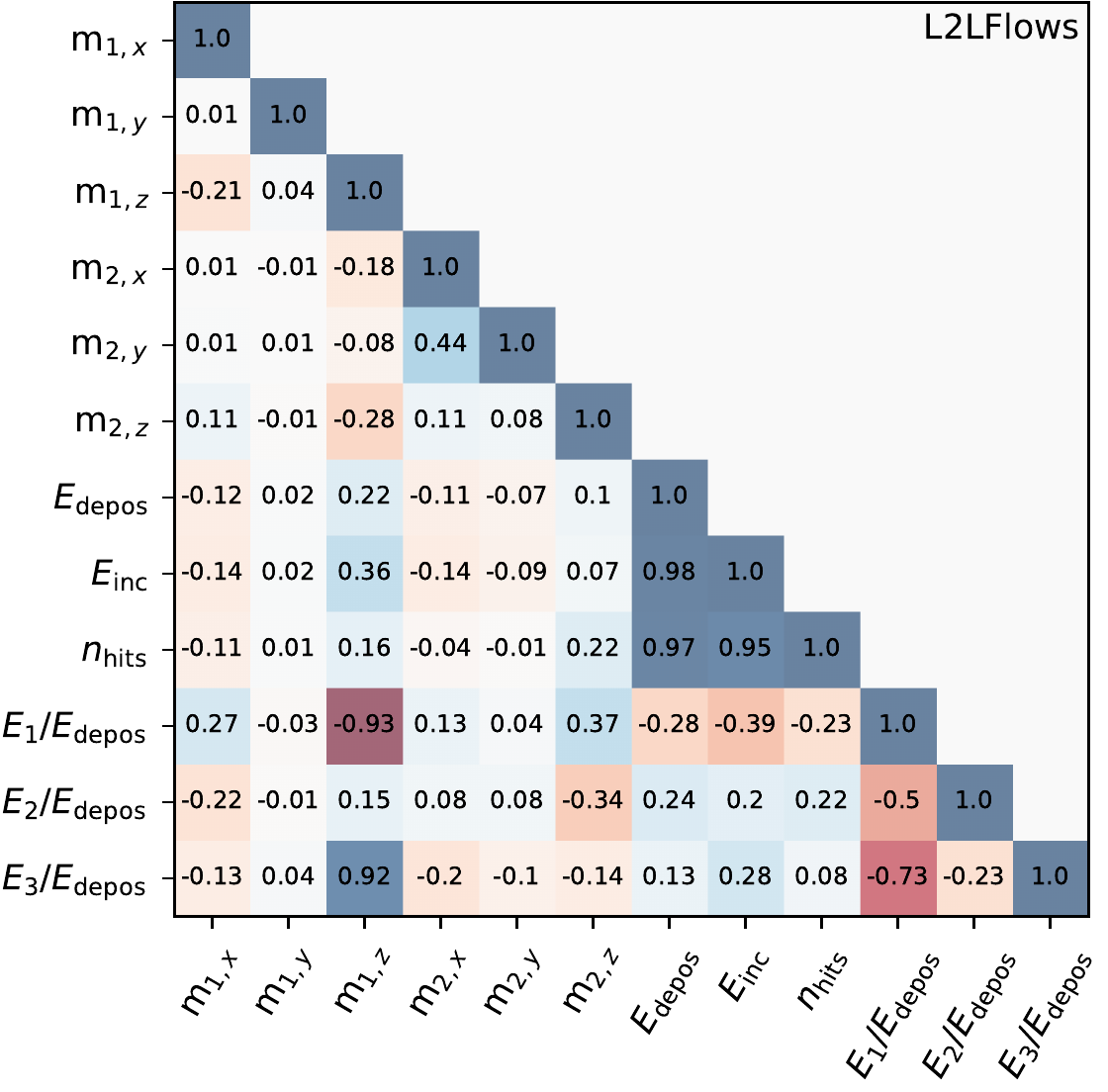}
    \end{minipage}\vspace{0.5cm}
    \begin{minipage}[c]{0.495\textwidth}
        \centering 
        \includegraphics[width=\textwidth]{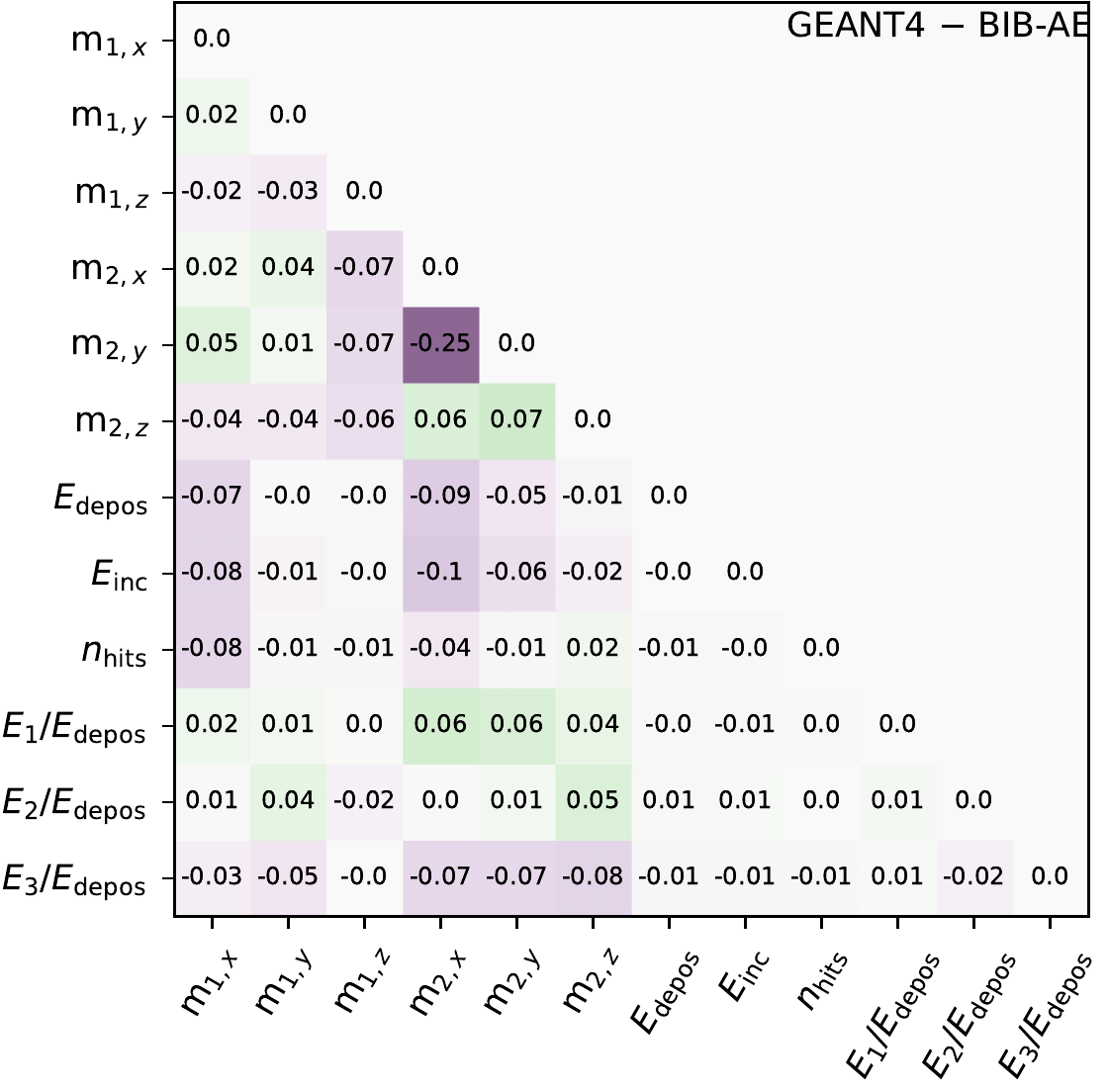}
    \end{minipage}
    \begin{minipage}[c]{0.495\textwidth}
        \centering 
        \includegraphics[width=\textwidth]{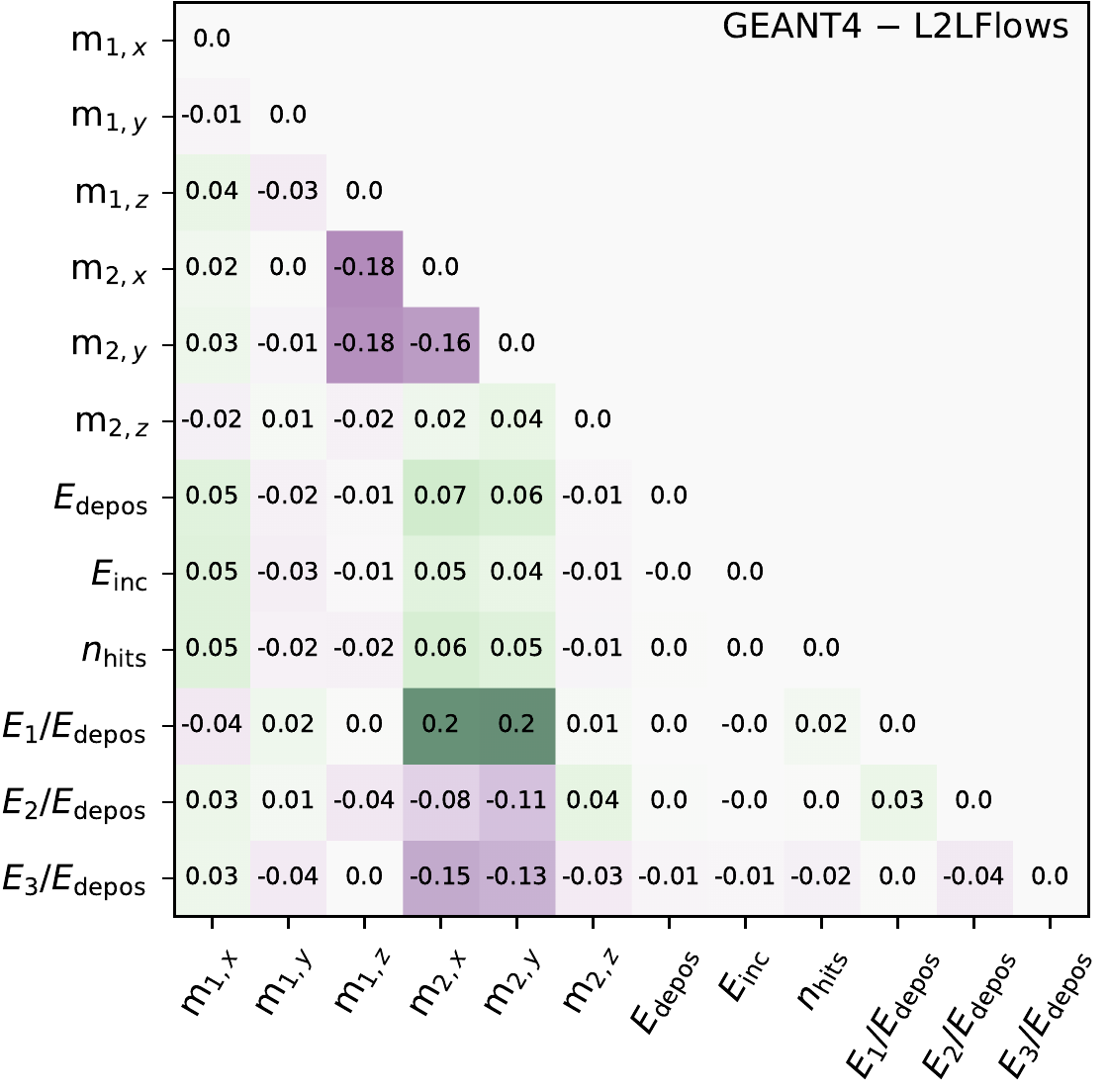} 
    \end{minipage}
    \caption{Pearson correlation matrices for \textsc{Geant4} (upper left), \textsc{L$2$LFlows} (upper right), the difference between the \textsc{Geant4} and BIB-AE correlations (bottom left) and the difference between the \textsc{Geant4} and the \textsc{L$2$LFlows} correlations (bottom right). For every simulator, $95$k showers are used. A description of the variables can be found in the text.}
    \label{corr_matrices}
\end{figure}

In addition, Fig.~\ref{corr_matrices} shows correlation matrices for pairwise Pearson correlation coefficients between several high-level observables for \textsc{Geant4} and the difference of \textsc{Geant4} to the BIB-AE and \textsc{L$2$LFlows}. The observables are, in order of appearance, the first and second moments along the $x$, $y$, and $z$ directions, the visible energy sum, the incident photon energy, the number of hits, and the energy fractions in the three thirds of the calorimeter along the $z$-directions. More details can be found in Ref.~\cite{Buhmann:2020pmy}. It can be seen that both generative models correctly describe a large number of the investigated pair-wise correlations. Both models do, however, struggle with specific correlations, involving the second moments in the $x$- and $y$-direction.

\begin{figure}[h!]
    \begin{minipage}[c]{0.495\textwidth}
        \centering
        \includegraphics[width=\textwidth]{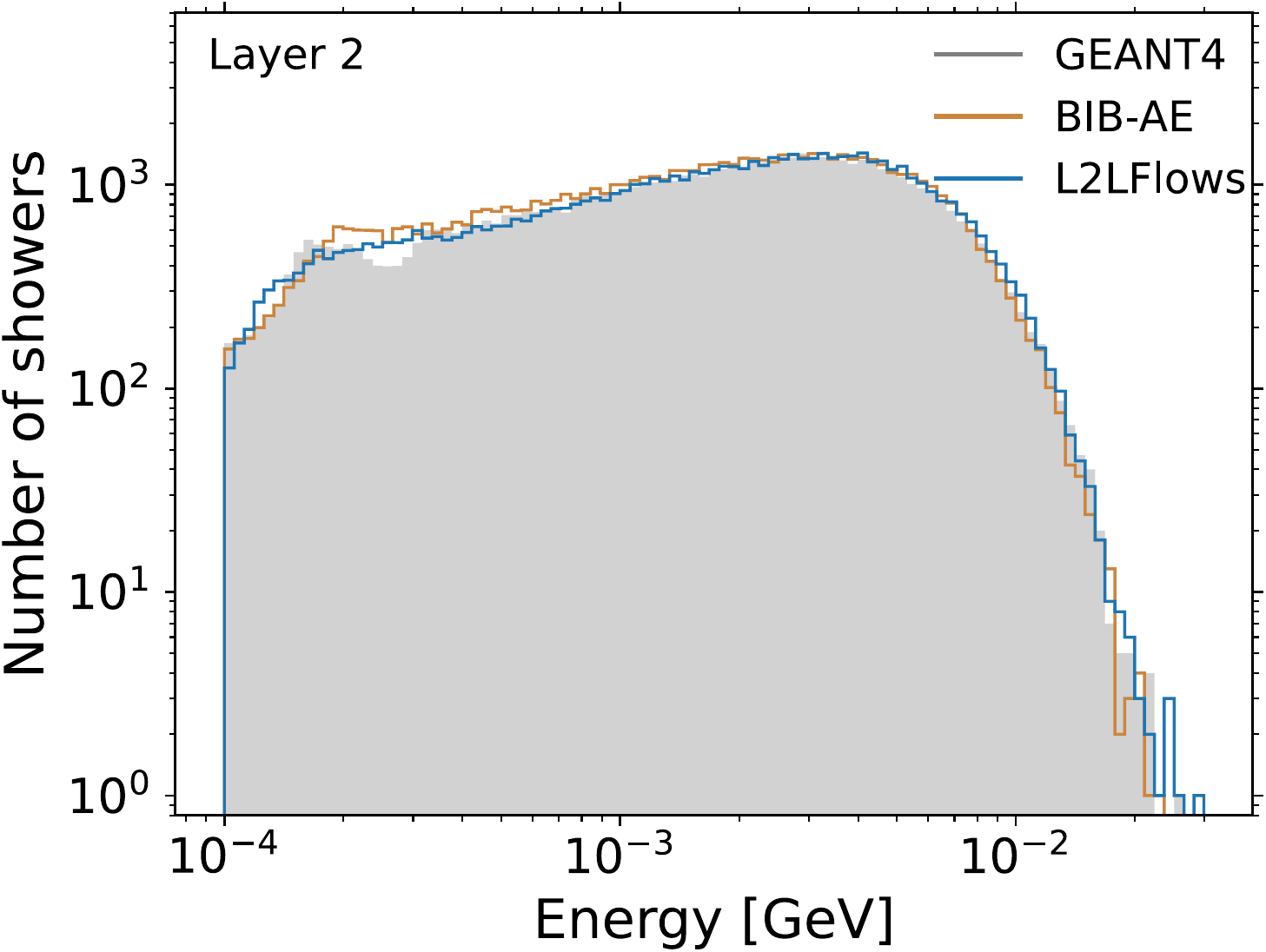}
    \end{minipage}
    \begin{minipage}[c]{0.495\textwidth}
        \centering
        \includegraphics[width=\textwidth]{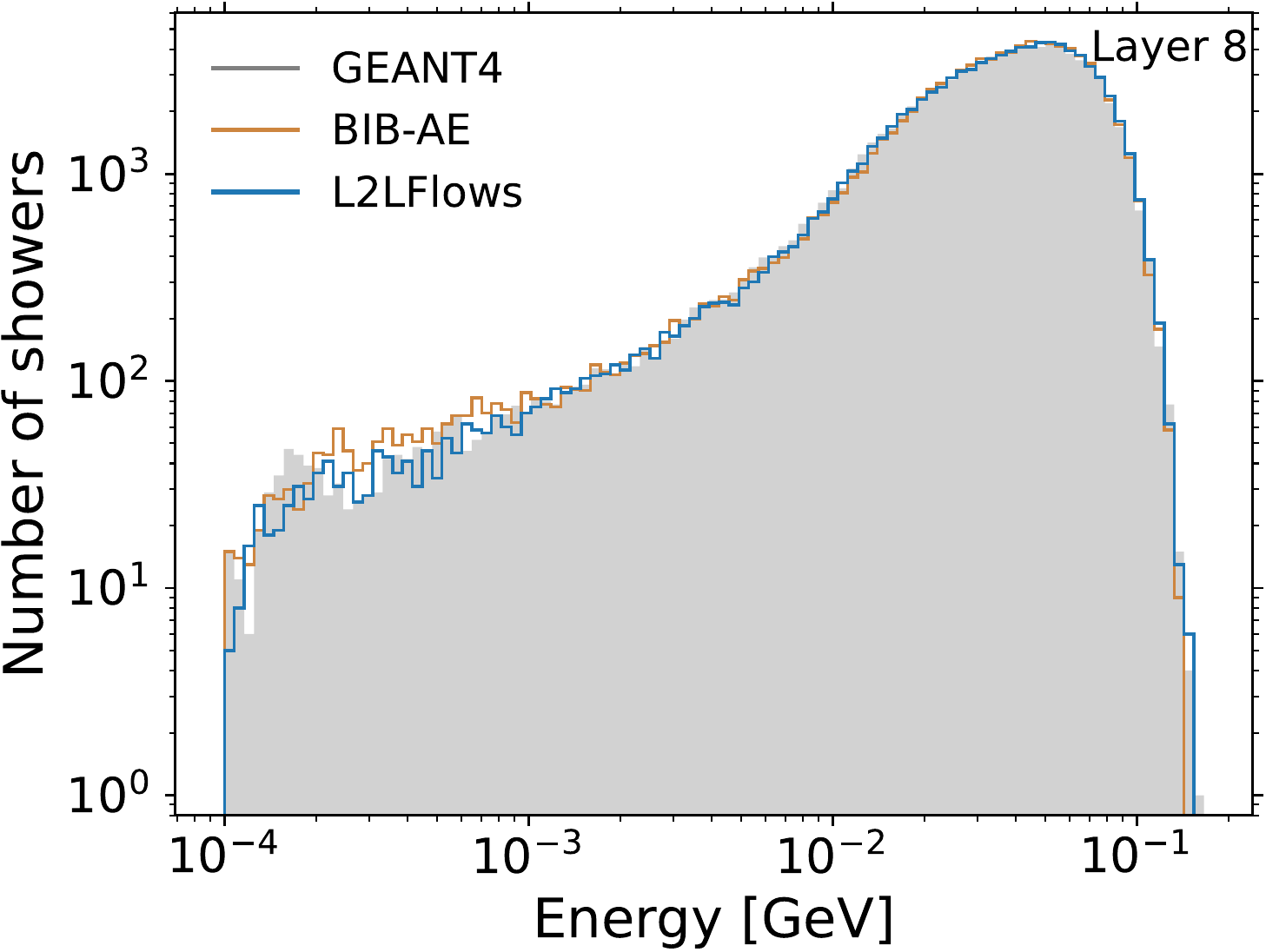}
    \end{minipage}\vspace{0.4cm}
    \begin{minipage}[c]{0.495\textwidth}
        \centering
        \includegraphics[width=\textwidth]{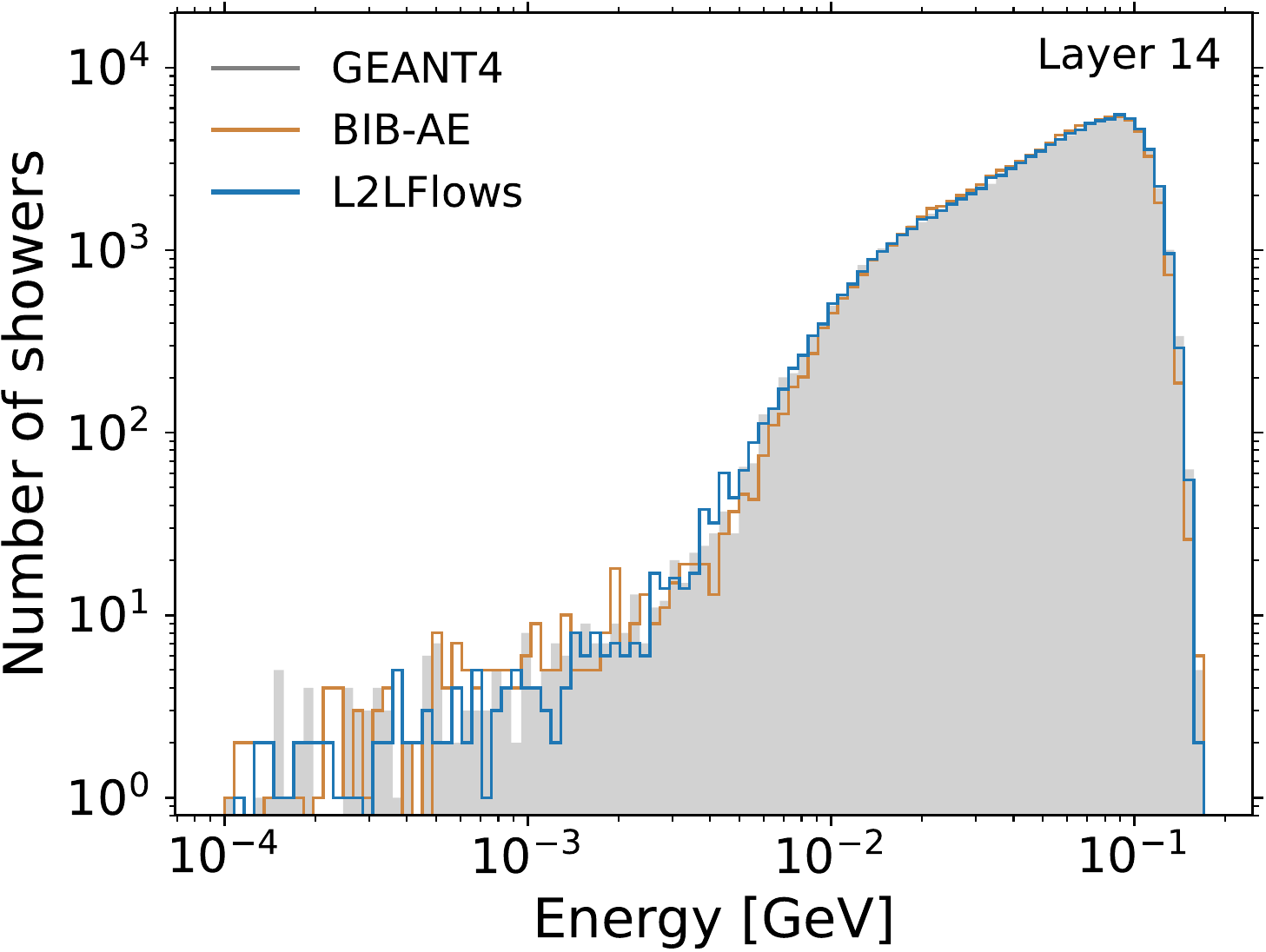}
    \end{minipage}
    \begin{minipage}[c]{0.495\textwidth}
        \centering
        \includegraphics[width=\textwidth]{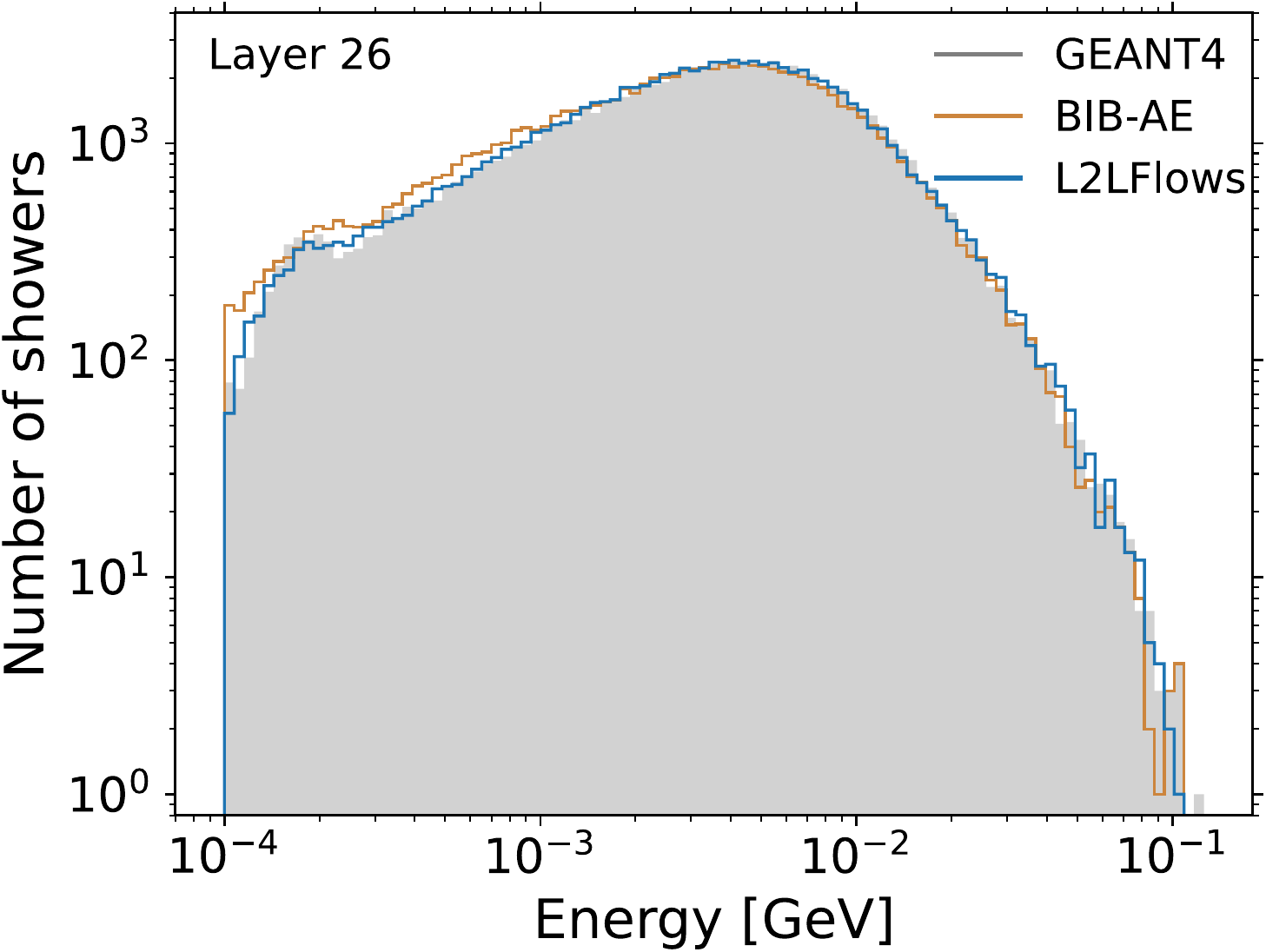}
    \end{minipage}
    \caption{Energy depositions per layer (summed over all ECal voxels) for \textsc{Geant4} (grey), the BIB-AE (orange) and \textsc{L$2$LFlows} (blue) in different layers. All plots are shown for the full spectrum with $95$k showers for every simulator.}
    \label{results:total_energy_depos}
\end{figure}

Finally, Fig.~\ref{results:total_energy_depos} shows the total energy depositions per layer for four selected layers. In layers $2$, $8$, $14$ and $26$, \textsc{L$2$LFlows} is at least comparable to the BIB-AE, if not better. 

Judging from the histograms and plots that have been shown so far, \textsc{L$2$LFlows} seems to outperform the BIB-AE in almost every single physics quantity, however it does slightly worse in capturing pairwise correlations.

In order to judge the performance more comprehensively in the full multivariate phase space, various metrics have been suggested in~\cite{Krause:2021ilc,Kansal:2022spb,Lim:2022nft}. For this comparison, we turn to classifier-based tests described in~\cite{Krause:2021ilc,Lim:2022nft} in the following subsection, and leave the exploration of other metrics suggested in~\cite{Kansal:2022spb} as a future research direction. 

\subsection{Classifier Tests}
\label{sec:classifier}

As in \cite{Krause:2021ilc,Krause:2021wez,Krause:2022jna}, we now turn to a classifier-based metric to evaluate the quality of the generated showers in the full $3000$-dimensional phase space. In total, two binary fully-connected classifiers are trained, one on \textsc{Geant4} vs BIB-AE generated showers, the other on \textsc{Geant4} vs \textsc{L$2$LFlows} generated showers. Both classifiers have the same architecture and make use of the same hyperparameters; details can be found in App.~\ref{appendix:classifier_tests}. The idea of the classifier metric is that if the classifier is optimal, then by the Neyman-Pearson lemma it directly computes the likelihood ratio $p_{\text{generated}}(x)/p_{\text{reference}}(x)$ in the full phase space. A perfect generative model should have $p_{\text{generated}} = p_{\text{reference}}$ and optimal classifier scores that are identically 0.5.\footnote{Indeed, we find an AUC of 0.5 when training on \textsc{Geant4} vs \textsc{Geant4} samples.} For an imperfect generative model, the optimal classifier should be the most powerful detector of any deviations from $p_{\text{generated}} = p_{\text{reference}}$. 
 
Of course, it is never possible, given finite samples and finite model capacity, to learn the truly optimal classifier. Therefore, the classifier metric we evaluate here is at best an approximate measure of model quality. At most, we could expect the classifier AUC score we obtain here to be a lower bound on the true AUC score that would be given by the optimal classifier. However, given identical model architectures and training set sizes, we expect the {\it relative} comparison of binary classifier scores between 
\textsc{Geant4} vs BIB-AE and \textsc{Geant4} vs \textsc{L$2$LFlows} to still be meaningful and informative. 

The results of $10$ classifier trainings are shown in Tab.~\ref{results:classifier_test}. As can be observed, the BIB-AE--generated showers allow for almost perfect classification, which reflects itself in an AUC close to $1$. The \textsc{L$2$LFlows}-generated showers, on the other hand, are much better able to fool such a classifier. However, we note that there is still some separation power to \textsc{Geant4}-generated showers, 
as the mean AUC of the classifiers is far away from $0.5$. 

\begin{table}[h!]
    \centering 
	\begin{tabular}{ c|c|c } 
		\hline
		\textbf{\# Showers per simulator} &  \textbf{AUC \textsc{Geant4} vs \textsc{L$2$LFlows}} & \textbf{AUC \textsc{Geant4} vs BIB-AE} \\
		\hline \hline 
	 	$95$k & $0.8518 \pm 0.0042$ & $0.9947 \pm 0.0025$  \\ \hline 
            $190$k & $0.8768 \pm 0.0029$ & $-$  \\ \hline 
            $380$k & $0.8962 \pm 0.0024$ & $-$ \\ \hline 
            $760$k & $0.9402 \pm 0.0011$ & $-$ \\ \hline 
	\end{tabular}
	\captionof{table}{Classifier results for different number of showers, where the left column shows the number of showers per simulator used for the classifier tests (a $60\%:20\%:20\%$ split is made to obtain training, validation and test showers of the classifiers). The middle and right columns show the mean and standard deviation of the AUC of $10$ independent runs for \textsc{Geant4} vs \textsc{L$2$LFlows} and \textsc{Geant4} vs BIB-AE classifiers. Since the mean AUC of the BIB-AE in $10$ independent runs is already very close to $1$ for $95$k showers, more showers are only used for the \textsc{Geant4} vs \textsc{L$2$LFlows} classifiers.}
    \label{results:classifier_test}
\end{table}

In Tab.~\ref{results:classifier_test}, we have also gone beyond previous works, to study the dependence of the classifier metric on training sample size. (We only studied the dependence on training sample size for \textsc{L$2$LFlows}, since the BIB-AE is very close to $1$ when trained on only $95$k showers.)
As becomes apparent, the mean AUC of \textsc{L$2$LFlows} worsens with more showers, which is unsurprising, as with more statistics, the classifier can find more differences between the \textsc{Geant4}- and \textsc{L$2$LFlows}-generated showers. At an even larger number of showers used for classifier training, we would expect the finite size of the generator training set to become an issue, too~\cite{Matchev:2020tbw,Butter:2020qhk,Bieringer:2022cbs}. Nevertheless, we observe that for a given number of showers, the BIB-AE showers are more separable from \textsc{Geant4} than the \textsc{L$2$LFlows} showers, indicating a better performance of \textsc{L$2$LFlows}. Also, even though the AUC scores for \textsc{Geant4} vs.\ \textsc{L$2$LFlows} are worsening with more training data (and may be asymptoting to $1$, there is insufficient training data to say for sure), the fact that they are not immediately close to $1$ (as is the case for \textsc{Geant4} vs.\ BIB-AE) is a further indication that the L$2$LFlows showers are of higher quality. 

To further test the relative quality of \textsc{L$2$LFlows} vs.~BIB-AE, we use the new {\it Multi-Model Classifier Metric} proposed in~\cite{Lim:2022nft}.
Instead of training separate binary classifiers between each generated model and the reference data, which can be constrained by limited amounts of the latter, we instead train a classifier (potentially multi-class) between the different generative models. This learns the probability that a shower came from each model. Then we evaluate this classifier on \textsc{Geant4}, BIB-AE and L$2$LFlows showers, and see which model the classifier prefers.
Note that while there are some limitations to the use of classifiers as an absolute metric discussed earlier, we expect the interpretation in a relative sense (as is done here) to be more straighforward. 

For this test, we use $760$k showers sampled from the BIB-AE and \textsc{L$2$LFlows} each. Just as for the \textsc{Geant4} vs BIB-AE/\textsc{L$2$LFlows} classifier, we make a $60\% : 20\%: 20\%$ split to obtain training, validation and test showers of the classifier. The AUC of the classifier on the test dataset (where we evaluate the checkpoint with highest validation accuracy) is $1.0000$, implying that a fully-connected classifier has no trouble distinguishing between BIB-AE and \textsc{L$2$LFlows} showers. The architecture and hyperparameters of this BIB-AE vs \textsc{L$2$LFlows} classifier are identical to the \textsc{Geant4} vs BIB-AE/\textsc{L$2$LFlows} classifiers; with the exception of the number of training epochs, see details in App.~\ref{appendix:classifier_tests}. 

\begin{figure}[h!]
    \centering
    \includegraphics[width=0.6\textwidth]{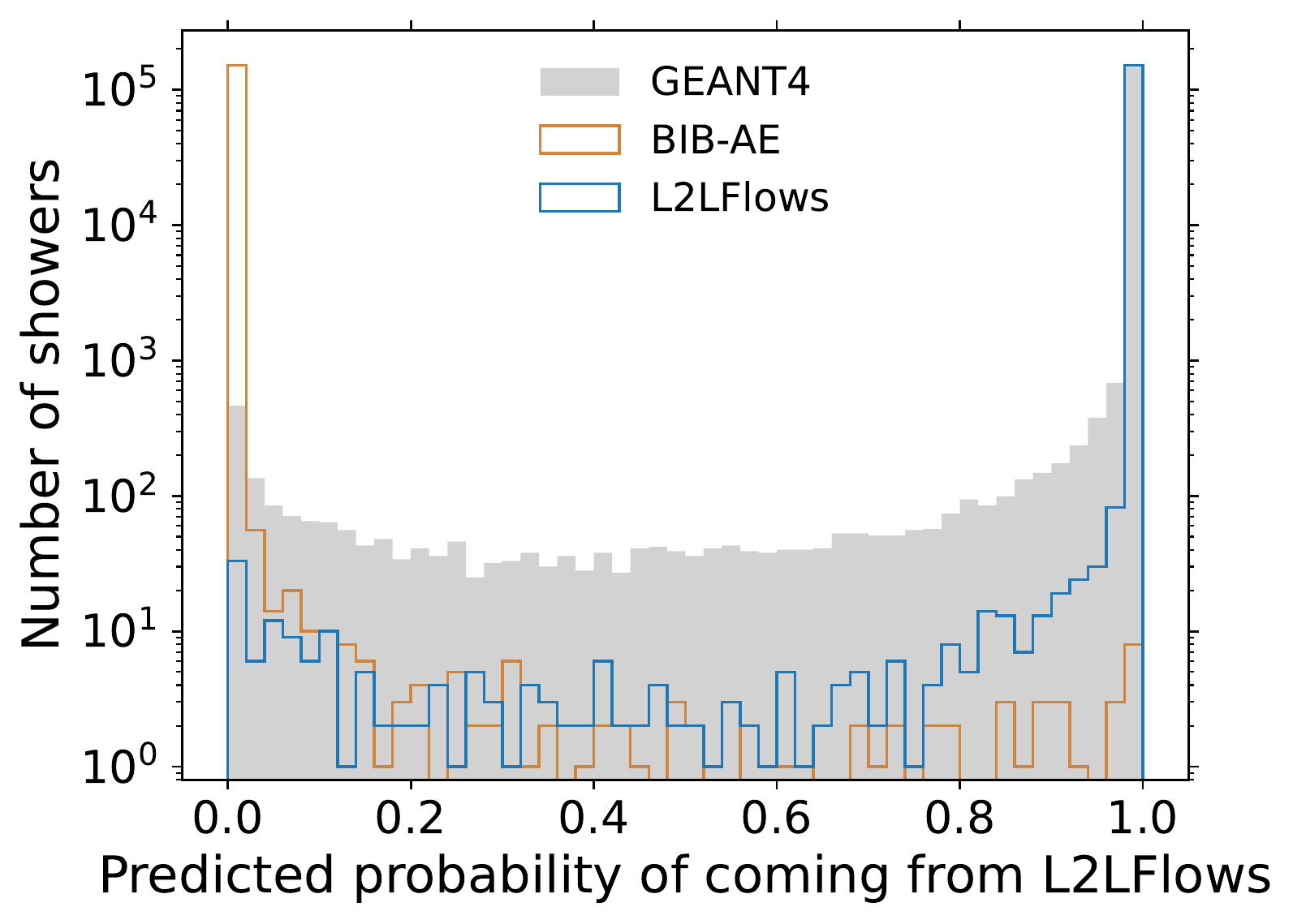}
    \caption{Predicted probability that showers come from \textsc{L$2$LFlows} for \textsc{Geant4} (grey), the BIB-AE (orange) and \textsc{L$2$LFlows} (blue) showers as input. Plot is shown for the full spectrum with $152$k showers for every simulator.}
    \label{results:multimodel_classif}
\end{figure}

For evaluation, we consider the test sets of the classifier containing $152$k showers for the BIB-AE and \textsc{L$2$LFlows} each and use $152$k \textsc{Geant4} test showers to compare the classifier outputs. The means of the output probabilities $p(\textsc{L$2$LFlows}|x)$ for $x$ coming from BIB-AE, \textsc{L$2$LFlows}, and \textsc{Geant4} are $0.03 \%$, $99.91 \%$, and $98.84 \%$ respectively. This indicates that \textsc{Geant4} and \textsc{L$2$LFlows}-generated showers are much closer to each other than \textsc{Geant4} and BIB-AE-generated showers are.  
To further visualize this result, we plot the predictions of the classifier on the test showers in Fig.~\ref{results:multimodel_classif}. This also shows us that \textsc{Geant4} showers are on average more likely to be identified as coming from \textsc{L$2$LFlows} than BIB-AE by the classifier. All of this strenghtens our conclusion that \textsc{L$2$LFlows} captures the underlying shower distribution of \textsc{Geant4} much better than the BIB-AE.

\subsection{Shower Generation Timings}
\label{sec:timing}

Table \ref{comput_times_bench} shows the mean sampling time per shower for \textsc{Geant4}, the BIB-AE and \textsc{L$2$LFlows}. 
For \textsc{L$2$LFlows}, the sampling times of the \flowOne\ are not accounted for, as they are negligibly small compared to the mean shower generation times of the \flowTwo. For \textsc{Geant4}, the same number as in Ref.~\cite{Buhmann:2020pmy} is taken, as cropping the dataset from $30\times 30\times 30$ to $30\times 10\times 10$ was done once the \textsc{Geant4} showers were simulated in the full ECal prototype.\footnote{Simulating the showers in $30\times10 \times 10$ would be unphysical, since this would not take into account backscattering for example. Also, we do not expect a large difference between generation timings for showers simulated in a $30\times 30\times 30$ cube or a $30\times 10\times 10$ cuboid, since we focused ourselves on the core of the showers, where most energy depositions happen.} We note that, in contrast to \textsc{Geant4}, shower generation times for \textsc{L$2$LFlows} and the BIB-AE do not depend on the incident energy. 
 
\begin{table}[h!]
    \sisetup{
    separate-uncertainty=true,
    table-format=5.3(5)
    }
    \centering 
	\begin{tabular}{c|c|r|S| l}
		\hline
		Simulator & Hardware & Batch size & {$10$ -- $100$ GeV [ms]} & 
            Speedup
		\\ \hline \hline \makecell{\textsc{Geant4} \\ ($30\times 30\times 30$)} & CPU & $/$ & 4081.53 \pm 169.92 & $/$ 
		\\ \hline \textsc{L$2$LFlows}           & CPU   & $1$ & 19617.24 \pm 894.08 & $\times 0.2$ 
		\\ 			($30\times 10\times 10$)	&     & $10$   & 3130.25 \pm 104.74 & $\times 1.3$ 
		\\                       &     & $100$  & 1395.52 \pm 26.55 & $\times 2.9$ 
		\\                       &     & $1000$ & 1338.13 \pm 24.03 & $\times 3.1$ 
		\\ \hline BIB-AE & CPU & $1$   & 102.25 \pm 0.64 & $\times 40$
		\\ 			($30\times 10\times 10$) &     & $10$   & 37.81 \pm 0.13 & $\bm{\times 110}$
		\\                       &     & $100$  & 48.51 \pm 0.01 & $\times 84$
		\\                       &     & $1000$ & 48.19 \pm 0.01 & $\times 85$
		\\ \hline \hline \textsc{L$2$LFlows}   & GPU & $1$ & 22560.34 \pm 263.00 & $\times 0.2$ 
		\\   ($30\times 10\times 10$) &     & $10$   & 2103.58 \pm 18.36 & $\times 1.9$ 
		\\             &     & $100$ & 
        213.38 \pm 0.23 & $\times 19$
		\\             &     & $1000$ & 
        23.14 \pm 0.16 & $\times 180$
		\\             &     & $2000$ & 13.70 \pm 0.03 & $\times 300$
		\\             &     & $8000$ & 
        9.61 \pm 0.01 & $\times 420$
		\\ & & $128000$ & 8.62 \pm 0.02 & $\times 470$
		\\ \hline BIB-AE & GPU & $1$ & 74.22 \pm 3.18 & $\times 55$ 
		\\ ($30\times 10\times 10$)	&     & $10$  & 6.85 \pm 0.25 & $\times 600$ 
		\\                       &     & $100$  & 0.91 \pm 0.02 & $\times 4500$
		\\                       &     & $1000$ &  0.249 \pm 0.002 & $\bm{\times 16000}$
		\\                       &     & $2000$ &  0.248 \pm 0.001 & $\bm{\times 16000}$
		\\ \hline 
	\end{tabular}
	\caption{For $25$ runs, the mean and the standard deviation of the sampling time per shower as well as the obtained speedup in comparison with \textsc{Geant4} are shown for different batch sizes and hardware during sampling for \textsc{Geant4}, the BIB-AE and \textsc{L$2$LFlows}. The GPU is an NVIDIA\textsuperscript{\textregistered} A100\textsuperscript{\textregistered} with $40$ GB VRAM. For the CPU, an Intel\textsuperscript{\textregistered} Xeon\textsuperscript{\textregistered} E5-2640 v4 was chosen, and the value for \textsc{Geant4} is taken from Ref.~\cite{Buhmann:2020pmy}, where the simulated showers have a shape of $30\times 30\times 30$.} 
	\label{comput_times_bench}
\end{table} 

Since the generation times of a MAF scale with the dimensionality $d$ of the input samples, one can expect the sampling times for \textsc{L$2$LFlows} to worsen by a factor of $9$ when going from the $30\times 10\times 10$ to the full $30\times 30\times 30$ data, while the \textsc{Geant4} run time would stay the same\footnote{For the BIB-AE, the mean sampling times on the full dataset can be found in Ref.~\cite{Buhmann:2020pmy}.}. The main bottleneck is not our autoregressive treatment of the ECal layers, but more the MAFs with which we model every single ECal layer. 

The speedups obtained on the cropped dataset are for \textsc{L$2$LFlows} up to a factor of $200$ slower than the BIB-AE (with a batch size of $1$ on the CPU), and in comparison to \textsc{Geant4}, \textsc{L$2$LFlows} is only a factor of $3$ faster on the CPU (with a batch size of $1000$). On the GPU, \textsc{L$2$LFlows} is about $470$ times faster than \textsc{Geant4} (with a batch size of $128000$), whereas the BIB-AE can obtain a speedup of about $16000$ (with a batch size of $2000$). Reference \cite{Krause:2021wez} also observed mean sampling times that were much slower than their GAN baseline network from Ref.~\cite{Paganini:2017hrr,Paganini_2018}, and to combat this, a MAF-IAF setup using probability density distillation, inspired by Ref.~\cite{https://doi.org/10.48550/arxiv.1711.10433}, was used. The obtained speedup was a factor of $\mathcal O(d)$, with a negligible loss in shower quality. Here, IAF refers to the inverse autoregressive flow~\cite{iaf_kingma}, an alternative architecture for autoregressive flows that we detail in App.~\ref{appendix:models}. 
Applying the same MAF-IAF concept to this work is an interesting future research direction; if it works, a speedup $\mathcal O(100)$ can be expected.  
This implies that \textsc{L$2$LFlows} has the potential to outperform the BIB-AE not just in the fidelity, but also in the speed with which the generated showers are obtained.


\section{Conclusions and Outlook}
\label{sec:conclusion}
This work built on Ref.~\cite{Krause:2021ilc} and demonstrated for the first time that NFs can be used to generate high-fidelity showers in a  
highly-granular sampling calorimeter. Showers were generated in a two-step approach, where the \flowOne\ first learned the energy depositions per ECal layer. Then, $30$ NFs (one per layer) --- which we dubbed \flowTwo\ --- were used to learn the voxel distributions, while being conditioned on the total deposited energy in that layer, the incident energy of the photons, and the voxel energies of the previous $5$ layers. 
The use of fully-connected embedding networks, which distill the conditioning features, cf.~Tab.~\ref{context_multiple_flows_n_5}, further reduces the number of parameters with no loss in performance. It was found that for all considered distributions in Sec.~\ref{subsec:distributions}, \textsc{L$2$LFlows} either outperforms the BIB-AE or is as good as it, with the exceptions of correlations, where the BIB-AE performs slightly better. 

Further,  \textsc{L$2$LFlows} has a much better AUC than the state-of-the-art network on the dataset --- the BIB-AE --- in the classifier tests. The classifiers used in this work took as input both \textsc{Geant4}-simulated and neural network-generated showers as well as the incident energies of the photons: The BIB-AE yields an AUC of $0.9947 \pm 0.0025$, whereas \textsc{L$2$LFlows} leads to an AUC of $0.8518 \pm 0.0042$. We also trained a classifier directly on \textsc{BIB-AE} and \textsc{L$2$LFlows} showers. As shown, when taking \textsc{Geant4} showers as input, such a classifier is much more likely to label it as an \textsc{L$2$LFlows} instead of a BIB-AE shower, further indicating the superior quality of the proposed approach. It was further shown that \textsc{L$2$LFlows} outperforms the BIB-AE in almost every considered physics distribution. 

The two-step approach, which was first introduced in Ref.~\cite{Krause:2021ilc}, can also be applied to other generative networks. For example, adding an \flowOne\ to the BIB-AE approach may also improve the fidelity of the generated showers there.

One bottleneck of the developed approach, however, is the required sampling time per shower. 
The problem is generally that a MAF is slow during generation, as it sequentially calculates each output dimension of a sample during generation. If the MAF-IAF approach from Ref.~\cite{Krause:2021wez} also succeeds in training the much faster IAF for \textsc{L$2$LFlows}, then a potential speedup factor of $\mathcal O(100)$ could be obtained. This would result in an NF architecture that could be used to sample faster than the BIB-AE.

Although the dataset has a uniform number of voxels in each layer, \textsc{L$2$LFlows} could straightforwardly generalize to non-uniform cases. In addition, splitting the learning of the shower shape into several NFs also has the advantage that the training can be parallelized on several GPUs. However, one pays a price that, even when employing an IAF setup, $30$ individual NFs evaluations are required. This limits the speedup of the proposed IAF setup to $\mathcal O(100)$, as opposed to the $\mathcal O(3000)$ speedup that could be achieved by a single-flow approach.

Further, we believe our NF architecture to have applications beyond the use in high-energy physics. \textsc{L$2$LFlows} could in principle also be studied for image or video generation. For example, Ref.~\cite{glow} uses an NF to generate high-fidelity images, yet for large images, a batch size of $1$ was used during training. To mitigate these memory constraints, it might be possible to use not only a single NF, but several of them, where each NF sees only a subset of the pixels, yet is conditioned on the previous pixels. The NFs could learn the full image in a top-bottom approach, where the first NF learns the first set of pixels, the second NF learns the second set while being conditioned on the first set, and so on. Image or video generation would then happen sequentially. To the best of our knowledge, such an approach has not been considered in the literature yet, and since each NF can be trained separately, a higher batch size can be chosen. This proposed approach is very similar to autoregressive models such as PixelRNN \cite{oord2016pixel}, PixelCNN \cite{oord2016conditional}, PixelCNN++ \cite{salimans2017pixelcnn} or PixelSNAIL \cite{chen2017pixelsnail}, but instead of generating the image pixel by pixel, chunks of pixels would be generated at once. 

As a proof of concept that NFs also scale to higher-dimensional datasets, this work cut down the $30\times 30\times 30$ projection to a $30\times 10\times 10$ projection. An extension of this work to the full projection is believed to be straightforward, as every NF would then have to learn a $900$- instead of a $100$-dimensional PDF, which should be feasible to tackle computationally with \textsc{L$2$LFlows}. In addition, this work can be extended by not just studying photon showers in the ILD ECal, but also pion showers in an HCal prototype for the ILD, which was done for the BIB-AE in Ref.~\cite{Buhmann:2021caf}, where the HCal was projected to a cuboid of size $48\times 25\times 25$. With \textsc{L$2$LFlows}, this would result in $48$ NFs, where each NF has to learn a $625$-dimensional distribution. 

It is also important to perform angular conditioning studies in the future, as the dataset used in this work shot the photons perpendicularly into the ECal. And just as Ref.~\cite{Buhmann:2021caf} considered the output of state-of-the-art reconstruction algorithms on the output of the neural network--generated showers, it would be interesting to do the same once \textsc{L$2$LFlows} has been extended to the full $30\times 30\times 30$ dataset of the ILD ECal prototype. Last but not least, \textsc{L$2$LFlows} can be studied for the three different datasets from the CaloChallenge~\cite{calo_challenge}.

\begin{center} \textbf{Acknowledgments} \end{center}

\noindent The authors would like to thank Katja Kr\"uger for valuable feedback on the draft of this paper. SD is funded by the Deutsche Forschungsgemeinschaft under Germany’s Excellence Strategy – EXC 2121  Quantum Universe – 390833306. The work of CK and DS was supported by DOE grant DOE-SC0010008. This research was supported in part through the Maxwell computational resources operated at Deutsches Elektronen-Synchrotron DESY, Hamburg, Germany. CK would like to thank the Baden-W\"urttemberg-Stiftung for support through the program \textit{Internationale Spitzenforschung}, project \textsl{Uncertainties --- Teaching AI its Limits} (BWST\_IF2020-010). 

 
\appendix

\section{Model Details}\label{appendix:models}

\textsc{L$2$LFlows} uses MAFs~\cite{2017arXiv170507057P} as generative models. A MAF is a bijective function, connecting the data distribution on one side with a Gaussian latent distribution on the other side. The transformation is given by a set of rational quadratic splines~\cite{NEURIPS2019_7ac71d43,10.1093/imanum/2.2.123} for each dimension. The parameters of the splines (like location of bin edges or derivative values at the knots) are given by an autoregressive neural network, ensuring that the parameters of the transformation of $x_i$ only depend on $x_{<i}$. In practice, such an autoregressive structure can be realized by masking the network weights with zeros in appropriate places, as done in the MADE block~\cite{2015arXiv150203509G}. A single pass through the network will then give the full set of autoregressive parameters needed for the entire transformation. The parameters for the inverse transformation are, however, harder to obtain. Passing random input through the MADE block will only give the correct parameters of the transformation of the first coordinate, since these do not depend on any other $x_i$. Using these parameters to correctly invert $x_i$, we can get the correct parameters to invert $x_2$ after a second pass through the MADE block. In total we therefore need $d$ passes through the MADE block to fully construct the inverse transformation. The MAF now uses the fast pass to compute the log-likelihood, allowing for a fast training at the price of slow sampling. The IAF would allow for faster sampling, but only at the expense of much slower (or even impossible because of memory constraints) training. More details about MAFs and IAFs and their use for calorimeter simulations can be found in~\cite{Krause:2021wez}.

The details of the \flowOne\ are summarized in the left column of Tab.~\ref{tab:hyperparams}. The MADE blocks make use of fully-connected layers, and their total number per MADE block is given by the input layer, the number of hidden layers and the output layer. 
The hidden layers inside the MADE blocks make use of the ReLU activation function. The minimum bin width and height refer to the minimum values of an RQS bin, and the minimum derivative to the minimum derivative value at the knots of a bin; and the permutation hyperparameter refers to the permutation of the input features before they are passed to the MADE block. Since the transformations parameterized by each MADE block are autoregressive in nature, the permutation layers help to increase the expressivity of the normalizing flow. Here we differentiate between ``random'', which denotes a randomly determined permutation, and ``reverse'',  where the permutation inverts the ordering of the data dimensions. 
During generation, double precision parameters are used, as we found that using only float precision parameters leads to numerical instabilities, which arose when the RQS solved a quadratic equation for the inverse~\cite{10.1093/imanum/2.2.123,NEURIPS2019_7ac71d43}. During training, however, there is almost no advantage of using double precision, as we checked in an ablation study. Since up to $50 \%$ of the memory can be saved that way, we use single float precision during training. 

The hyperparameters of the \flowTwo\ can be found in the right column of Tab.~\ref{tab:hyperparams}. Just as for the \flowOne, an ablation study showed barely any advantage of double precision parameters for training, hence we use only float precision (generation still happens with double precision parameters).  

\begin{table}[h!]
	\centering 
	\begin{tabular}{c|c|c} 
		\hline 
		\multirow{2}{*}{\textbf{Hyperparameter}} & \multicolumn{2}{c}{\textbf{Value}} \\
  & \;\flowOne\; & \; \flowTwo \; \\
		\hline \hline
		Learning rate & $6\cdot 10^{-5}$ & $6\cdot 10^{-4}$  \\ \hline 
            Optimizer & ADAM & ADAM\\ \hline
		Batch size & $256$ & $1024$ \\ \hline 
		$\#$ Epochs & $200$ & $200$ \\ \hline 
		$\#$ MADE blocks & $4$ & $4$\\ \hline 
		$\#$ Hidden layers & $1$ & $1$ \\ \hline
		$\#$ Hidden nodes & $64$ & $128$ \\ \hline 
		$\#$ RQS bins & $8$ & $8$ \\ \hline 
		Min. bin width/height & $10^{-6}$ & $10^{-6}$ \\ \hline 
		Min. derivate & $10^{-6}$ & $10^{-6}$ \\ \hline 
		Cutoff value $\alpha$ & $10^{-6}$ & $10^{-6}$ \\ \hline  
		Permutations & \enquote{random} & \enquote{reverse}\\ \hline 
            Dtype training & float32 & float32\\ \hline 
            Dtype generation & float64 & float64\\ \hline 
            Noise & Gaussian & Uniform \\ & ($\mu = 1$ keV, $\sigma = 0.2$ keV) &  ($a = 0$ keV, $b = 1$ keV) \\ \hline 
            Base distribution & $30$-dim. multivar. Gaussian & $100$-dim. multivar. Gaussian \\ & ($\mu = 0$, $\Sigma = \text{Id}$) &  ($\mu = 0$, $\Sigma = \text{Id}$) \\ \hline 
	\end{tabular}
	\caption{Most important hyperparameters for the training of \textsc{L$2$LFlows}. For the \flowTwo, the hyperparameters are identical for every NF $i$ learning layer $i$. The noise hyperparameter refers to the elementwise noise that is added during training. 
 }
    \label{tab:hyperparams}
\end{table}

\section{Pre- and Postprocessing}\label{new_pp_appendix}
\subsection{Preprocessing \flowOne\ }

The training input of the \flowOne\ consists of the energies per layer $E_{i}$. These energies are preprocessed before they are passed to the \flowOne. In the first step, the
layer energies $E_{i}$ are smeared using an additive Gaussian noise term, with mean $\mu = 1$ keV and standard deviation $\sigma = 0.2$ keV\footnote{We clip all negative values to $0$.}. This helps the \flowOne\ learn the marginal distributions of $E_{i}$ and results in a noticeable performance increase compared to training without the added noise.

The smeared energies per layer $E_{i}$, are then further processed using
\begin{align}
    E_{i}^{\text{proc}} :=\ell_{\alpha}(E_{i}) \ \forall i\in \{0, \dots, 29\},
\end{align}
where the smeared $E_{i}$ is given in units of GeV and $\ell_{a}$ describes a modified logit function~\cite{2017arXiv170507057P}: 
\begin{align}\label{logit_trafo}
    \ell_{\alpha}(x)= \ln\frac{\alpha + (1 - 2\alpha)x}{1 - \alpha - (1 - 2\alpha) x}, 
\end{align} 
with the scalable hyperparameter $\alpha$, chosen to be $\alpha = 10^{-6}$. Note that the logit transformation is only defined for $x\in [0,1]$, however, as the energies per calorimeter layer in the ILD ECAL do not exceed $1$ GeV in the photon shower data set, this requirement is inherently fulfilled. 

The \flowOne\ is conditioned on the energy of the incident particle $E_{\text{inc}}$, which is processed according to
\begin{align}\label{incid_energies_cond} 
    E_{\text{inc}}^{\text{proc}} := 2 \cdot \log_{10}\left(E_{\text{inc}}\right) - 3 \in [-1, 1], 
\end{align}
where $E_{\text{inc}}$ is given in units of GeV. 

\subsection{Preprocessing \flowTwo}

The inputs to the individual NFs of the \flowTwo\ consist of the energy depositions in the voxels of a layer, $\mathcal I_{j}$, with $j \in [0, 29]$, and $\mathcal I_{j} \in \mathbb R^{100}$. Before being passed to the NFs, the inputs are preprocessed. The first step consists of adding noise (sampled uniformly from $[0,1]^{100}$~keV) to each $\mathcal I_{j}$. The resulting values are then normalized and logit-transformed in accordance with
\begin{align}
    \mathcal I_{j}^{\text{logit}} := \ell_{\alpha}\left(
        \frac{\mathcal I_{j}}{\text{max}}
    \right)\forall j,
\end{align}
where $\text{max}$ denotes the maximum voxel energy taken over all training data  
and $\ell_{\alpha}$ is the logit transformation defined in Eq.~\eqref{logit_trafo}. 

As described in Sec.~\ref{subsec:multiple_flow}, NF $i$ is conditioned on the voxel energies of the previous $5$ layers, as well as on the energy in their respective layer $E_{i}$ and the energy of the incident particle $E_{\text{inc}}$. The preprocessing of $E_{\text{inc}}$ is identical to what is given by Eq.~\eqref{incid_energies_cond}. During training, the layer energies $E_{i}$ are derived from the voxel energies in the $i$-th layer $\mathcal I_{i}$ using
\begin{align}\label{eq:Ei_def}
    E_{i} := \left[\sum_{i} (\mathcal I_{i}^{\text{cut}} + \text{noise})\right] \;, 
\end{align} 
where $\mathcal I_{i}^{\text{cut}}$ are the voxel energies after the threshold cut, defined by
\begin{align}
    \mathcal I_{i}^{\text{cut}} := (\mathcal I_{i} \geq \text{threshold}) \; , 
\end{align} 
and the noise in Eq.~\eqref{eq:Ei_def} refers to the Gaussian noise added during training, cf.~Tab.~\ref{tab:hyperparams}. 

The $E_{i}$ are then further processed according to
\begin{align}\label{eq:Ei_preproc}
    E_{i}^{\text{proc}} := \log_{10}\left(E_{i} + \epsilon \right) + 1 \; ,
\end{align} 
with $\epsilon = 10^{-6}$. During generation, the $E_{i}$ come from the \flowOne\ and they are also processed according to Eq.~\eqref{eq:Ei_preproc}.

\subsection{Postprocessing \flowTwo}

The postprocessing used in Ref.~\cite{Krause:2021ilc} ensures via a renormalization of the $\mathcal I_i$ that the energies per layer of the returned showers are approximately  those that the \flowOne\ dictates. Usually, one is interested in showers that have an energy threshold applied, since the inherent electronic noise of the detector implies that too small energies cannot be converted into a signal that can be read out. This thresholding reduces the deposited energy per layer below the energy dictated by the \flowOne. However, the study of Ref.~\cite{Krause:2021ilc} is based on the dataset from Ref.~\cite{Paganini_2018}, where the energies both from the passive absorber layers and the active detector layers are assumed to be available, and applying an energy threshold on the generated showers barely made a difference in the energies per layer. 

This work uses a realistic sampling calorimeter, where the energies from the passive absorber layers are unavailable. For this reason, the voxel depositions are much smaller, and the cut makes a non-negligible difference. Hence, a new postprocessing was needed for this work. In a nutshell, the new postprocessing checks how many of the dimmest voxels need to be set to zero such that the renormalized remaining voxels are all above the threshold.

\begin{figure}[h!]
    \begin{minipage}[c]{0.495\textwidth}
        \centering
        \includegraphics[width=\textwidth]{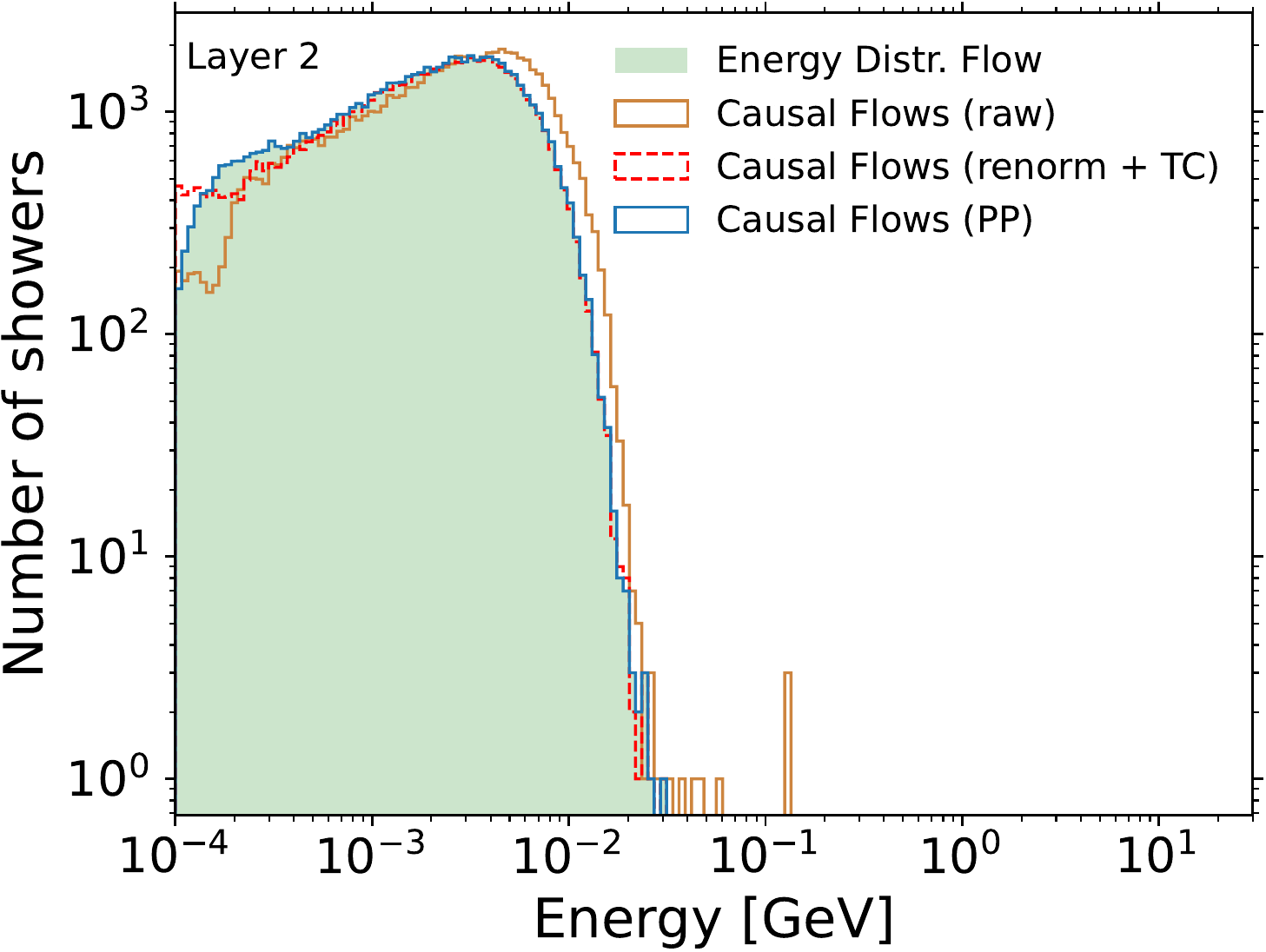}
    \end{minipage}
    \begin{minipage}[c]{0.495\textwidth}
        \centering
        \includegraphics[width=\textwidth]{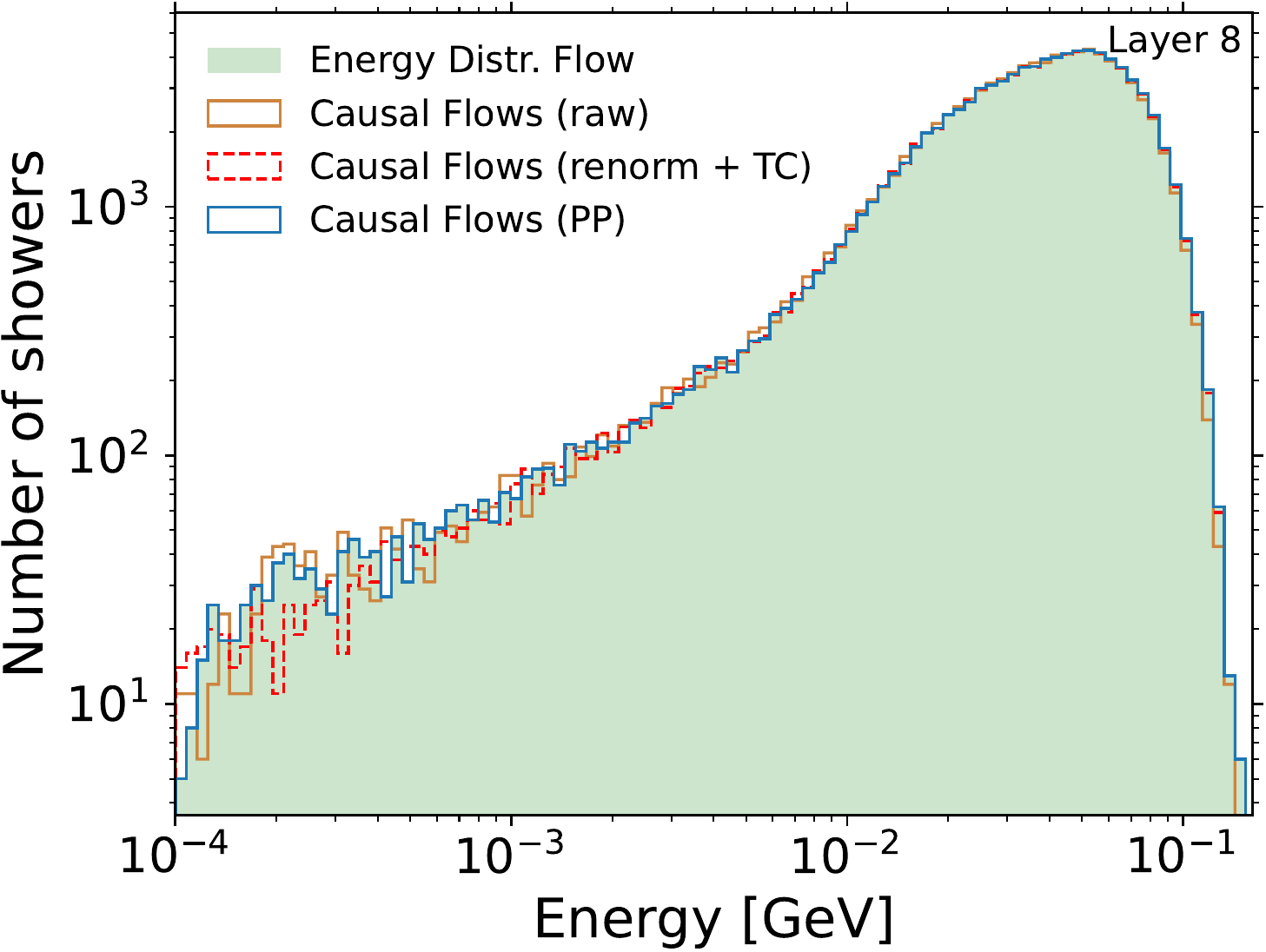}
    \end{minipage}\vspace{0.4cm}
    \begin{minipage}[c]{0.495\textwidth}
        \centering
        \includegraphics[width=\textwidth]{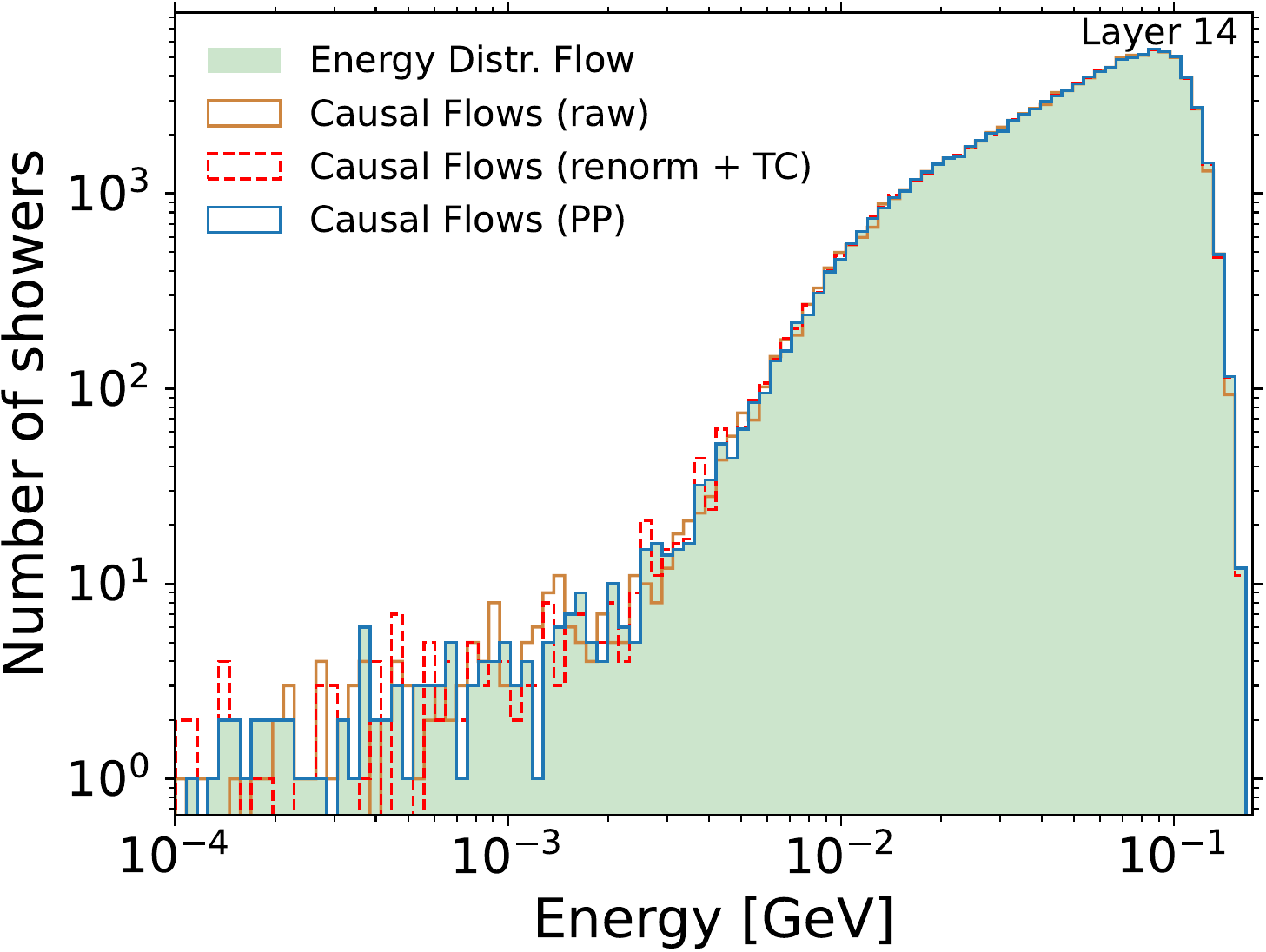}
    \end{minipage}
    \begin{minipage}[c]{0.495\textwidth}
        \centering
        \includegraphics[width=\textwidth]{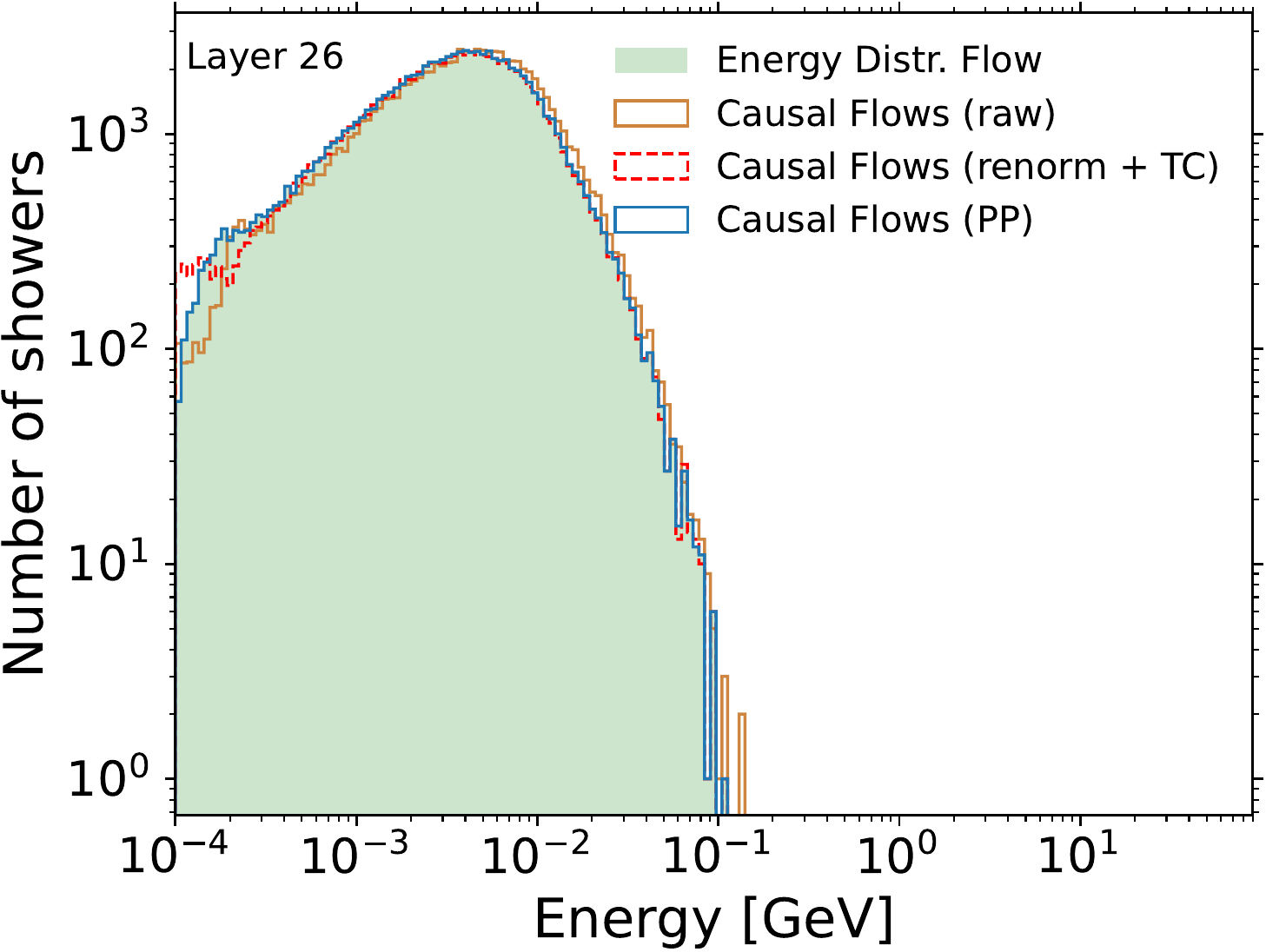}
    \end{minipage}
    \caption{Effect of different postprocessing steps on the distribution of $E_i$ in different layers. Raw showers without a threshold cut (\enquote{\flowTwo\ (raw)}) are smeared around the $E_i$ as given by the \flowOne. The CaloFlow postprocessing of renormalization and a subsequent threshold cut (\enquote{\flowTwo\ (renorm + TC)}) tends to under- or overshoot the $E_i$ of the \flowOne\ around half the MIP cutoff of $10^{-4}$ GeV in the initial and final layers of the ECal. Our postprocessing ensures that the distribution of $E_i$ exactly follows the one of the \flowOne.}
    \label{fig:app:pp}
\end{figure}

We define $S(\mathcal{I}_{i}) := \sum_i \mathcal{I}_{i}$, the sum over voxels $\mathcal{I}_{i}$ in layer $i$ and $\mathcal{I}_{i, \geq t}$ as the voxels $\mathcal{I}_{i}$ thresholded by $t$, i.e.~all voxels less than $t$ in layer $i$ are set to zero. The postprocessed voxels are then given by 
\begin{equation}
 \mathcal{I}_{i}^{\text{pp}} := E_i \cdot \mathcal{I}_{i, \geq t} / S(\mathcal{I}_{i, \geq t}),  
\end{equation}
where $t$ is set by requiring 
\begin{equation}
 E_i \cdot t / S(\mathcal{I}_{i, \geq t})  = \text{desired threshold}.
\end{equation}
As a result, all generated voxel energies of layer $i$ sum to $E_i$ \textit{after} the threshold cut is applied. Figure~\ref{fig:app:pp} illustrates the effect of the postprocessing. There, we see the distribution of energies in layers $2$, $8$, $14$, and $26$ as given by different algorithms: In light green, we see the distributions as given by the \flowOne; in orange, the distributions of raw, generated showers without a threshold cut (which only scatters around the $E_i$ of the \flowOne, even though it was conditioned on it); in red, the distributions of raw showers after the CaloFLow postprocessing (renormalization and subsequent threshold cut); and in blue, the distributions of the same showers using our postprocessing. We clearly see that a simple application of the threshold cut after the renormalization distorts the distribution towards lower energies. This effect is stronger in the outer layers of the calorimeter, where the overall scale of energy depositions is smaller. 

\section{Classifier Tests: Architectures and Details}\label{appendix:classifier_tests}

Both the \textsc{Geant4} vs BIB-AE as well as the \textsc{Geant4} vs \textsc{L$2$LFlows} classifiers are fully-connected neural networks with the same architecture and hyperparameters: They consist of four hidden layers with $4096$, $512$, $64$ and $8$ nodes with the LeakyReLU activation function, using a slope of $0.01$ for input that is smaller than $0$. The output layer has $2$ nodes, and we use the cross entropy loss. Each output node can be interpreted as the likelihood of a given sample belonging to \textsc{Geant4} or the BIB-AE/\textsc{L$2$LFlows}. Therefore, the likelihood ratio is available in the binary classification setup. In total, every classifier has $14.4$M parameters. The classifier is trained on input with double precision. The learning rate is set to $10^{-4}$, the batch size to $256$ and both classifiers are trained for in total $50$ epochs. The final model is chosen that has the highest validation accuracy. 

Unlike Ref.~\cite{Krause:2021ilc}, which considers the ECal voxel energies, (log-transformed) deposited energies per ECal layer as well as the (log-transformed) incident energy as input to the classifiers, we believe the use of the deposited energies per layer to be redundant, as they are already encoded in the ECal voxel energies. Thus, we make use of $3001$ input features, where the incident energies are log-transformed as in Eq.~\eqref{incid_energies_cond}. The voxel energies have half the MIP cutoff applied. 

The only difference between the \textsc{Geant4} vs BIB-AE/\textsc{L$2$LFlows} and \textsc{BIB-AE} vs \textsc{L$2$LFlows} classifier is that for convergence reasons of the validation accuracy, the latter is trained for $100$ epochs. 

\bibliographystyle{JHEP}
\bibliography{literature}
\end{document}